\documentclass[a4paper,11pt]{article}

\usepackage{jheppub} 
\usepackage{graphicx}
\usepackage{bm}
\usepackage{bbold}
\usepackage{amssymb,amsmath,mathrsfs}
\usepackage{cancel}
\usepackage{slashed}
\usepackage{hyperref}
\usepackage{verbatim}
\usepackage{xspace}
\usepackage{tcolorbox}
\usepackage[export]{adjustbox}

\newcommand{\TeV}{\mathrm{TeV}}
\newcommand{\GeV}{\mathrm{GeV}}
\newcommand{\MeV}{\mathrm{MeV}}

\newcommand{\beq}{\begin{equation}}
\newcommand{\eeq}{\end{equation}}
\newcommand{\br}{\mathrm{BR}}

\newcommand{\Npot}{N_{\mathrm{POT}}}
\newcommand{\Nevt}{N_{\mathrm{evt}}}

\newcommand{\gag}{g_{a\gamma}}
\newcommand{\cGG}{c_{GG}}
\newcommand{\pythia}{\texttt{PYTHIA}\xspace}

\DeclareMathOperator\tr{tr}

\newcommand{\vev}[1]{\langle #1 \rangle}
\newcommand\UVIC{Department of Physics and Astronomy, University of Victoria, Victoria, BC V8P 5C2, Canada}

\newcommand\FNAL{Fermi National Accelerator Laboratory, Batavia, IL 60510, USA}
\newcommand\MSU{Department of Physics and Astronomy,
Michigan State University, East Lansing, Michigan 48824, USA}
\newcommand\UM{Physics Department, University of Michigan, Ann Arbor, MI 48109 USA}

\begin{document}
%\title{Axion-like Particle Production in Proton Beam-dumps}
\title{Axion-like Particle Searches at DarkQuest}
\author[a,b]{Nikita Blinov}
\emailAdd{nblinov@uvic.ca}
\author[b,c]{, Elizabeth Kowalczyk}
\emailAdd{kowalc40@msu.edu}
\author[b,d]{, Margaret Wynne}
\emailAdd{megwynne@umich.edu}
\preprint{FERMILAB-PUB-21-749-V}

\affiliation[a]{\UVIC}
\affiliation[b]{\FNAL}
\affiliation[c]{\MSU}
\affiliation[d]{\UM}

\abstract{
    Axion-like particles (ALPs) interacting with the Standard Model can be abundantly produced in proton beam fixed-target experiments. Looking for their displaced decays is therefore an effective search strategy for ALPs with a mass in the MeV to GeV range. Focusing on the benchmark models where the ALP interacts dominantly with photons or gluons, we show that the proposed DarkQuest experiment at Fermilab will be able to test parameter space which has been previously inaccessible. We pay particular attention to the self-consistency of gluon-coupled ALP production and decay calculations, which has been recently shown to be a problem in many existing predictions.
    We also apply these results to explore existing constraints in the ALP parameter space.
}

\maketitle

\section{Introduction}

Axions and axion-like particles (ALPs) are a generic feature of many 
theories of beyond-Standard Model (BSM) physics. Originally introduced to 
resolve the strong CP problem~\cite{Peccei:1977hh,Peccei:1977ur,Wilczek:1977pj,Weinberg:1977ma}, axions occur as pseudo-Nambu-Goldstone bosons (pNGBs) of spontaneously-broken global symmetries, or as zero-modes of higher-dimensional 
gauge fields~\cite{Arias:2012az,Svrcek:2006yi,Arvanitaki:2009fg,Cicoli:2012sz}. The pNGB nature of axions ensures that they are technically natural, even if the 
physics associated with the global symmetry lies at a very high scale. 
As a result, these particles can be the first messengers of the ultraviolet (UV) that can be accessible 
at experiments. 

The nature of UV physics (i.e., the gauge charges and couplings of the 
heavy BSM states) dictates the couplings of the ALP to SM particles. For example, 
the defining coupling of the QCD axion is its interaction with gluons which is generated by 
new particles with strong interactions. Depending on other charges of these heavy states, couplings to other SM gauge bosons are induced. In this minimal model the axion obtains a mass from the QCD condensate, and, as a result, its mass and coupling are fixed by a single parameter. Axion-\emph{like} particles are a generalization of such a scenario where the mass and coupling can be varied individually. 

The shift symmetry of axions also ensures that they interact with SM particles 
through dimension five operators suppressed by the high scale associated with the 
UV physics. 
Depending on the size of this scale and the mass of the ALPs, they can provide 
a cosmologically stable dark matter candidate~\cite{Abbott:1982af,Dine:1982ah,Preskill:1982cy,Arias:2012az}. 
Alternatively, if their lifetime 
is short on cosmological time scales, one can produce and detect them 
in astrophysical systems or in terrestrial experiments. A wide range of observations 
has been carried out across decades in mass and coupling resulting in stringent constraints 
on the axion and axion-like particle parameter space -- see, e.g., Refs.~\cite{Ertas:2020xcc,Agrawal:2021dbo}. 
In this paper we study the sensitivity of the proposed high-intensity proton beam-dump experiment DarkQuest at Fermilab and show that it can probe previously inaccessible parameter space. Many other dark sector models have also been studied in the context of DarkQuest -- see, e.g.,  Refs.~\cite{Gardner:2015wea,Berlin:2018tvf,Berlin:2018pwi,Choi:2019pos,Dobrich:2019dxc,Tsai:2019buq,Darme:2020ral,Batell:2020vqn}.

Several collider and fixed target experiments with different beams and targets have searched for ALPs over the last 30+ years. 
The non-renormalizable nature of the ALP coupling 
naturally leads to long lifetimes on experimental length scales. 
As a result, if these particles are produced they can travel a macroscopic distance before decaying, giving rise to a 
displaced energy deposition in a detector downstream of the target. 
While certain regions of parameter space have been excluded by searching for this signature, 
large swaths remain untested. These are typically associated either with masses that are kinematically inaccessible, or with lifetimes that lead to decays in the shielding. The DarkQuest experiment provides a unique opportunity to access precisely these regions of parameter space with a compact set-up and a high-energy proton beam. While the proton beam also enables the study of many different couplings of ALPs, in this work we focus on the their interactions with gluons or photons. 

Even though the hadronically-coupled ALP is an extensively studied model, its interactions at low energies (below the QCD phase transition) are 
still not fully understood due to the interplay with non-perturbative QCD dynamics. In particular, Refs.~\cite{Bauer:2020jbp,Bauer:2021wjo} have recently pointed out that many existing calculations of ALP reaction rates miss important contributions that are needed for theoretical consistency (to be made precise in the following sections). For some processes, this leads to rate changes of more than an order of magnitude~\cite{Bauer:2021wjo}. Therefore, in our calculations we ensure that these theoretical consistency conditions are satisfied. This amounts to tracking the cancellation of unphysical parameters in physical amplitudes describing ALP production and decay.

This paper is organized as follows. In Sec.~\ref{sec:dq_overview} we 
describe the proposed DarkQuest experiment and discuss potential backgrounds 
for the search for ALP decays to photons $a\to \gamma\gamma$ (this decay channel is generally present both in the dominantly photon- and gluon-coupled models). We then discuss ALP interactions below the QCD phase transition in Sec.~\ref{sec:alp_interactions}. Following Refs.~\cite{Bauer:2020jbp,Bauer:2021wjo} we discuss the cancellation of unphysical parameters in chiral perturbation theory, extending some of their findings to three flavours and other interactions. We then use these results to study ALP production channels in proton beam-dump experiments in Sec.~\ref{sec:alp_production_from_photons} for photon-coupled ALPs and in Sec.~\ref{sec:alp_production_from_gluons} for gluon-coupled ALPs.\footnote{The photon case has already been studied in Ref.~\cite{Berlin:2018pwi}; here we perform an extended analysis but find quantitatively similar results.}
We apply these calculations in Sec.~\ref{sec:sensitivity} to project the sensitivity of DarkQuest and to study existing constraints. 
Our main results for the photon coupling are summarized in Fig.~\ref{fig:dq_reach_and_other_projections} and for the gluon coupling in Fig.~\ref{fig:dq_gluon_reach}. Some constraints on the gluon-coupled model are re-evaluated in Fig.~\ref{fig:gluon_coupled_alp_constraints_only}.
We conclude in Sec.~\ref{sec:conclusion}.

\section{DarkQuest Experiment}
\label{sec:dq_overview}
\begin{figure}
    \centering
    \includegraphics[width=0.7\textwidth]{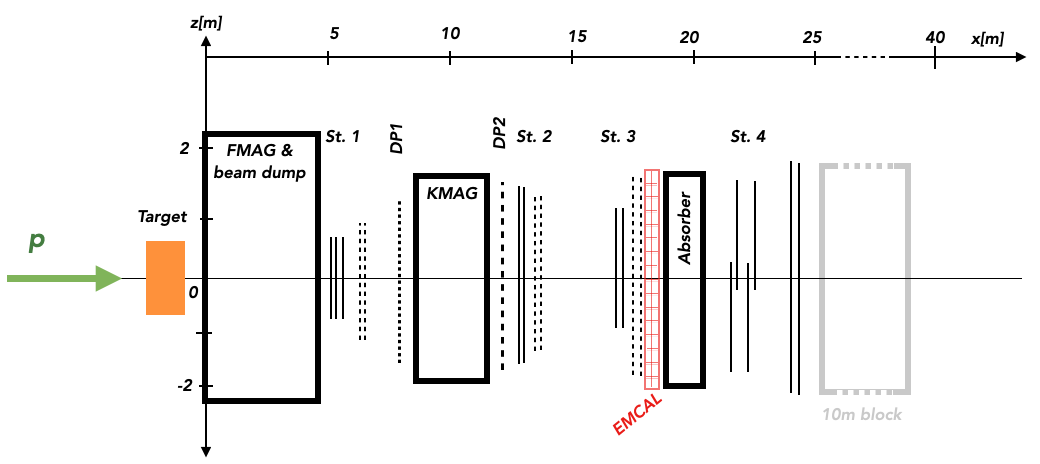}
    \caption{Layout of the SpinQuest experiment along with the additional 
    ECAL module at $z\approx 19$ m for the proposed DarkQuest upgrade. Graphic adapted from the \href{https://www.snowmass21.org/docs/files/summaries/RF/SNOWMASS21-RF6_RF0_Nhan_Tran-025.pdf}{DarkQuest SNOWMASS LOI}.}
    \label{fig:dq_setup}
\end{figure}

DarkQuest is a proposed upgrade of the existing SeaQuest/SpinQuest experiment 
at Fermilab~\cite{SeaQuest:2017kjt}, which collides a 120 GeV proton beam with a thin target. The set-up 
of the experiment is shown in Fig.~\ref{fig:dq_setup}. The current 
spectrometer is optimized for measuring energetic muons emanating from the target. 
The target is followed by a 5 m magnetized iron block, the FMAG, which serves as a beam dump and also sweeps soft charged particles out of the detector acceptance. 
The FMAG is followed by a series of detector subsystems for tracking and analyzing charged particles; we will focus on ALP decays to photons and therefore will not make use of these.

The DarkQuest upgrade will add an ECAL at $z\approx 19$ m and improve data acquisition systems, 
enabling sensitivity to electrons and photons. This version of the detector will make possible a multitude of searches for various dark sector particles~\cite{Gardner:2015wea,Berlin:2018tvf,Berlin:2018pwi,Choi:2019pos,Dobrich:2019dxc,Tsai:2019buq,Darme:2020ral,Batell:2020vqn}. The high beam energy and relatively compact geometry of the dump make this an ideal experiment to 
study production and decay of long-lived particles. In addition, the ECAL modules will be repurposed from the PHENIX experiment~\cite{Aphecetche:2003zr} allowing for rapid and cost-effective construction with a physics run possible as early as  2023. The Fermilab accelerator complex can provide $10^{18}$ protons on target ($\Npot$) in about two years of running. 
We will refer to this benchmark as phase 1 of DarkQuest. 
Future upgrades of the complex in the on-going Proton Improvement Plan will enable even more intense beams~\cite{Shiltsev:2017mle}. We assess the gains from increased luminosity by considering a second phase with $\Npot = 10^{20}$. 

Unlike searches for long-lived particle decays to charged particles, photon final states are subject to additional backgrounds which we describe below. These necessitate additional shielding after FMAG, or preshower detectors to enable tracking. 

\subsection{Backgrounds}
\label{sec:bg}
Certain SM processes can mimic the displaced decay signal of ALPs. 
Below we consider two such processes: production of short-lived particles in the back of the dump and decays of SM long-lived particles. 
    
Proton-nucleus interactions that take place in the back of the beam dump can produce particles that make it out to the decay volume. Production of $\pi$ and $\eta^{(\prime)}$ mesons and their decay into $\gamma\gamma$ can therefore potentially mimic our signal if it occurs deep enough in FMAG. 
The number of mesons produced in the last interaction length of the $5$ m iron FMAG is
\beq
N_{M} \approx 3\times 10^{-13} n_M \Npot, 
\eeq
where the first factor is the attenuation from requiring the beam particle to make it this far in the iron without an inelastic scattering, $n_M$ 
is the number of mesons produced per interaction ($n_{\pi^0} \approx 2.5$ 
and $n_\eta \approx 0.3$). The photons from the decays of these mesons
have a high acceptance rate and would correspond to $\gtrsim 10^5 \; (10^7)$ background events in phase 1 (2). These backgrounds can be mitigated 
by placing an additional $14-17$ interaction lengths of shielding before 
or after FMAG. Alternatively, if the segmentation of the ECAL 
allows a rough vertexing of the decay photons, events that originate 
from the end of FMAG can be rejected. Following Ref.~\cite{Berlin:2018pwi} 
we will use $z\gtrsim 7$ m as our fiducial decay region, allowing 
up to two meters after FMAG for additional shielding or vertexing. 
    
Standard model long-lived particles can also mimic the ALP displaced decay signal. In particular, neutral kaons $K_{S,L}$ and 
the $\Lambda$ baryon can have lab-frame lifetimes comparable to the 
size of the DarkQuest experiment. These hadrons have decay modes into 
neutral final states that include $\pi_0$'s which can produce photons, e.g.,  $K_S\rightarrow \pi^0\pi^0$, 
$K_L\to \pi^0\pi^0\pi^0$ and $\Lambda \to n \pi^0$. We use \pythia to 
estimate the multiplicity $n_H$ and typical boosts $\langle\gamma_H\rangle$ of $H=K^0$ and $\Lambda$. The multiplicities are $n_K \approx 0.36$ (this number 
includes both $K^0$ and $\overline{K}^0$ and is consistent 
with measurements of the combined $K^\pm$ multiplicity at NA49~\cite{Anticic:2010yg}), $n_\Lambda \approx 0.13$ (consistent with the recent measurements from NA61/SHINE~\cite{Aduszkiewicz:2015dmr}), while for mean boosts we find $\langle\gamma_K\rangle \approx 18$ and $\langle\gamma_\Lambda\rangle \approx 17$. We use these to estimate the 
number of hadrons that traverse the FMAG and decay into final states containing photons beyond:
\beq
N_H \sim \Npot \times \begin{cases}
10^{-19} & H = K_S\\
10^{-15} & H = K_L \\
10^{-16} & H = \Lambda
\end{cases}.
\eeq
In computing the above we assumed that the hadrons are produced in the 
first interaction length of FMAG, and accounted for the attenuation of 
their flux from inelastic interactions and decays in the dump. 
We see that displaced $K_L$ and $\Lambda$ decays can produce significant 
backgrounds for the ALP search. As above, the simplest solution is 
additional shielding beyond FMAG. The same amount of shielding would 
reduce the rates for these displaced decays to negligible levels in 
phase 1 and 2. Additionally event selections based on the number of photons 
detected in the ECAL can be used to eliminate processes with multiple $\pi^0$'s such as $K_L\to \pi^0\pi^0\pi^0$.

\section{ALP Interactions}
\label{sec:alp_interactions}
The general ALP Lagrangian depends on the ultraviolet (UV) completion of the model~\cite{Bauer:2020jbp,Bauer:2021wjo}; it contains direct couplings to quarks, leptons and gauge bosons. Moreover, even if the UV completion only generates one of these couplings at a high scale, renormalization group (RG) evolution will generate a myriad of other interactions. 

In this work we focus on two simplified ALP models in which the dominant coupling just above the QCD scale is either to photons or to gluons.\footnote{This simplifying assumption ensures that we do not generate other important interactions through RG evolution. Such couplings can induce many additional signals and constraints -- see Ref.~\cite{Bauer:2021mvw}.} 
Following the notation of Ref.~\cite{Bauer:2020jbp} we write the ALP Lagrangian as 
\beq
\mathscr{L} \supset \frac{1}{2} (\partial a)^2 - \frac{1}{2}m^2_{a,0} a^2 + c_{\gamma\gamma}\frac{\alpha}{4\pi} \frac{a}{f}F_{\mu\nu} \widetilde{F}^{\mu\nu} + \cGG\frac{\alpha_s}{4\pi} \frac{a}{f} G_{\mu\nu}^a \widetilde{G}^{a,\,\mu\nu},
\label{eq:alp_gluon_and_photon_int}
\eeq
where $m_{a,0}$ is the bare ALP mass (without including QCD effects), $f$ is the ALP decay constant (with dimensions of energy) and $c_{\gamma\gamma}$ and $\cGG$ are dimensionless constants, which are typically $\mathcal{O}(1)$.\footnote{Parametrically larger coefficients can be obtained in non-minimal constructions 
such as clockwork~\cite{Farina:2016tgd}, or through mixing of multiple ALPs~\cite{Hook:2018dlk}.} For the QCD axion $m_{a,0} = 0$ and the mass comes entirely from QCD (we will not make this assumption). 
The benchmark models with $c_{\gamma\gamma} \neq 0$ or $\cGG\neq 0$ have also been explored in the CERN PBC study~\cite{Beacham:2019nyx}.
When discussing the model with dominantly-photon interaction we will also use the notation $\gag = c_{\gamma\gamma} \alpha /(\pi f)$.

The ALP-gluon coupling leads to interactions with hadrons below the QCD phase transition. These interactions can be derived within chiral perturbation theory ($\chi$PT), as we describe in the following section. As first pointed in Ref.~\cite{Bauer:2020jbp}, this procedure is fraught with the possibility of missing quantitatively important terms. We therefore pay particular attention to ensure that final results are physical (in the sense to be made precise below).

\subsection{Hadronic Interactions at Low Energies}
We now discuss the interactions of ALPs in the $c_{\gamma\gamma} = 0,\;\cGG \neq 0$ model.
The ALPs that can be discovered at DarkQuest have masses below about a GeV, where the appropriate description of ALP couplings in Eq.~\ref{eq:alp_gluon_and_photon_int} is in terms 
hadrons rather than gluons due to confinement. In order to derive this description it is convenient to perform a field redefinition on the quarks
\beq
q\to \exp\left(-i \bm{\kappa}\gamma_5\cGG \frac{a}{f}\right)q,
\label{eq:quark_chiral_rot}
\eeq
 where $\bm{\kappa}$ is a matrix in flavour space (we will consider both two and three-flavour limits). The rotated Lagrangian contains~\cite{Bauer:2020jbp,Bauer:2021mvw}
\begin{align}
\mathscr{L} & \supset \cGG\frac{\alpha_s}{4\pi}\left(1 - \tr \bm{\kappa}\right) \frac{a}{f} G_{\mu\nu}^a \widetilde{G}^{a,\,\mu\nu} 
+ \hat{c}_{\gamma\gamma}\frac{\alpha}{4\pi}\frac{a}{f} F_{\mu\nu} \widetilde{F}^{\mu\nu} 
\\ 
& + \frac{\partial_\mu a}{2f} \bar{q} \hat{\bm{c}}_{qq} \gamma^\mu \gamma_5 q - \bar q e^{-i \bm{\kappa} \gamma_5 \cGG a/f}\bm{m_q}e^{-i \bm{\kappa} \gamma_5 \cGG a/f} q,
\label{eq:alp_gluon_int_after_rotation}
\end{align}
where the hatted quantities include the effects of the chiral rotation:
\beq
\hat{c}_{\gamma\gamma} = - 2 N_c \cGG \tr \bm{Q}^2 \bm{\kappa},\;\;\hat{\bm{c}}_{qq} = (2\cGG) \bm{\kappa}.
\label{eq:hat_constant_defs}
\eeq
The first term in Eq.~\ref{eq:alp_gluon_int_after_rotation} is the original gluon coupling combined with a contribution that arises due to the chiral anomaly; the second term is the induced coupling to photons that also arises due to the chiral anomaly; the third term comes from the quark kinetic term, and the last term is the modified quark mass term. At this point the matrix $\bm{\kappa}$ is arbitrary, but we can see that if we choose 
\beq
\tr \bm{\kappa} = 1, 
\label{eq:gluon_coupling_cancellation_condition}
\eeq
the gluon term is eliminated. One also usually chooses $\bm{\kappa}$ to be diagonal in the same 
basis as $\bm{m_q}$, i.e., $\bm{\kappa} = \mathrm{diag}(\kappa_u,\kappa_d)$. The key point emphasized in Refs.~\cite{Bauer:2021wjo,Bauer:2020jbp} is that $\bm{\kappa}$ is completely arbitrary up to the condition in Eq.~\ref{eq:gluon_coupling_cancellation_condition}; as a result, physical observables cannot depend on $\bm{\kappa}$. Cancellation of these unphysical parameters in Feynman diagram calculations therefore provides a non-trivial check of the results. This was test was carried out for several processes involving ALPs in~\cite{Bauer:2021wjo,Bauer:2020jbp}.

We now derive the interactions with mesons and nucleons starting from Eq.~\ref{eq:alp_gluon_int_after_rotation}.

\subsubsection{Interactions with Mesons and Photons}
\label{sec:interactions_with_mesons_and_photons}

We focus on the two flavour theory for simplicity first. We will have to study three-flavour case in order to incorporate interactions with $\eta$ and $\eta'$ mesons. 
The second line of Eq.~\ref{eq:alp_gluon_int_after_rotation} becomes the following 
leading-order chiral Lagrangian:
\begin{subequations}
\begin{align}
\mathscr{L} & \supset \frac{1}{2}(\partial a)^2 - \frac{m_{a,0}^2}{2}a^2  \\ 
&+  \frac{f_\pi^2}{4} \tr (\partial_\mu \Sigma) (\partial^\mu \Sigma)^\dagger
+ \frac{f_\pi^2}{2}B \tr \left(\Sigma \bm{m_q}^\dagger(a) + \bm{m_q}(a)\Sigma^\dagger\right) \\
& + \frac{i f_\pi^2}{2} \frac{\partial_\mu a}{2f}\tr \hat{\bm{c}}_{qq} \left(\Sigma \partial^\mu \Sigma^\dagger- \Sigma^\dagger \partial^\mu \Sigma\right).
\end{align}
\label{eq:chiral_lagrangian}
\end{subequations}
The first line is the ALP kinetic and mass terms, the second term is the meson kinetic and mass terms ($\bm{m_q}^\dagger(a)$ is the quark mass matrix with the two exponential factors in Eq.~\ref{eq:alp_gluon_int_after_rotation} but without the $\gamma_5$); the third line comes from matching the axial current in the quark theory to the axial current in the meson theory.\footnote{The meson coupling to the isovector axial current in the last line of Eq.~\ref{eq:chiral_lagrangian} can also be derived using the external current formalism of Ref.~\cite{Scherer:2002tk}. The derivatives in the kinetic term are promoted to covariant derivatives with $D_\mu \Sigma \equiv \partial_\mu \Sigma + i \Sigma r_\mu - i l_\mu \Sigma$. The external currents $r_\mu$ and $l_\mu$ are such that the axial current is $a_\mu = (r_\mu - l_\mu)/2$; choosing $a_\mu = \partial_\mu a / (2f) \times \hat{\bm{c}}_{qq}$ gives the correct interaction.} In the two-flavour theory (with a decoupled $\eta^\prime$), the matrix meson field is
\beq
\Sigma = \exp(2 i \Pi /f_\pi),\; \Pi =\frac{1}{2} \begin{pmatrix}
  \pi_0& \sqrt{2}\pi^+  \\
  \sqrt{2}\pi^- & - \pi_0 
\end{pmatrix},
\label{eq:sigma_and_pi_def}
\eeq
and we use $f_\pi \approx 93\;\MeV$~\cite{Scherer:2002tk,Scherer:2005ri}. The three-flavour model is considered in Appendix~\ref{sec:three_flavour_mixing}.
Note that our normalization conventions differ slightly from Ref.~\cite{Bauer:2021wjo,Bauer:2020jbp}: our $f_\pi$ is a factor of $\sqrt{2}$ smaller than theirs, which accounts for various differences between their formulas and ours.

The Lagrangian in Eq.~\ref{eq:chiral_lagrangian} contains off-diagonal terms involving the ALP and $\pi^0$ (and $\eta^{(\prime)}$ in the three-flavour theory):
\begin{align}
\mathscr{L} & \supset \frac{1}{2}(\partial a)^2 - \frac{m_a^2}{2}a^2 + \frac{1}{2}(\partial \pi^0)^2 - \frac{m_\pi^2}{2}(\pi^0)^2 \\ 
& + \frac{(\hat{c}_{uu}-\hat{c}_{dd})f_\pi}{2f}(\partial a)(\partial \pi^0)
- \frac{2\cGG (m_u \kappa_u - m_d \kappa_d) m_\pi^2 f_\pi}{(m_u + m_d) f} a \pi^0, 
\label{eq:2f_pion_alp_lagrangian}
\end{align}
where
\beq
m_\pi^2 = B (m_u + m_d),\;\; m_a^2 = m_{a,0}^2 + \frac{4B\cGG^2 f_\pi^2 (m_u \kappa_u^2 +m_d \kappa_d^2)}{f^2}.
\eeq
This mixing can be accounted for either perturbatively for each process involving ALPs, or by performing a complete diagonalization of the meson-ALP kinetic and mass terms at the Lagrangian level. The former provides a simple explicit illustration of how unphysical $\bm{\kappa}$ dependence cancels; the latter is more straightforward to use in the three-flavour case discussed in Appendix~\ref{sec:three_flavour_mixing}.

In the perturbative approach, a generic ALP production process can be 
schematically represented as 
\beq
%\feynmandiagram [baseline=(b.base), horizontal=a to b] {
%a [blob]  -- [scalar, momentum={[arrow style=red]\(p\)}] b [particle=\(a^{\mathrm{phys}}\)] 
%}; = 
%\feynmandiagram [baseline=(b.base), horizontal=a to b] {
%a [blob]  -- [scalar] b [particle=\(a\)] 
%}; + 
%\feynmandiagram [baseline=(b.base), horizontal=a to c] {
%a [blob]  --  [scalar,insertion=0.5,edge label=\(\pi^0\)] c [particle=\(a\)]
%};
\includegraphics[valign=c, raise=0.9ex]{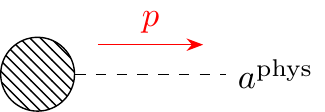}
\quad = \quad
\includegraphics[valign=c]{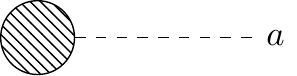}
\quad+\quad
\includegraphics[valign=c, raise=0.5ex]{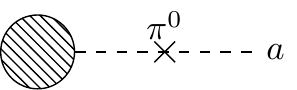}
\eeq
There are two contributions: one from direct ALP production and a second one from $\pi^0$ mixing with an ALP; both are needed to demonstrate the cancellation of the unphysical $\kappa$ parameters. 
The mixing contribution to the processes requires the following Feynman rule 
which follows from Eq.~\ref{eq:2f_pion_alp_lagrangian}
\beq
%\feynmandiagram [baseline=(b.base), horizontal=a to b] {
% a [particle=\(\pi^0\)]  --  [scalar,insertion=0.5, momentum={[arrow style=red]\(p\)}] b [particle=\(a\)]
% };
\includegraphics[valign=c,raise=1.5ex]{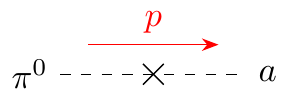}
 = i\left[ \frac{(\hat{c}_{uu}-\hat{c}_{dd})f_\pi}{2f} p^2 
 - \frac{2\cGG (m_u \kappa_u - m_d \kappa_d) m_\pi^2 f_\pi}{(m_u + m_d) f}
 \right],
 \label{eq:alp_pion_2f_mixing_fr}
\eeq
where the first term arises from the kinetic mixing, and the second term is the mass mixing. 

An immediate application of this is to derive the physical ALP-photon amplitude Ref.~\cite{Bauer:2020jbp}:
\begin{subequations}
\begin{align}
    i\mathcal{A}(a\to \gamma\gamma) & \propto \frac{\alpha}{4\pi}\left(\frac{i\hat{c}_{\gamma\gamma}}{f} + i\left[ \frac{(\hat{c}_{uu}-\hat{c}_{dd})f_\pi}{2f} m_a^2 
 - \frac{2\cGG (m_u \kappa_u - m_d \kappa_d) m_\pi^2 f_\pi}{(m_u + m_d) f}
 \right]\frac{i}{p^2 - m_\pi^2} \left(\frac{-i}{f_\pi}\right)\right) \\
& = -\frac{i\alpha \cGG}{4\pi f}\left(\frac{5}{3} + \frac{m_\pi^2\delta_I}{m_\pi^2 - m_a^2}\right).
\end{align}
\label{eq:two_photon_amp_2f}
\end{subequations}
where
\beq
\delta_I = \frac{(m_d - m_u)}{(m_d + m_u)} 
\eeq
parametrizes isospin breaking due the quark mass matrix.
In the last line of Eq.~\ref{eq:two_photon_amp_2f} we used Eq.~\ref{eq:hat_constant_defs} and Eq.~\ref{eq:gluon_coupling_cancellation_condition}; note that the result is now independent of $\kappa$, so it gives the physical coupling of ALPs to photons. 
The equivalent three-flavour result that accounts for mixing with $\eta$ and $\eta^\prime$ mesons is given in Eq.~\ref{eq:two_photon_amp_3f}.

Now we use the chiral Lagrangian to derive amplitudes relevant for 
low-mass ALP decays to photons and mesons, as well as ALP production in meson decays. Some of these results are known (in particular, see Refs.~\cite{Bauer:2020jbp,Bauer:2021mvw,Bauer:2021wjo}) in the two-flavour $\chi$PT. We extend these calculations to three flavours when relevant.  
For ALPs near the $\rho$ mass and above, virtual vector mesons can play an important role in determining amplitudes of certain decay channels~\cite{Aloni:2018vki,Cheng:2021kjg}, which are not included in our calculation. It would be interesting to extend their calculations while keeping track of $\bm{\kappa}$ in pseudoscalar \emph{and} vector meson interactions. This can be done within the hidden local symmetry framework~\cite{Harada:2003jx}, for example.

The amplitudes that are relevant for sub-GeV ALP decays are  
\begin{itemize}
    \item Three-pion final states that follow from the chiral Lagrangian
      \begin{subequations}
      \begin{align}
    \mathcal{A}(a\to \pi^0\pi^0\pi^0) &= \frac{\cGG \delta_I m_\pi^2 \left(-2 m_a^4+m_a^2 \left(4 m_K^2-3 \
m_\pi^2\right)+m_\pi^4\right)}{f f_\pi \left(m_a^2-m_\pi^2\right) \
\left(3 m_a^2-4 m_K^2+m_\pi^2\right)}\\
    \mathcal{A}(a\to \pi^0\pi^+\pi^-) &= \frac{\cGG \delta_I m_\pi^2 \left(m_a^4+m_a^2 \left(7 m_\pi^2-9 s\right)-12 m_K^2 \left(m_\pi^2-s\right)+4 m_\pi^4-3 m_\pi^2 s\right)}{3 f f_\pi \left(m_a^2-m_\pi^2\right) \left(3 m_a^2-4 m_K^2+m_\pi^2\right)}
  \end{align}
\end{subequations}
    In the above expressions we have decoupled the singlet meson, but still worked in three flavours. In our numerical results we retain $\eta-\eta'$ mixing with 
    their physical masses. 
    Note that the two-flavour result of Ref.~\cite{Bauer:2020jbp} is obtained by taking $m_K\to \infty$ in the two amplitudes above; in this limit the $3\pi^0$ amplitude vanishes if $m_a$ is small, but this is not so in the three-flavour model.

    In the mass window where $a\to \eta \pi\pi$ is allowed it is clearly not appropriate to decouple the singlet meson. 
    For these and other three-flavour calculations we work in the simplified $\eta-\eta^\prime$ mixing scheme (in which we approximate $\sin\theta_{\eta\eta'} \approx -1/3$) unless stated otherwise; this produces compact expressions for the amplitudes.\footnote{The cancellation of $\kappa$ dependence does not rely on this assumption.}
    The $a \to \eta \pi^0 \pi^0$ amplitude is 
    \beq
    \mathcal{A}(a \to \eta \pi^0 \pi^0) = \frac{\sqrt{\frac{2}{3}} \cGG m_\pi^2 \left(3 m_a^2-2 m_\eta^2-m_{\eta^\prime}^2\right) \left(2 m_\eta^2-5 m_{\eta^\prime}^2+3 m_\pi^2\right)}{27 f f_\pi \left(m_a^2-m_\eta^2\right) \left(m_a^2-m_{\eta^\prime}^2\right)}    
    \label{eq:a_to_eta_pi0_pi0}
    \eeq
    To leading order in isospin violation the amplitude to charged pions is the same
    \beq
    \mathcal{A}(a \to \eta \pi^+ \pi^-) \approx  \mathcal{A}(a \to \eta \pi^0 \pi^0)
    \label{eq:a_to_eta_pip_pim}
    \eeq
    The ALP mass scale for which these amplitudes are relevant is in the regime where standard $\chi$PT becomes unreliable. In order to improve the 
    accuracy of our estimates, we follow Ref.~\cite{Aloni:2018vki} and use ``$k$-factors''  which rescale the resulting partial widths. The $k$-factors 
    are obtained by comparing our calculations of the $\eta' \to \eta\pi\pi$ amplitudes to measurements~\cite{Tanabashi:2018oca}.
    \item The decay to $\pi\pi \gamma$ arises from the anomalous Wess-Zumino-Witten Lagrangian~\cite{Donoghue:1992dd}
    \beq
    \mathscr{L}_{\mathrm{WZW}} \supset \frac{N_c e}{48\pi^2}  \epsilon^{\mu\nu\alpha\beta}A_\mu \tr \bm{Q} (\mathcal{R}_\nu \mathcal{R}_\alpha \mathcal{R}_\beta + \mathcal{L}_\nu \mathcal{L}_\alpha \mathcal{L}_\beta),
    \eeq
    where $\mathcal{R}_\mu = (\partial_\mu \widetilde{\Sigma}^\dagger) \widetilde{\Sigma}$, $\mathcal{L}_\mu = \widetilde{\Sigma} \partial_\mu \widetilde{\Sigma}^\dagger$ and $\widetilde\Sigma$ is the axially-transformed meson field:
    \beq
    \widetilde{\Sigma} = \exp\left(i \bm{\kappa}\cGG \frac{a}{f}\right)\Sigma\exp\left(i \bm{\kappa}\cGG \frac{a}{f}\right).
    \eeq
    This transformed field gives rise to direct couplings of the ALP to $\pi\pi \gamma$ and plays a key role in the cancellation of the unphysical 
    $\kappa$ dependence (along with the meson-mixing contributions).
    The resulting amplitude is 
    \beq
    \mathcal{A}(a\to \gamma \pi^+\pi^-) = \frac{\cGG e N_c \epsilon_\mu^* k_{1\nu} k_{2\alpha} k_{3\beta}  \epsilon^{\mu\nu\alpha\beta}\left(3 m_a^2-2 m_\eta^2-m_{\eta^\prime}^2\right) \left(2 m_\eta^2-5 m_{\eta^\prime}^2+3 m_\pi^2\right)}{324 \pi ^2 f f_\pi^2 \left(m_a^2-m_\eta^2\right) \left(m_a^2-m_{\eta^\prime}^2\right)},
    \eeq 
    where $k_{1,2,3}$ are the four momenta of $\pi^-$, $\pi^+$ and $\gamma$, respectively; $\epsilon_\mu^*$ is the photon polarization. Similar calculations for $\eta^{(\prime)}\to \pi^+ \pi^- \gamma$ reproduce the SM results~\cite{Holstein:2001bt}.
\end{itemize}

\begin{figure}
    \centering
    \includegraphics[width=0.45\textwidth]{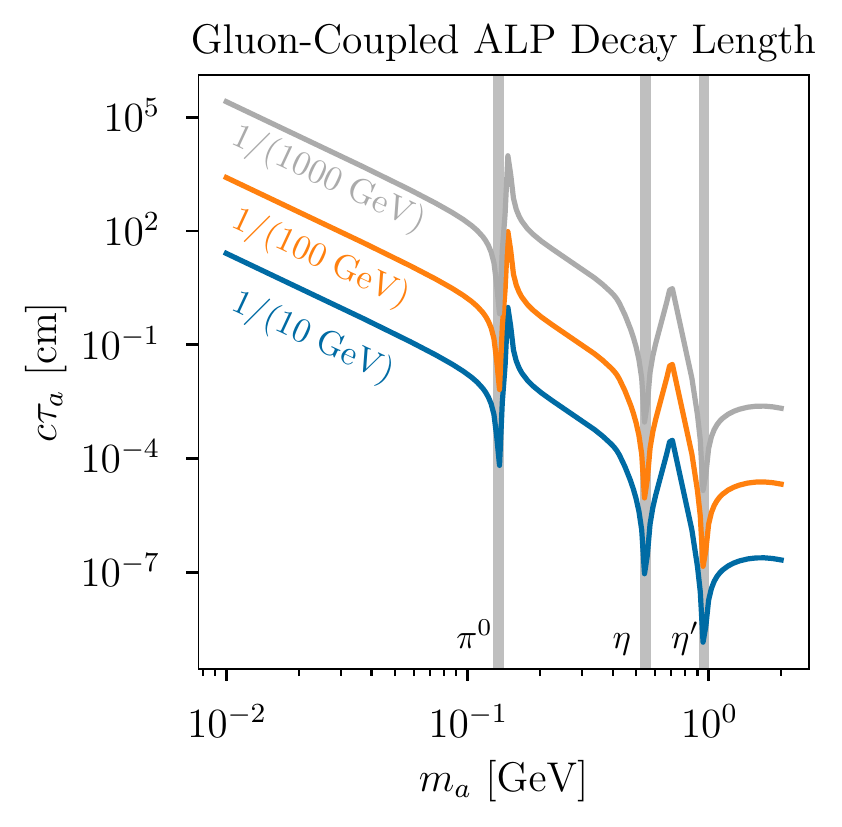}
    \includegraphics[width=0.45\textwidth]{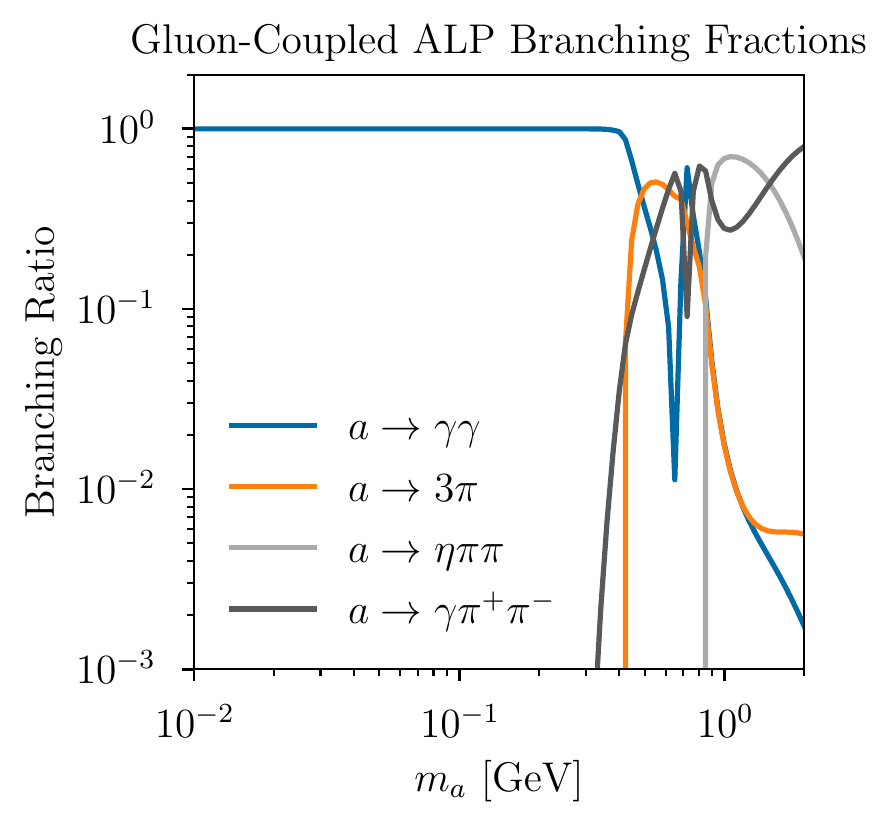}
    \caption{ALP decay length (left panel) and branching fractions (right panel) as a function of ALP mass in the dominantly-gluon-coupled ALP model. The decay length is shown for $\cGG/f = 1/(10\;\GeV)$, $1/(100\;\GeV)$, $1/(1000\;\GeV)$. When the ALP mass is close the masses of the neutral mesons, mixing is enhanced; these regions are highlighted by the vertical gray bands. In the right panel, the lines labelled $a\to 3\pi$ and $a\to \eta\pi\pi$ combine 
    partial widths to charged and neutral pions.}
    \label{fig:gluon_coupled_alp_lifetime_and_branching_fractions}
\end{figure}

The ALP decay calculations are summarized in Fig.~\ref{fig:gluon_coupled_alp_lifetime_and_branching_fractions} in which we show the ALP decay length and branching fractions to various final states. We note that for $m_a \lesssim 1\;\GeV$ the partial width to $\gamma\gamma$ remains appreciable in most of the parameter space, making this a good decay channel to search for even in the dominantly-gluon-coupled ALP model; other decays result in more complex final states but can be used to recover some signal rate if $a\to\gamma\gamma$ is suppressed.

At low masses the dominant ALP production mechanisms are rare decays of $\pi^\pm$, $\eta^{(\prime)}$, $K_{L,S}$ and $K^\pm$ mesons.
We find
\begin{itemize}
    \item The $\eta \to a \pi \pi$ decays are simply related to $a \to \eta \pi^0 \pi^0$ by crossing symmetry:
    \beq
    \mathcal{A}(\eta \to a \pi^0 \pi^0) = \mathcal{A}(a \to \eta \pi^0 \pi^0),
    \label{eq:eta_to_a_pi0_pi0}
    \eeq
    \beq
    \mathcal{A}(\eta \to a \pi^+ \pi^-) = \mathcal{A}(a \to \eta \pi^+ \pi^-),
    \label{eq:eta_to_a_pip_pim}
    \eeq
    where the right-hand sides are given in Eqs.~\ref{eq:a_to_eta_pi0_pi0} and~\ref{eq:a_to_eta_pip_pim}.
    The approximate $\eta-\eta'$ mixing scheme allows to relate these simply to the corresponding $\eta^\prime$ amplitudes:
    \beq
    \mathcal{A}(\eta^\prime \to a \pi^0 \pi^0) = \frac{1}{\sqrt{2}}\mathcal{A}(\eta \to a \pi^0 \pi^0)
    \label{eq:etap_to_a_pi0_pi0}
    \eeq
    \beq
    \mathcal{A}(\eta^\prime \to a \pi^+ \pi^-) = \frac{1}{\sqrt{2}}\mathcal{A}(\eta \to a \pi^+ \pi^+)
    \label{eq:etap_to_a_pip_pim}
    \eeq
    The derivation of these amplitudes (and more details about cancellation of $\kappa$) is discussed in Appendix~\ref{sec:rare_eta_decays}.
    \item The $\pi^\pm$ and kaon rare decays arise from electroweak interactions. We focus here on kaon decays because the ALP mass range relevant for $\pi^\pm$ is well-covered experimentally (the amplitude for $\pi^\pm \to \nu \mu^\pm a$ is given in Ref.~\cite{Bauer:2021mvw}). Following the notation of Ref.~\cite{Gori:2020xvq}, the EW Lagrangian relevant for $s\to d$ transitions after the chiral rotation, Eq.~\ref{eq:quark_chiral_rot}, is\footnote{We neglect the 27-plet operator for simplicity since its Wilson coefficient,  $G_{27}$, is 20 times smaller than $G_8$.}
    \beq
    \mathscr{L}_{s\to d} \supset G_8 f_\pi^4 \tr{R^\dagger \lambda R D_\mu \Sigma^\dagger D^\mu \Sigma} + \mathrm{h.c.}
    \eeq
    where $G_8  \approx -9\times 10^{-6}\;\GeV^{-2}$, 
    \beq
    D_\mu \Sigma = \partial_\mu\Sigma + \frac{i \partial_\mu a}{2f} \left(
    \Sigma^\dagger \hat{\bm{c}}_{qq} + \hat{\bm{c}}_{qq}\Sigma\right),\;\;\;     
    R = \exp\left(-i \bm{\kappa}\cGG \frac{a}{f}\right)
    \eeq
    and 
    \beq
    \lambda = \begin{pmatrix}
     0 & 0 & 0\\
     0 & 0 & 0\\
     0 & 1 & 0
   \end{pmatrix}.
    \eeq
    Linearising the right-handed chiral rotation $R$ allows one to extract terms relevant for various kaon decays. Both of the ALP terms appearing in $R$ and $D_\mu$ are necessary for $\kappa$-dependence to cancel. We find the following amplitudes:
    \beq
    \mathcal{A}(K_S \to \pi^0 a ) \approx \frac{8 i \cGG f_\pi^2 G_8 \left(m_a^2-m_K^2\right) \left(m_K^2-m_\pi^2\right)}{f \left(3 m_a^2-4 m_K^2+m_\pi^2\right)}
    \label{eq:KS_to_pi0_a}
    \eeq
    \beq
    \mathcal{A}(K_L \to \pi^0 a ) \approx -\epsilon_K A(K_S \to \pi^0 a ),
    \label{eq:KL_to_pi0_a}
    \eeq
    where $\epsilon_K \approx 2.23\times 10^{-3}$ is the CP-violating kaon mixing parameter.
    The charged kaon decay was already presented in Ref.~\cite{Bauer:2021mvw} in the $m_{\eta^\prime} \to \infty$ limit. Here we give the amplitude in the same limit, but use the full result in our numerics:
    \beq
    \mathcal{A}(K^\pm  \to \pi^\pm a) = 
  \frac{8 i \cGG f_\pi^2 G_8 \left(m_a^2-m_K^2\right) \left(m_K^2-m_\pi^2\right)}{f \left(3 m_a^2-4
  m_K^2+m_\pi^2\right)}.
    \label{eq:Kpm_to_pipm_a}
    \eeq
    These amplitudes are given in the limit of decoupled singlet meson and to leading order in isospin violation for brevity. 
    Their calculation, as well as subleading isospin-violating terms are discussed in Appendix~\ref{sec:rare_eta_decays}. 
    Note that the leading-isospin amplitudes satisfy $|\mathcal{A}(K_S \to \pi^0 a )| \approx  |\mathcal{A}(K^\pm  \to \pi^\pm a)|$.
\end{itemize}

\subsubsection{Interactions with Nucleons}
\label{sec:interactions_with_nucleons}
\begin{figure}
    \centering
    \includegraphics[width=0.8\textwidth]{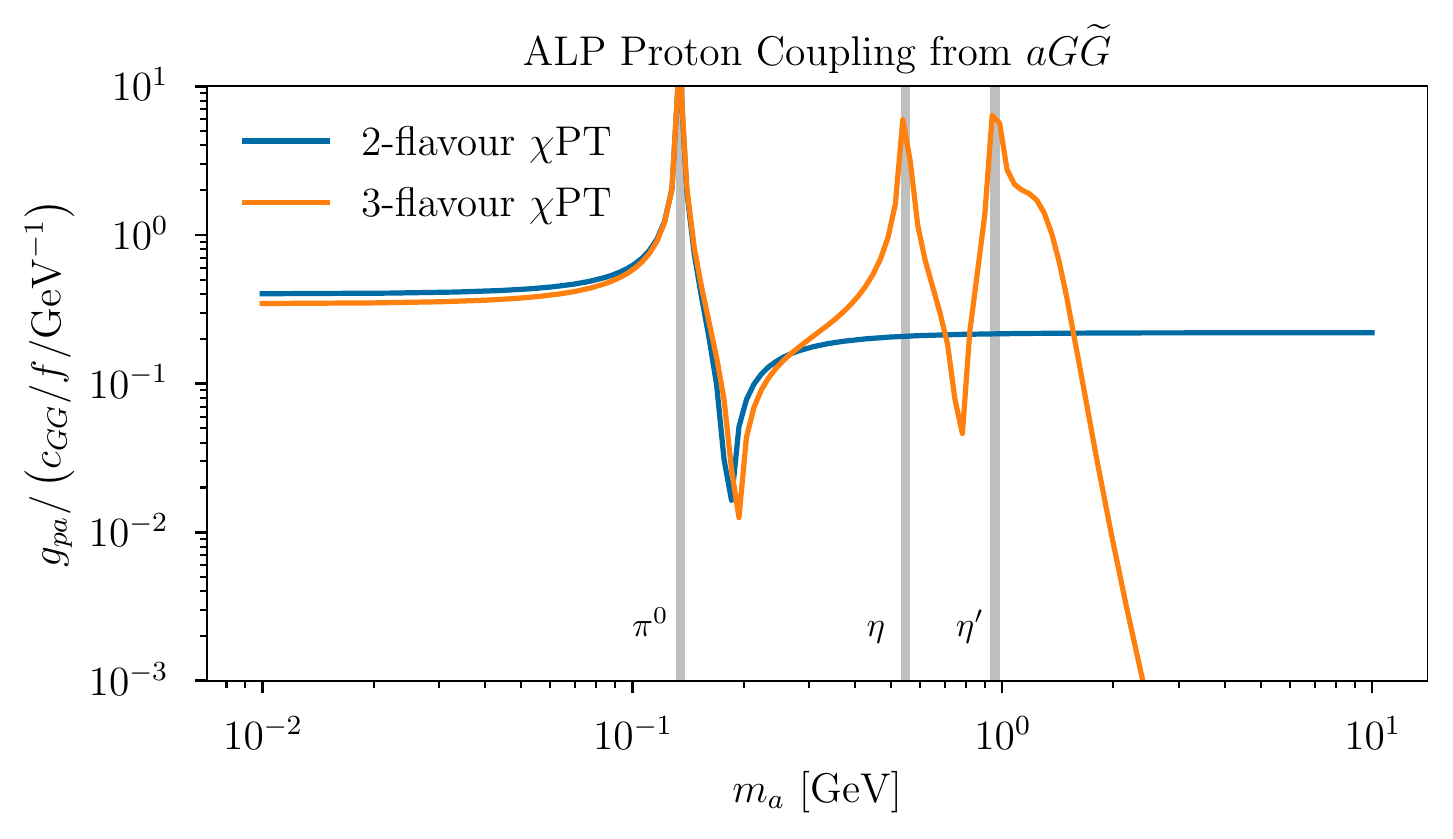}
    \caption{Coefficient of the ALP-proton coupling in two- and three-flavour chiral perturbation theory as a function of the ALP mass $m_a$. The couplings are enhanced when the $m_a$ is close to one of the neutral meson masses; these regions are highlighted by gray vertical bands.}
    \label{fig:alp_proton_coupling}
\end{figure}

Next we consider the nucleon-ALP interaction. In the low 
energy limit (such as that relevant for DM axion detection or astrophysical 
processes) it is enough to consider two-flavour $\chi$PT -- these results appear in many places, including Refs.~\cite{GrillidiCortona:2015jxo,Bauer:2021wjo}. In particular, Ref.~\cite{Bauer:2021wjo} has demonstrated the $\kappa$-independence of this interaction. We first repeat their analysis to illustrate this cancellation of unphysical parameters and then extend it to the three-flavour model. We will find that mixing with $\eta$ and $\eta'$ becomes very important near the GeV-scale. The two- and three-flavour results for the effective proton interaction $g_{pa}$
are compared in  Fig.~\ref{fig:alp_proton_coupling}.
\paragraph{Two-flavour Model}
The leading pion-nucleon Lagrangian is~\cite{Scherer:2002tk,Scherer:2005ri}
\beq
\mathscr{L}_{\pi N} = \overline{\Psi} \left(i\gamma^\mu D_\mu + \frac{g_A}{2}\gamma^\mu \gamma_5 u_\mu|_{\mathrm{vec}} + \frac{g_0}{2}\gamma^\mu \gamma_5 u_\mu|_{\mathrm{scal}}\right)\Psi,
\eeq
where $\Psi = (p,\;\; n)$ is the nucleon doublet. The restrictions to ``isovector'' or ``isoscalar'' parts will be explained shortly. 
The nucleons transform non-linearly under $SU(2)_L\times SU(2)_R$ but simply under the isospin subgroup. The vielbein $u_\mu|_\mathrm{vec}$ is given by
\beq
u_\mu|_\mathrm{vec} = i \left[\xi (\partial_\mu - i r_\mu|_\mathrm{vec} )\xi^\dagger - \xi^\dagger (\partial_\mu - i l_\mu|_\mathrm{vec} )\xi\right] \supset  2 \frac{\partial_\mu \Pi}{f_\pi} +2 a_\mu|_\mathrm{vec},
\eeq
where $\xi^2 = \Sigma$, $r_\mu$ and $l_\mu$ are right- and left-handed external currents, and 
\beq
a_\mu = \frac{1}{2}(r_\mu - l_\mu) = \frac{\partial_\mu a}{2f} \hat{\bm{c}}_{qq}
\eeq
is the axial current.\footnote{Note that our $u_\mu$ is slightly different from that of~\cite{Scherer:2002tk,Scherer:2005ri}, because we take $\Sigma$ to transform as $\Sigma\to L \Sigma R^\dagger$ (our $\xi$ is their $u^\dagger$).} In the above we have isolated contributions from isovector and isoscalar axial currents; the isoscalar pieces receive contributions from $\eta'$ and ALP interactions, but in the two-flavour limit the $\eta'$ is decoupled, so 
\beq
u_\mu|_\mathrm{scal} = i \left[( - i r_\mu|_\mathrm{scal} ) - (- i l_\mu|_\mathrm{scal} )\right] = 2 a_\mu|_\mathrm{scal}.
\eeq

Decomposing the coupling matrix $\hat{\bm{c}}_{qq}$ into isovector and isoscalar components gives:
\beq
\hat{\bm{c}}_{qq} = \frac{\hat{c}_{uu}-\hat{c}_{dd}}{2}\begin{pmatrix}1 & 0\\0 & -1\end{pmatrix} +  \frac{\hat{c}_{uu}+\hat{c}_{dd}}{2}\begin{pmatrix}1 & 0 \\0 & 1\end{pmatrix}
\equiv \hat{\bm{c}}_{qq}|_\mathrm{vec} + \hat{\bm{c}}_{qq}|_\mathrm{scal}.
\eeq
The leading ALP-proton interactions are therefore
\beq
\mathscr{L}_{\pi N} \supset 
\bar{p}\gamma^\mu \gamma_5\left[g_A\left(\frac{(\hat{c}_{uu}-\hat{c}_{dd})\partial_\mu a}{4f} 
+ \frac{\partial_\mu \pi^0}{2f_\pi}
\right) + g_0\left(\frac{(\hat{c}_{uu}+\hat{c}_{dd})\partial_\mu a}{4f}\right)\right]p
\label{eq:pseudoscalar_proton_lang_2f}
\eeq
Note that the isoscalar piece is proportional to $\tr\bm{\kappa}$ making them physical (i.e., no meson mixing contributions are required to cancel off unphysical $\kappa$ dependence). 

As with the physical ALP photon amplitude, the amplitude for 
$p \to p a$ involves mixing with $\pi^0$ as well as a direct coupling~\cite{Bauer:2021wjo}; their sum is $\kappa$-independent:
\beq
\mathcal{A}(p\to p a) =  \bar{u}(p') i\slashed{k} \gamma_5 u(p)\left( g_0 \left[\frac{\cGG}{2f}\right] +  g_A  \left[\frac{\cGG \delta_I m_\pi^2}{2f (m_\pi^2 - k^2)}\right]\right),
\label{eq:two_flavour_alp_proton_amplitude}
\eeq
where we made use of Eqs.~\ref{eq:gluon_coupling_cancellation_condition} and \ref{eq:alp_pion_2f_mixing_fr}. 

\paragraph{Three-flavour model}
Production of ALPs near the GeV scale is sensitive to mixing with heavier mesons, so it is useful to include $\eta$, $\eta'$ couplings to baryons, which necessitates we study the nucleon octet $B$.
In order to include the singlet $\eta$ we construct a $U(3)_L\times U(3)_R$ invariant Lagrangian following~\cite{Borasoy:1999nd,Bruns:2019fwi}
\begin{subequations}
\begin{align}
\mathscr{L} & \supset \tr{\bar{B} (i\slashed{D}-m_N) B} \\
& + \frac{D}{2}\tr{\bar{B}\gamma_\mu \gamma_5 \{u^\mu, B\}} +
\frac{F}{2}\tr{\bar{B}\gamma_\mu \gamma_5 [u^\mu, B]} + 
\frac{D_s}{2}\tr{\bar{B}\gamma_\mu \gamma_5 B} \tr{u^\mu}.
\end{align}
\label{eq:three_flavour_alp_baryon_lang}
\end{subequations}
We will see that linear combinations of the coupling constants $D$, $F$ and $D_s$ are related to $g_A$ and $g_0$ from the two-flavour formalism. For example, if we forget about the ALP the leading interactions are 
\beq
\frac{1}{2f_\pi}\left[(D+F)\partial_\mu \pi^0 - \frac{1}{\sqrt{3}}(D-3 F)\partial_\mu \eta_8  +\sqrt{\frac{2}{3}} (2D + 3D_s)\partial_\mu \eta_1\right]\bar{p} \gamma^\mu \gamma_5 p.
 \label{eq:proton_meson_lang}
\eeq
Comparing this with the two-flavour result, Eq.~\ref{eq:pseudoscalar_proton_lang_2f}, we therefore expect that $g_A \approx D + F$, and $g_0 = D + 2 D_s + F$. Refs.~\cite{Close:1993mv,Borasoy:1998pe} found that $D \approx 0.80$, $F \approx 0.46$ and 
Ref.~\cite{Bruns:2019fwi} fit $D_s$ in the range $[-0.6,-0.2]$ (but they do not provide a best fit value). These values are consistent with three-flavour extractions of $g_A = 1.283$ and $g_0 = 0.384$ from lattice QCD~\cite{Alexandrou:2020okk,Alexandrou:2021wzv} if we take $D_s \approx -0.41$. The additional ALP term that results from Eq.~\ref{eq:three_flavour_alp_baryon_lang} is
\beq
\frac{\partial_\mu a}{2f_a}\left[\hat{c}_{dd} D_s + \hat{c}_{ss} (D + D_s - F) + \hat{c}_{uu} (D + D_s + F)\right]\bar{p} \gamma^\mu \gamma_5 p.
\label{eq:proton_alp_lang}
\eeq
In rotating to the physical basis, the physical ALP-proton coupling receives contributions from both Eq.~\ref{eq:proton_meson_lang} and~\ref{eq:proton_alp_lang}; we write the result as 
\beq
\mathscr{L} \supset g_{pa} (\partial_\mu a^\mathrm{phys})\bar{p} \gamma^\mu \gamma_5 p,
\eeq
where
\begin{align}
g_{pa} & = \frac{1}{2f_a}\left[
\hat{c}_{dd} D_s + \hat{c}_{ss} (D + D_s - F) + \hat{c}_{uu} (D + D_s + F)
\right] \\ 
& + \frac{1}{2f_\pi}\left[(D+F)\vev{\pi a} - \frac{1}{\sqrt{3}}(D-3 F)\vev{\eta_8 a}  +\sqrt{\frac{2}{3}} (2D + 3D_s)\vev{\eta_1 a}\right]
\end{align}
where the quantities $\vev{\cdot a}$ are defined in Appendix~\ref{sec:three_flavour_mixing}. By inserting explicit expressions for these, one can check for the cancellation of $\kappa$ dependence. One can also take various limits, like decoupling the third flavour, or taking the mass of the singlet $m_{\eta_1}\to \infty$ to recover Eq.~\ref{eq:two_flavour_alp_proton_amplitude}. As before, working in the ``simplified $\eta-\eta'$ mixing'' scheme with $\sin\theta_{\eta\eta'} = -1/3$ gives
\begin{subequations}
\begin{align}
g_{pa} & = \frac{c_{GG} \left(2 m_\eta^2-5 m_{\eta^\prime}^2+3 m_{\pi}^2\right) \left(6
   D \left(m_a^2-m_\eta^2\right)+D_s \left(9 m_a^2-8
   m_\eta^2-m_{\eta^\prime}^2\right)+2 F
   \left(m_\eta^2-m_{\eta^\prime}^2\right)\right)}{54 f_a
   \left(m_a^2-m_\eta^2\right)
   \left(m_a^2-m_{\eta^\prime}^2\right)} \\
   & +\frac{c_{GG} \delta_I m_{\pi}^2
   (D+F) \left(2 m_\eta^2+m_{\eta^\prime}^2-3 m_{\pi}^2\right) \left(2
   m_\eta^2-5 m_{\eta^\prime}^2+3 m_{\pi}^2\right)}{54 f_a
   \left(m_a^2-m_{\pi}^2\right) \left(m_{\pi}^2-m_\eta^2\right)
   \left(m_{\pi}^2-m_{\eta^\prime}^2\right)}.
\end{align}
\end{subequations}
The result is reassuringly $\kappa$-independent and retains various enhancements in mixing with $\pi$, $\eta$ and $\eta'$ when the ALP mass is near the corresponding meson mass. We compare $g_{pa}$ in two- and three-flavour theories in Fig.~\ref{fig:alp_proton_coupling} (the two-flavour $g_{pa}$ is just the quantity in the parentheses of Eq.~\ref{eq:two_flavour_alp_proton_amplitude}). We see important differences at large ALP masses, where the coupling is suppressed for $m_a > m_{\eta'}$ (although at these large masses one should start including higher meson resonances). At low masses the two results are not quite equal because $m_\pi/m_{\eta^{(\prime)}} \neq 0$.

In the discussion above we took the hadrons as point-like. In realistic calculations we must introduce form-factors to account for their extended size and substructure, which we will discuss in Sec.~\ref{sec:alp_production_from_gluons}. 

\section{ALP Production and Signals From the Photon Coupling}
\label{sec:alp_production_from_photons}
In this section we focus on the photon-coupled ALP model and set $\cGG = 0$. 
We mostly follow the notation of Ref.~\cite{Dobrich:2015jyk} 
and write the interaction as
\beq
\mathscr{L} \supset \frac{\gag}{4} a F_{\mu\nu} \widetilde{F}^{\mu\nu},
\label{eq:alp_photon_int}
\eeq
where $\gag = \alpha c_{\gamma\gamma}/(\pi f)$ in Eq.~\ref{eq:alp_gluon_and_photon_int}.

\begin{figure}
    \centering
    \includegraphics[width=5cm]{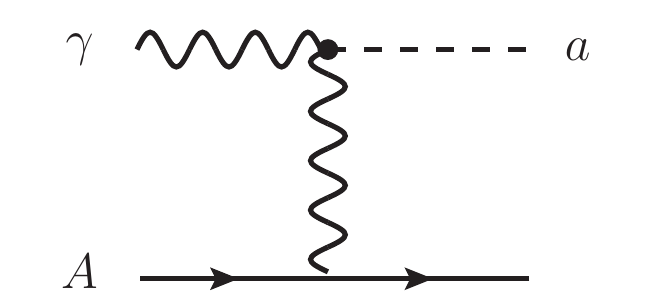}\hspace{1.5cm}
    \includegraphics[width=5cm]{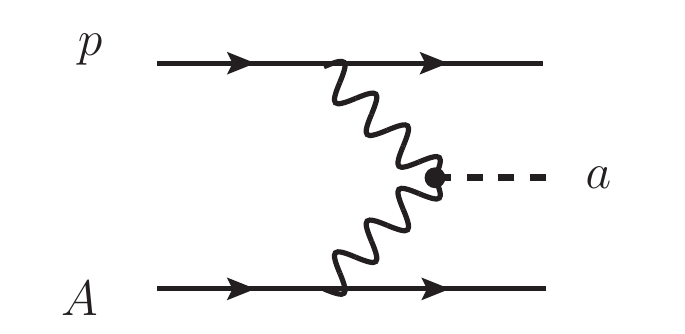}
    \caption{Dominant production mechanisms of photon-coupled axion-like particles 
    in proton beam dump experiments. In the left panel, a real secondary photon 
    from the decay of a meson collides with a nucleus in the dump. 
    The right panel shows the fusion of two virtual photons coherently radiated 
  off the beam proton and target nucleus.}
    \label{fig:photon_production_mechanisms}
\end{figure}

\begin{figure}
    \centering
    \includegraphics{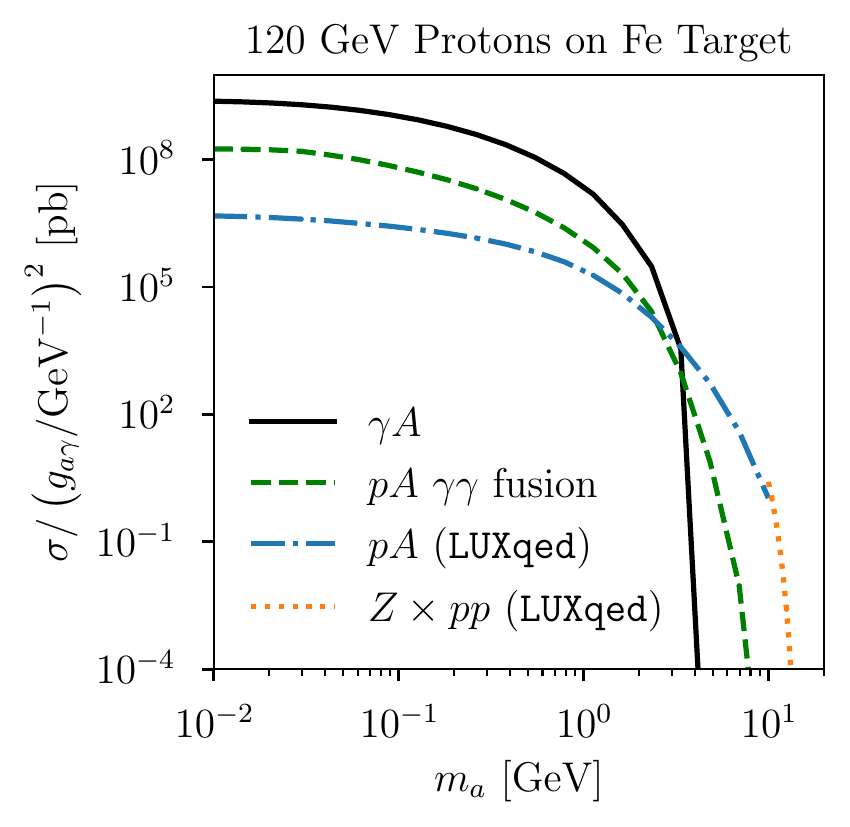}
    \caption{Cross-sections for various photon-coupled ALP production channels in collisions of a 120 GeV 
    proton beam with an iron target. Primakoff ($\gamma A$) and coherent photon fusion ($pA$) 
    processes are used to estimate the sensitivity of DarkQuest. 
    The latter channel is restricted to proton momentum transfers of $Q^2 \leq 1\;\GeV^2$ for 
    computational simplicity. Cross sections for processes with higher momentum 
    transfers are obtained using the \texttt{LUXqed} photon PDF, but end up being subdominant in the parameter space accessible at DarkQuest.}
    \label{fig:cross_sections}
\end{figure}

The interaction in Eq.~\ref{eq:alp_photon_int} enables the decay of ALPs into photons with width
\beq
\Gamma_a = \frac{\gag^2 m_a^3}{64\pi},
\eeq
and a variety of production channels in proton fixed target experiments.
The most important of these are Primakoff production $\gamma A\rightarrow a A$ (with $\gamma$ a secondary photon from, e.g., meson decay) and quasi-elastic ``photon fusion'' processes 
$p A \rightarrow a p A$, which are illustrated in Fig.~\ref{fig:photon_production_mechanisms}. We discuss these in more detail below. While the large flux of secondary photons typically dominates the production of ALPs, photon fusion can produce ALPs with a slightly higher boost, enabling sensitivity to shorter lifetimes. 
For both processes the production is coherent over the nucleus for 
ALP masses $\lesssim \GeV$. Incoherent processes (such as analogues of 
deep inelastic scattering) can in principle produce more massive ALPs, but their cross-sections are typically very suppressed as we discuss below. In Fig.~\ref{fig:cross_sections} we compare the cross-sections for different ALP production processes in the DarkQuest beam and target configuration. We describe their calculation next. 

\subsection{Primakoff Production: \texorpdfstring{$\gamma A \rightarrow a A$}{gamma A to a A}}
Proton-nucleus collisions produce a large multiplicity of neutral mesons such 
as $\pi^0$ and $\eta$  
which decay to photons, giving rise an intense secondary photon flux. These 
photons can undergo a Primakoff-type interaction with a nucleus in the dump 
producing an ALP as shown in the left panel of Fig.~\ref{fig:photon_production_mechanisms}. 
This mechanism has been identified as the dominant production mode for photon-coupled ALPs in proton beam dump~\cite{Berlin:2018pwi,Dobrich:2019dxc} and (primary) photon beam experiments~\cite{Aloni:2019ruo} across a broad range of parameter space. Below 
we describe how we simulate this production channel at the DarkQuest experiment.

First, we need to estimate the rate, spectrum and angular distributions 
of photons produced in proton-nucleus collisions. Following Refs.~\cite{Berlin:2018pwi,Dobrich:2019dxc} we use \pythia 8.240 to model meson production 
and their subsequent decay into photons using the \texttt{SoftQCD} flag.
Ref.~\cite{Dobrich:2019dxc} has validated the meson multiplicity and angular distributions 
at several energies relevant for proton beam dump experiments, finding 
reasonable agreement of simulated transverse momentum and small longitudinal momentum 
distributions with data. However \pythia underestimates meson production at larger $p_z/p_z^{\mathrm{max}}$. These highly-boosted mesons produce the most energetic photons, 
and the most boosted ALPs (see Eq.~\ref{eq:ALP_energy} below). Since our simulation underestimates this 
region of phase space, we underestimate the sensitivity to shorter lifetimes and heavier 
ALPs. We leave an improved analysis utilizing more realistic 
distributions (such as those obtained from Ref.~\cite{Bonesini:2001iz}) for future work. 
Since \pythia 
simulates $pp$ collisions, we rescale the total cross-section to the 
experimentally fitted $p A$ value
\beq
\sigma_{pA} \approx (49.2\text{ mb}) A^{0.77},
\label{eq:tot_proton_nucleus_xsec}
\eeq
where $A$ is the nuclear mass number and we used the fit from Fig. 1 of Ref.~\cite{Carvalho:2003pza} (based on data of Ref.~\cite{RamanaMurthy:1975vfu}).~\footnote{ 
  Note that Fig. 3 in Ref.~\cite{Carvalho:2003pza} has a different fit to higher energy experiments which yields slightly larger cross-sections. This result was 
  used in Ref.~\cite{Dobrich:2019dxc}.}
The \pythia simulation 
produces an average of approximately 6 photons per $pp$ interaction with a mean 
energy of $\sim 3\;\GeV$; the typical transverse momentum of these photons 
is given by the QCD scale $\sim 0.2\;\GeV$.

The secondary photons produced in $pA$ collisions can interact 
the nuclei in the dump to produce ALPs.
The differential cross-section for Primakoff production follows 
from the interaction in Eq.~\ref{eq:alp_photon_int}
\beq
\frac{d \sigma_{\gamma A\rightarrow a A}}{dt} = 
%\frac{2\pi \alpha \gag^2 Z^2 F(|t|)^2}{16\pi (s-m_N^2)^2 t^2} 
%\left(m_a^2 t (m_N^2+s)- m_a^4 m_N^2-t((m_N^2-s)^2 + st)\right),
\frac{\gag^2 \alpha  Z^2 F(|t|)^2 s}{8(s-m_N^2)^2 t^2}(t_0 - t)(t-t_1)
\label{eq:primakoff_prod_xsec}
\eeq
which is in agreement with the expression in Ref.~\cite{Aloni:2019ruo} 
up to a different parametrization of the photon-nucleus interaction (this result is also equivalent to the expressions in Refs.~\cite{Dobrich:2015jyk,Dobrich:2019dxc,Dusaev:2020gxi} 
in the limit of $m_a$, $\sqrt{|t|} \ll E_a$, $m_A$). 
Here $s$ and $t$ are Mandelstam variables with $s=(p_\gamma + p_A)^2$, $t = (p_\gamma - p_a)^2$, with $t_{0,1}$ being the kinematic boundaries of $t$ given 
in Eq.~\ref{eq:t_boundary} below; 
$Z$ and $m_N$ are the target nucleus charge and mass. The electromagnetic form-factor $F$ is 
modelled using the Helm parametrization~\cite{Lewin:1995rx}
\beq
F(Q^2) = \frac{3j_1(\sqrt{Q^2}R_1)}{\sqrt{Q^2}R_1}\exp\left(-\frac{1}{2} \sqrt{Q^2} \tilde{s}\right)
\eeq
where $\tilde s = 0.9$ fm and 
\beq
R_1 = \sqrt{(1.23A^{1/3} -0.6\;\mathrm{fm})^2 + 7\pi^2(0.52\;\mathrm{fm})^2/3 - 5\tilde{s}^2}.
\eeq
This parametrization is also used in Refs.~\cite{Dobrich:2015jyk,Dobrich:2019dxc}. Finally, 
note that the kinematically-allowed range of $t$ is $[t_1,t_0]$ where~\cite{Tanabashi:2018oca}
\beq
t_0(t_1) = \frac{m_a^4}{4s} - (|\vec{p}_{\gamma,cm}| \mp |\vec{p}_{a,cm}|)^2
\label{eq:t_boundary}
\eeq
and
\beq
  |\vec{p}_{\gamma,cm}| = \frac{s - m_N^2}{2\sqrt{s}},
  \;\;\;\;\;\;\;
  |\vec{p}_{a,cm}| = \sqrt{\left(\frac{s + m_a^2 - m_N^2}{2\sqrt{s}}\right)^2 - m_a^2}.
\eeq
For a given incident photon, the ALP energy is given by 
\beq
E_a = E_\gamma + \frac{t}{2m_N}.
\label{eq:ALP_energy}
\eeq
For small ALP masses, the distribution in Eq.~\ref{eq:primakoff_prod_xsec} peaks 
at $t\sim -m_a^4/(2E_\gamma^2)$, so from Eq.~\ref{eq:ALP_energy} it is clear that typically ALPs inherit most of the incident photon energy.

We use the secondary photons from \pythia to produce a sample of ALP events by drawing from the differential distribution in Eq.~\ref{eq:primakoff_prod_xsec}. These ALPs are displaced and decayed into photons. We then estimate the total event yield from this mechanism via
\beq
\Nevt^{\mathrm{Prim}} \approx (\Npot n_A T^{(p)}) \frac{\sigma_{pA}}{N_{mc,pA}}\sum_{\{p_\gamma\}} 
(n_A T^{(\gamma)}) 
\frac{\sigma_{\gamma A}(s_{\gamma A})}{N_{mc,\gamma A}}\sum_{\{p_a\}} \mathcal{C}(p_a),
\label{eq:mc_primakoff_rate}
\eeq
where the outer sum is over the secondary photons and the inner sum is over the ALP momenta; 
$s_{\gamma A}$ is the Mandelstam variable for the $\gamma A$ system; the function $\mathcal{C}$ implements ALP decay to $\gamma\gamma$, and computes the probability weight associated with the various experimental cuts on the location of the 
vertex and properties of the photons which are described in Sec.~\ref{sec:sensitivity} for DarkQuest; $N_{mc,pA}$ is the number 
of $pA$ collisions simulated in \pythia and $N_{mc,\gamma A}$ is the number of 
ALP events generated per secondary photon. We will only consider secondary photon production in the first interaction length of FMAG ($T^{(p)} = 16.77$ cm) and ALP production in the first 
radiation length ($T^{(\gamma)} =  1.757$ cm); this simplification allows us to neglect attenuation of 
the initial proton and secondary photon beams. For ease of discussion below, we will 
separate the event yield into a total cross section $\sigma_{\mathrm{Prim}}$ and a dimensionless acceptance $\mathcal{A}_{\mathrm{Prim}}$
\beq
\Nevt^{\mathrm{Prim}} \approx \Npot n_A T^{(p)} \sigma_{\mathrm{Prim}} \mathcal{A}_{\mathrm{Prim}},
\label{eq:prim_yield_factorized}
\eeq
where 
\beq
\sigma_{\mathrm{Prim}} = \frac{\sigma_{pA}}{N_{mc,pA}}\sum_{\{p_\gamma\}} 
n_A T^{(\gamma)} \sigma_{\gamma N}(s_{\gamma N})
\label{eq:prim_total_crosssection}
\eeq
and $\mathcal{A}_{\mathrm{Prim}}$ follows from these two equations.

Representative kinematic distributions for ALP decay products for the Primakoff
production mechanism are shown in Figs.~\ref{fig:decay_photon_distribution} and~\ref{fig:decay_photon_separation}, 
while the total cross-section is compared to other processes in Fig.~\ref{fig:cross_sections}.

The general features of ALP production and decay via the Primakoff process are evident in the 
distributions in the left column of Fig.~\ref{fig:decay_photon_distribution}. 
Since the $\gamma A$ interaction peaks at small momentum transfers (see Eq.~\ref{eq:primakoff_prod_xsec} and Eq.~\ref{eq:ALP_energy}), it follows that ALPs and their decay products roughly inherit the properties 
of the secondary photons that produce them. As a result, the typical energies and angles are 
similar for different ALP masses up to threshold effects at small photon energies.

\subsection{Photon Fusion: \texorpdfstring{$\gamma^* \gamma^*\rightarrow a $}{gamma gamma to a} } %\texorpdfstring{$p A\rightarrow a p A$}{p A to a p A} }
\label{sec:fusion}
Axion-like particles can also be produced directly in the fusion of two virtual photons from the incoming proton and the nucleus as illustrated in 
the right panel of Fig.~\ref{fig:photon_production_mechanisms}. Several methods for evaluating this production rate have been considered before~\cite{Dobrich:2015jyk,Harland-Lang:2019zur}. Here we implement 
a version of the equivalent photon approximation in which we treat 
the incoming proton as a beam of nearly-on-shell 
photons~\cite{Dobrich:2015jyk}, enabling the 
following factorization of the production cross-section~\cite{Budnev:1974de}:
\beq
d\sigma_{pA\rightarrow apA} \approx \frac{d\sigma_{\gamma A}}{dt_2} dt_2 dn_\gamma(x,Q^2), 
\label{eq:EPA_xsec}
\eeq
where $d\sigma_{\gamma A}/dt_2$ is the differential cross-section in Eq.~\ref{eq:primakoff_prod_xsec} (we have relabelled $t\rightarrow t_2$ to emphasize that $t_2$ is the Mandelstam parameter of the $2\to2$ sub-process $\gamma^* A \rightarrow a A$, 
not of the actual $2\to 3$ reaction); $d n_\gamma$ is the (proton) virtual photon spectrum given by~\cite{Budnev:1974de}
\begin{align}
\frac{d^2 n_\gamma}{dx dQ^2} = \frac{\alpha}{\pi} \frac{1}{x} 
\frac{1}{Q^2}
\left[(1-x)D(Q^2) + \frac{x^2}{2} C(Q^2) - (1-x) \frac{Q^2_{\mathrm{min}}}{Q^2} D(Q^2)\right]
\end{align}
where $Q^2 = -q^2$ and $q = (\omega, \vec{q})$ denotes the four-momentum of the virtual photon, $x =  \omega/E_p$ is the fraction of proton beam energy the photon carries away. 
The momentum transfer $Q^2$ can be related to the magnitude of 
the transverse momentum $q_t$ of the virtual photon: 
\beq
Q^2 \approx \frac{q_t^2 + m_p^2 x^2}{1-x},
\eeq
where we took the limit $m_p/E_p\ll 1$. We see from this 
relation that $Q^2_{\mathrm{min}} \approx m_p^2/(1-x)$.
The form-factors $D$ and $C$ encode the non-point-like nature of the 
proton; these are fit to electron-proton scattering data and are given by 
\beq
D(Q^2) = \frac{4m_p^2 G_E^2(Q^2) + Q^2 G_M^2(Q^2)}{4m_p^2 + Q^2},\;\;\; C(Q^2) = G_M^2(Q^2),
\eeq
where
\beq
G_E(Q^2) = \frac{1}{(1 + Q^2/Q_0^2)^2},\;\;\; G_M(Q^2) = \frac{\mu_p}{(1 + Q^2/Q_0^2)^2},
\label{eq:form_factors}
\eeq
with $\mu_p^2 = 7.78$, $Q_0^2  = 0.71\; \GeV^2$. These ``dipole'' form-factors 
are valid at the $\sim 10\%$ level for momentum transfers $Q^2 \lesssim 1\;\GeV^2$~\cite{Bernauer:2013tpr}.

The factorization in Eq.~\ref{eq:EPA_xsec} is enabled by several approximations 
discussed in general in Ref.~\cite{Budnev:1974de} and scrutinized in the context 
of ALP production in Ref.~\cite{Harland-Lang:2019zur}. For example, the use of 
the two body sub-process cross-section in Eq.~\ref{eq:primakoff_prod_xsec} assumes that 
$Q^2 \ll m_a^2$, an approximation that breaks down for small ALP masses. Another 
simplification used above was taking $m_p/E_p \ll 1$. The quality of these 
approximations depends on the ALP mass. For the DarkQuest beam energy and 
the mass range of interest our calculation typically overestimates the exact photon fusion rate by a factor of at most $\sim 1.5 - 2$~\cite{Harland-Lang:2019zur} (our method is termed ``photon absorption'' in this paper).
We will show that photon fusion is sub-dominant to Primakoff production in rate by about an order of magnitude, so we will not pursue refinements of these approximations. 

We generate events based on the distribution of Eq.~\ref{eq:EPA_xsec} using 
\texttt{vegas}~\cite{Lepage:1977sw,peter_lepage_2020_3897199} to perform importance sampling on $x$, $q^2$ and $t_2$; these are transformed into an ALP four momentum which is then decayed into $\gamma\gamma$. We use these events to compute the total event yield, cross section and acceptance in analogy to Eqs.~\ref{eq:mc_primakoff_rate} and ~\ref{eq:prim_yield_factorized}, ~\ref{eq:prim_total_crosssection} for the Primakoff process.

\begin{figure}
    \centering
    \includegraphics[width=0.47\textwidth]{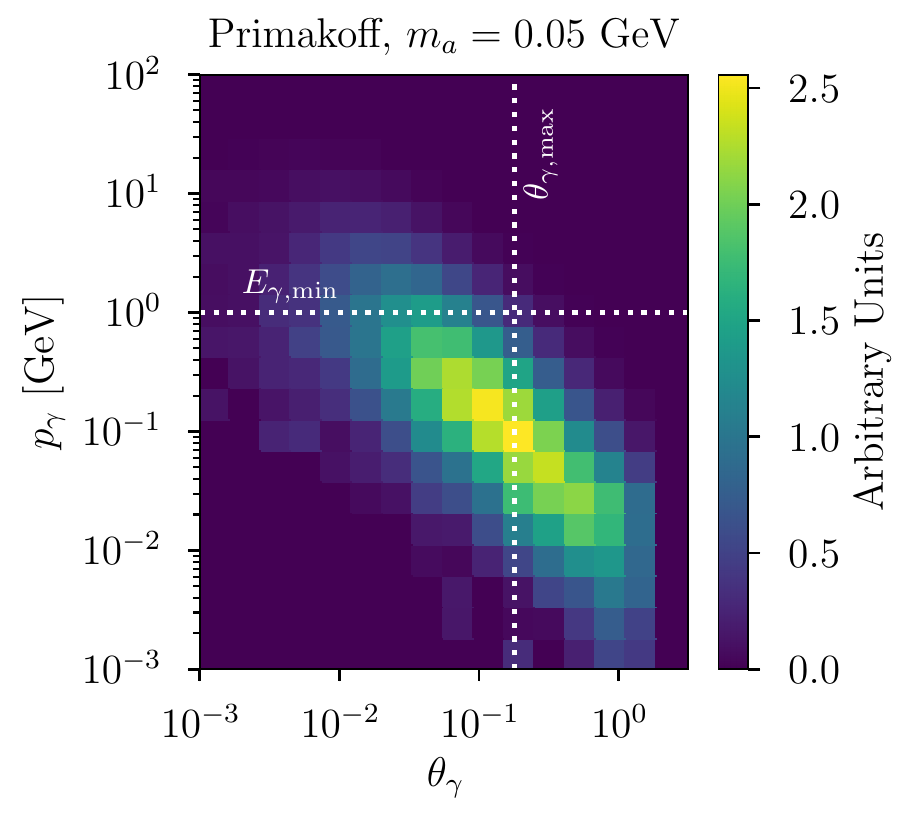}
    \includegraphics[width=0.47\textwidth]{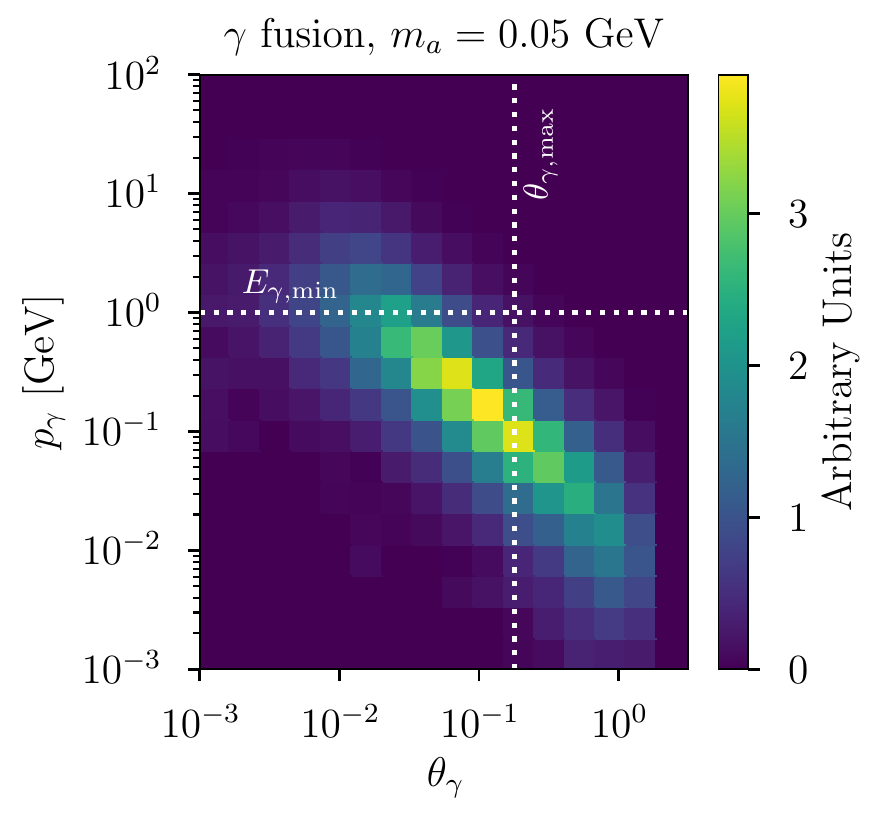}\\
    \includegraphics[width=0.47\textwidth]{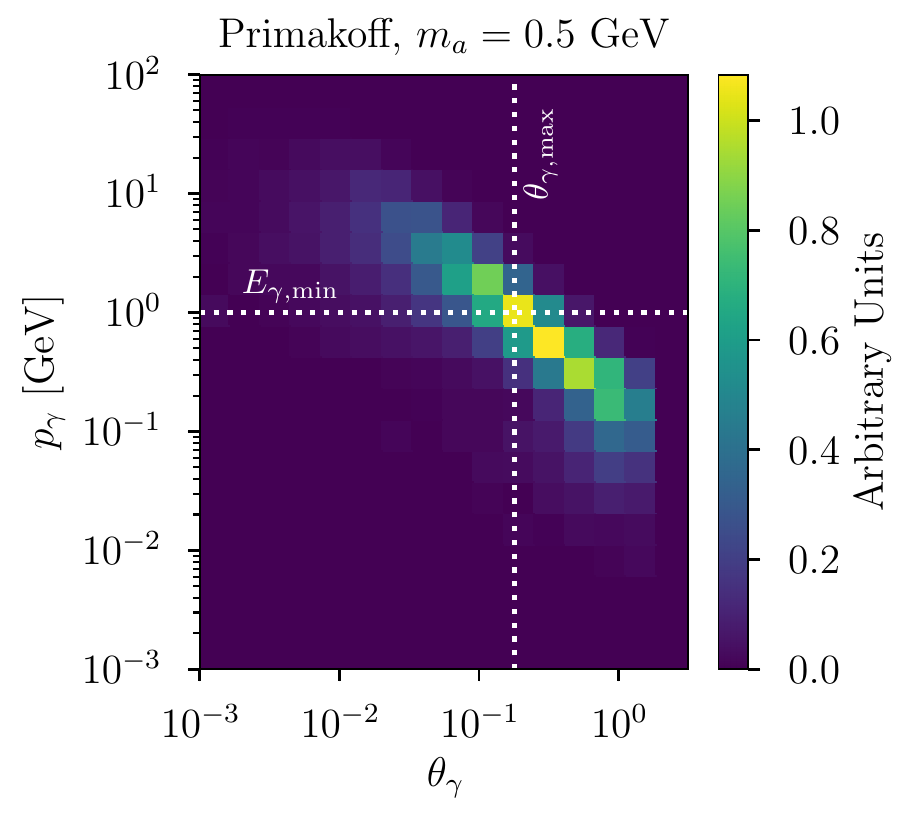}
    \includegraphics[width=0.47\textwidth]{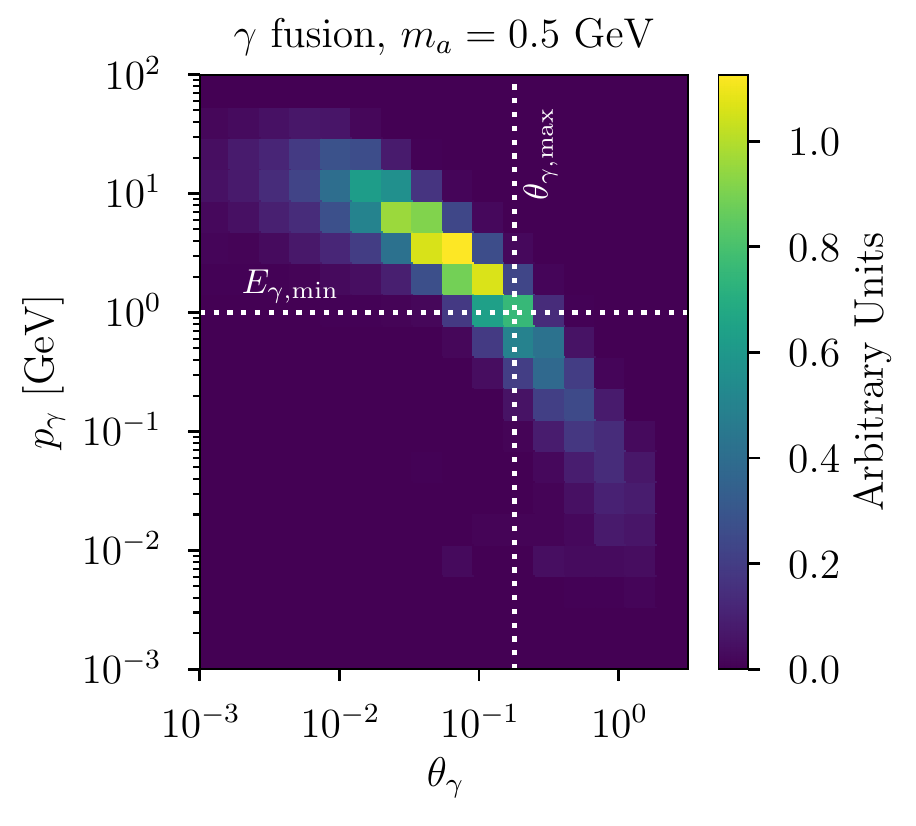}
    \caption{Distributions of ALP decay photon angles with respect to beam axis and momenta for two benchmark masses $m_a = 0.05$ and $0.5$ GeV (upper and lower rows, respectively) and the two main production mechanisms (left and right columns) in the dominantly-photon-coupled ALP model. The dotted 
    lines indicate the approximate selections imposed in the final analysis on the photon angle (assuming the decays happen at $z=8$ m and taking a $2\; \text{m}\times 2\;\text{m}$ detector at $z=19$ m) and energy. The histograms are normalized to unity.}
    \label{fig:decay_photon_distribution}
\end{figure}

\begin{figure}
    \centering
    \includegraphics[width=0.47\textwidth]{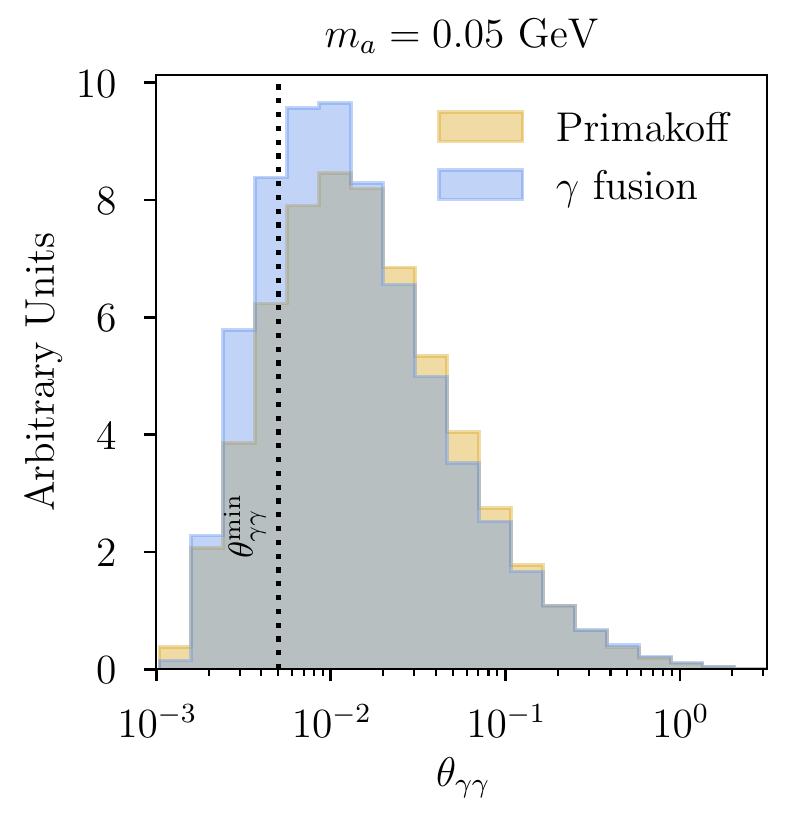}
    \includegraphics[width=0.47\textwidth]{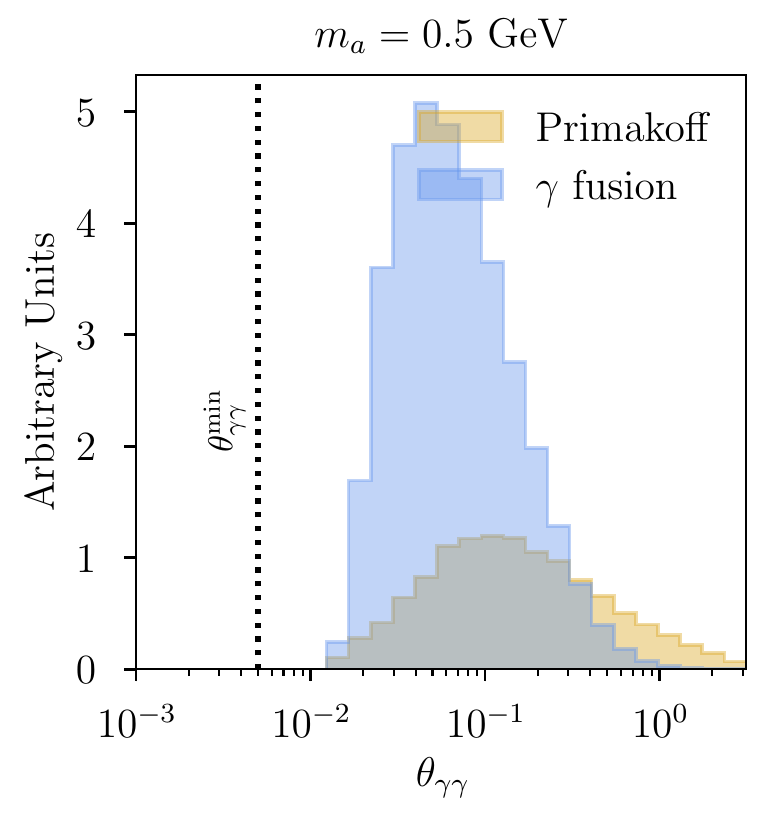}
    \caption{Distributions of angular separation between the photons in $a\rightarrow \gamma\gamma$ for two benchmark masses $m_a = 0.05$ and $0.5$ GeV (left and right panels) in the photon-coupled ALP model. In each plot we show the distributions for the Primakoff and photon fusion production processes as the yellow and blue histograms, respectively. The histograms are normalized to unity. The dotted line indicates the approximate minimum photon separation assuming 
    the decay happens at $z = 8$ m and the detector is at $z = 19$ m. Lighter ALPs are more 
    boosted and result in more collimated photons.}
    \label{fig:decay_photon_separation}
\end{figure}

Various kinematic distributions for ALP decay products for the photon fusion production mechanism are shown in Figs.~\ref{fig:decay_photon_distribution} and~\ref{fig:decay_photon_separation}. The acceptance 
function and total cross section are shown in Figs.~\ref{fig:cross_sections} and~\ref{fig:acceptance_cut_flow_photon_production}. Compared to the Primakoff process, heavier ALPs produced in 
photon fusion tend to be more boosted. This results in decay photons that are more forward and energetic as shown in the right column of Fig.~\ref{fig:decay_photon_distribution}. 
In Fig.~\ref{fig:decay_photon_separation} we show that this effect leads to 
more collimated photons for larger ALP masses; however these are still typically 
well separated enough. 

\subsection{Other Channels}
The production channels discussed above feature coherent scattering 
of protons or secondary photons off the iron nuclei in the dump, such that their rates are enhanced by $Z^2$. Production of heavier ALPs leads to larger momentum transfers and loss of this coherence. As a result processes like $\gamma p \rightarrow a + X$ have a much 
smaller rate; moreover since the initial secondary photon is the same as in 
coherent $\gamma A$ reactions, this channel does not give any new kinematic reach. 
Naively, larger masses should be accessible in $pA$ or $pp$ interactions, 
but we have restricted 
ourselves to momentum transfers $Q^2 \leq 1 \;\GeV^2$ in Sec.~\ref{sec:fusion} so that we 
could use simple dipole expressions for the proton form factors. At larger momentum transfers one must consider inelastic contributions and generation of photons from parton evolution of 
the proton constituents. These effects have been included in the \texttt{LUXqed} 
photon parton distribution function (PDF)~\cite{Manohar:2016nzj,Manohar:2017eqh}.
We use \texttt{LUXqed} to estimate the cross-sections for $p A \rightarrow a A + X$ with 
$Q^2 \geq 10\;\GeV$ (the minimum momentum transfer allowed in that PDF) and $pp \rightarrow a + X$. The former is calculated using \texttt{MadGraph}~\cite{Alwall:2014hca}, while the 
latter is computed directly using \texttt{LUXqed} and the narrow width approximation via 
\beq
\sigma(pp \to a  + X) = \frac{\gag^2 m_a^2}{8 s}\int^1_{m_a^2/s}\frac{dz}{z} f_\gamma(z, m_a^2) f_\gamma(\frac{m_a^2}{s z},m_a^2)
\eeq
where $f_\gamma(x,Q^2)$ is the photon PDFs evaluated using \texttt{lhapdf}~\cite{Buckley:2014ana}.
The results are shown in Fig.~\ref{fig:cross_sections}, where we scaled the $pp$ cross section 
by $Z=26$. 
We see that the ALP production rates are indeed dominated by Primakoff ($\gamma A$) and 
coherent photon fusion processes ($pA$ with $Q^2\leq 1\;\GeV^2$) for $m_a$ below a few GeV.  
We will see that the couplings $\gag$ required for ALP events to fall into the DarkQuest 
acceptance are $\lesssim 10^{-6}\;\GeV^{-1}$ for $m_a\gtrsim 1\GeV$; for such tiny $\gag$ 
we expect less than one ALP to be produced even in phase 2 from hard $pA$ and $pp$ interactions. 
We therefore neglect these processes in our analysis.

\section{ALP Production and Signals From the Gluon Coupling}
\label{sec:alp_production_from_gluons}
\begin{figure}
    \centering
    \includegraphics[width=5cm]{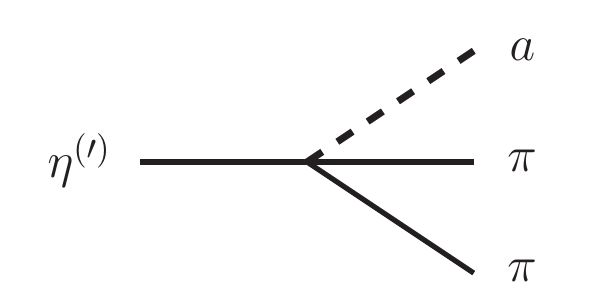}\hspace{1.5cm}
    \includegraphics[width=5cm]{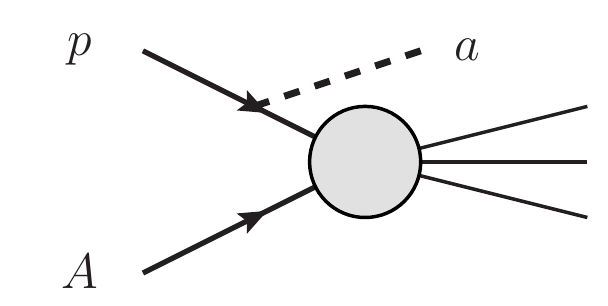}
    \caption{Representative production mechanisms of gluon-coupled axion-like particles 
      in proton beam-dump experiments. In the left panel, an ALP is produced in a rare meson decay (other production channels 
      of this kind include rare pion and kaon decays).
    The right panel shows the bremsstrahlung of an ALP from an incoming proton beam that undergoes a scattering off a nucleus.}
    \label{fig:gluon_production_mechanisms}
\end{figure}

Gluon-coupled ALPs feature a wide array of possible production and decay 
channels. In addition to the Primakoff and photon fusion ALP production (which are enabled by the induced ALP-photon coupling discussed in Sec.~\ref{sec:interactions_with_mesons_and_photons}), rare meson decay and other hadronic processes are allowed.  
We will show that the latter dominate because they are not suppressed by the electromagnetic coupling. Two representative processes are shown schematically in 
Fig.~\ref{fig:gluon_production_mechanisms}.

It is not obvious how to best model ALP 
production in hadronic interactions; various approaches have been used including emission in a parton shower~\cite{Aielli:2019ivi} or hadronization~\cite{Aielli:2019ivi,Kelly:2020dda}. While using the gluon coupling directly in a parton shower is manifestly $\kappa$-invariant, allowing the production of ALPs in hadronization by mixing with mesons can lead to $\kappa$-dependent results. This is because this is usually done by replacing neutral mesons from a Monte Carlo simulation of $pp$ collisions by an ALP and reweighing the cross-section by the corresponding ($\kappa$-dependent) mixing. 
We therefore consider a simpler alternative to clearly track $\kappa$ dependence: we instead compute ALP bremsstrahlung from a proton, following the recent results of Ref.~\cite{Foroughi-Abari:2021zbm}. It would be interesting to compare these different methods while ensuring cancellation of all unphysical parameters.

\begin{figure}
    \centering
    \includegraphics[width=0.9\textwidth]{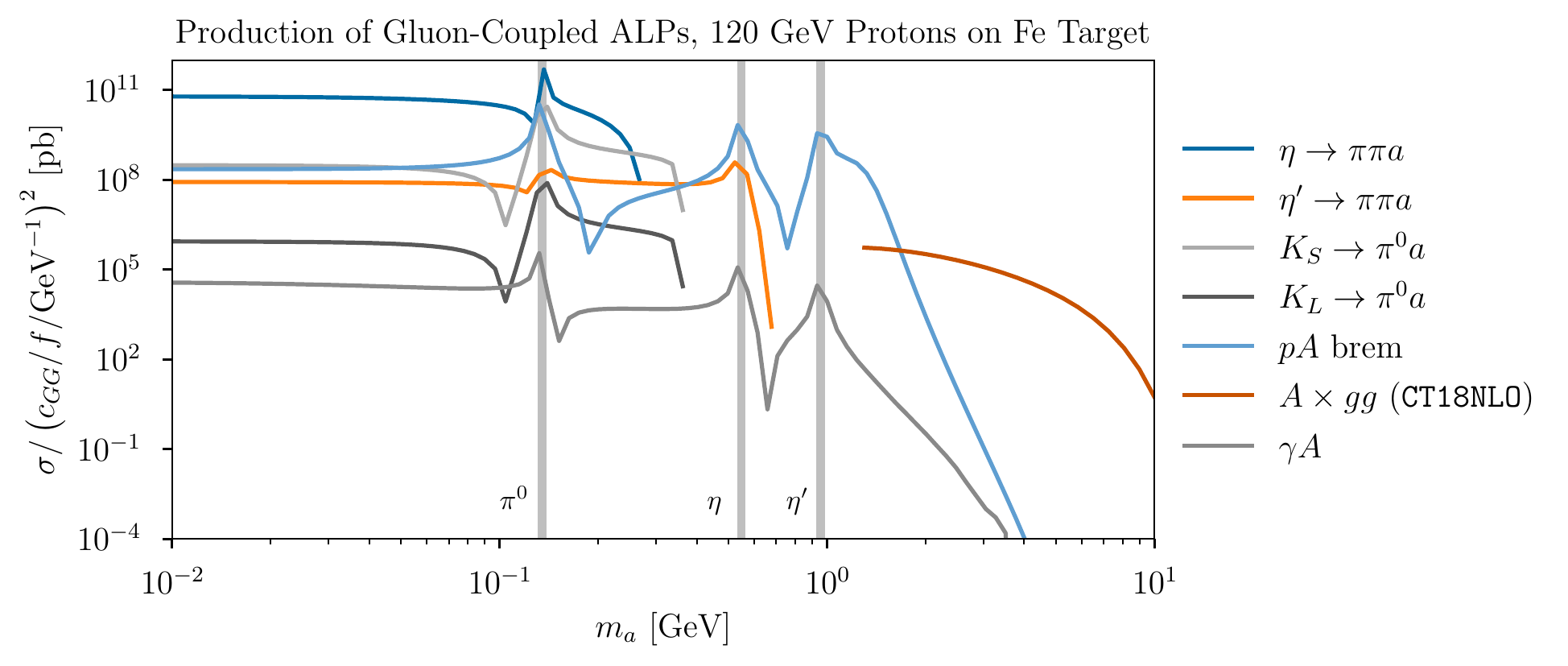}
    \caption{Cross-sections of various gluon-coupled ALP production processes in collisions of a 120 GeV 
      proton beam on an iron target. For rare meson decays, the cross-section is the meson production cross-section times its branching fraction 
    into ALPs. The line labelled $A\times gg$ refers to the gluon-gluon fusion process in proton-nucleon collisions scaled by the atomic number of the target.} 
    \label{fig:cross_sections_from_gluon_coupling}
\end{figure}

In Fig.~\ref{fig:cross_sections_from_gluon_coupling} we show the cross-sections for the DarkQuest beam and target configuration. We see that below a GeV the ALP production rate will be dominated by rare meson decays and proton bremsstrahlung.\footnote{For $m_a < m_\pi - m_\mu$ the ALP can be produced in $\pi^+ \to \nu \mu^+ a$; we do not consider this production channel only because that parameter space is well covered by existing searches.} In this plot, as in all other rate calculations we assume that the meson production occurs in the first interaction length of the iron dump.  

\begin{figure}
    \centering
    \includegraphics[width=0.47\textwidth]{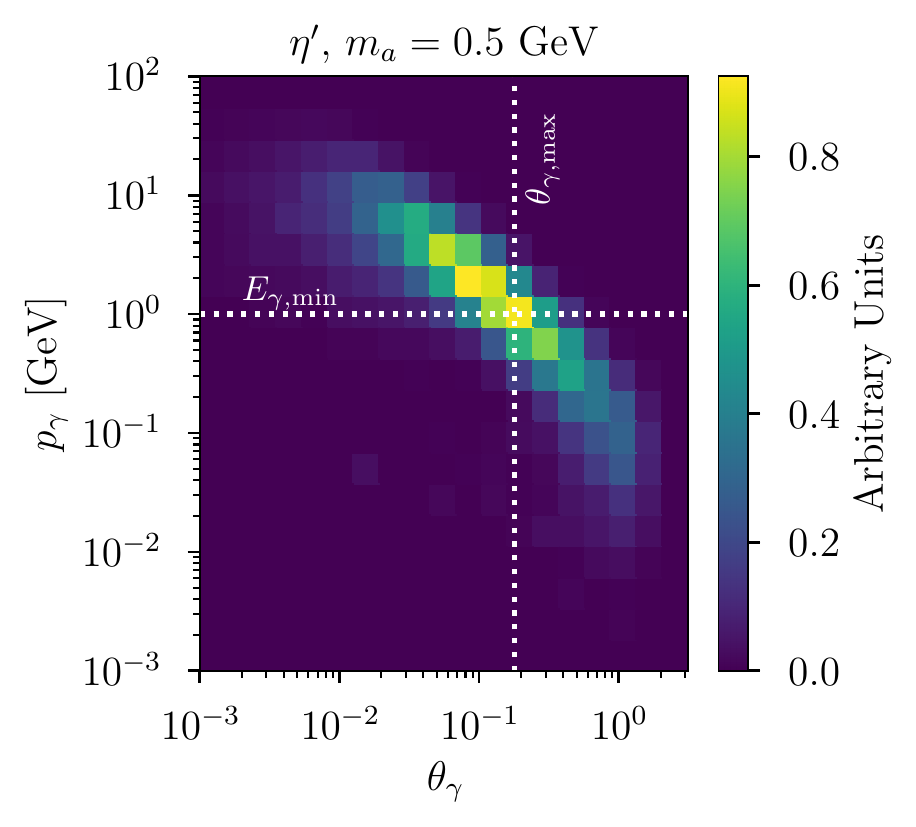}
    \includegraphics[width=0.47\textwidth]{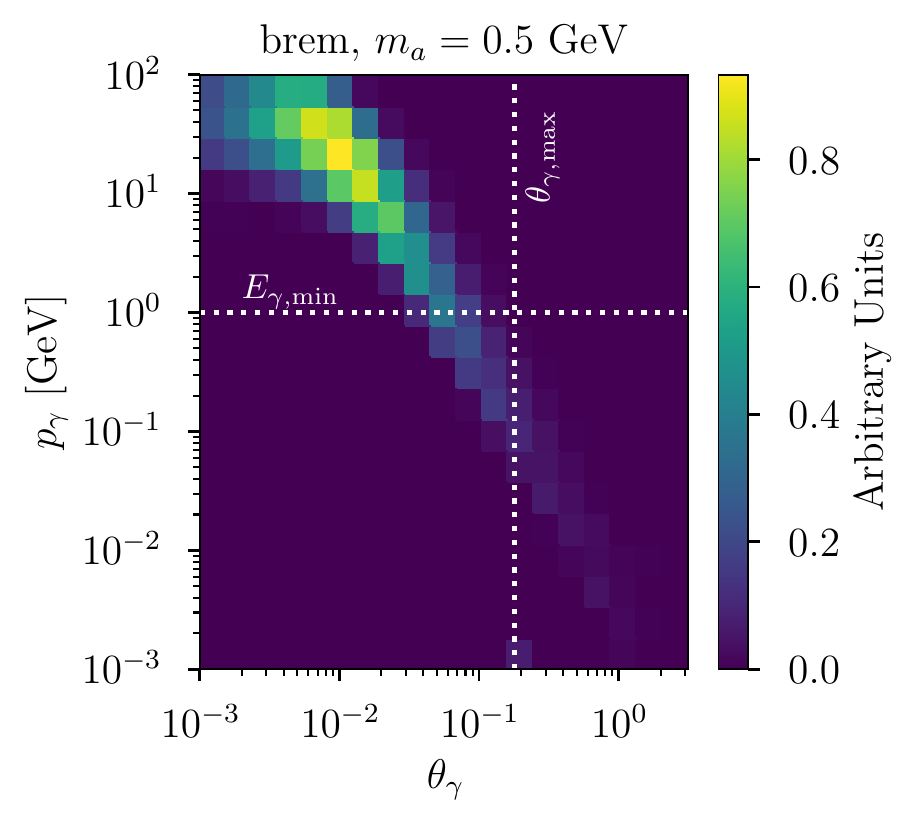}
    \caption{Distributions of ALP decay photon angles with respect to beam axis and momenta for two of the main production mechanisms of gluon-coupled ALPs, rare $\eta'$ decays and proton bremsstrahlung (left and right columns), with $m_a = 0.5\;\GeV$. The dotted 
    lines indicate the approximate selections imposed in the final analysis on the photon angle (assuming the decays happen at $z=8$ m and taking a $2\; \text{m}\times 2\;\text{m}$ detector at $z=19$ m) and energy. The histograms are normalized to unity.}
    \label{fig:decay_photon_distribution_gluon_coupled_alp}
\end{figure}

\begin{figure}
    \centering
    \includegraphics[width=0.47\textwidth]{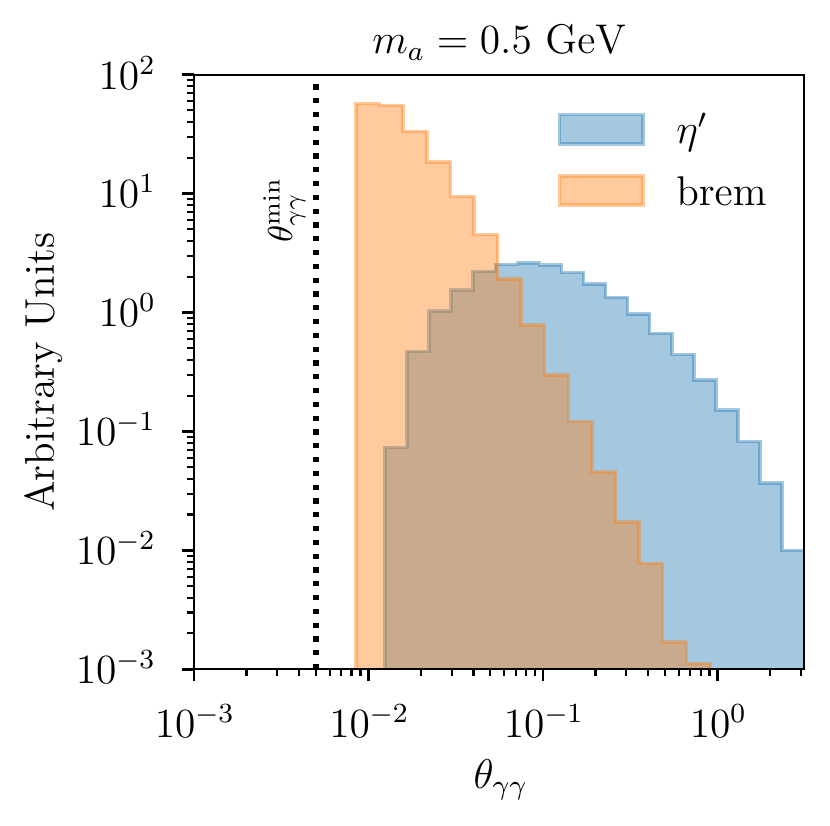}
    \caption{Distributions of angular separation between the photons in $a\rightarrow \gamma\gamma$ for two gluon-coupled ALP production mechanisms (rare $\eta'$ decays and proton bremsstrahlung) for $m_a = 0.5\;\GeV$.  The dotted line indicates the approximate minimum photon separation assuming 
    the decay happens at $z = 8$ m and the detector is at $z = 19$ m. The sharp edge in the bremsstrahlung histogram is due to the limited range of validity of the ``quasi-real'' approximation (see Sec.~\ref{sec:proton_brem}) which 
    prevents the ALP from carrying away too much of the beam momentum (and therefore cuts off the very collimated part of the phase space).
  }
    \label{fig:decay_photon_separation_gluon_coupled_alp}
\end{figure}

\subsection{Rare Meson Decays: \texorpdfstring{$M \to a + X$}{M to a X} }
ALPs can be abundantly produced in rare decays of $\pi^\pm$, $\eta$, $\eta'$, $K_{L,S}$ and $K^\pm$ mesons. 
We do not consider the $\pi^\pm$ production channel here only because that parameter space is well covered by existing searches. 
As in Sec.~\ref{sec:alp_production_from_photons}, we use \pythia 8.240 to model meson production rate and kinematics.
We find that the number $\eta^{(\prime)}$ and $K$ mesons produced per $pp$ interaction is 
\beq
n_{\eta} \approx 0.30, \;\;  n_{\eta'} \approx 0.034
\eeq
and 
\beq
n_{K^+} \approx 0.24,\;\; n_{K^-} \approx 0.15,\;\; n_{K_L} \approx 0.18,\;\;n_{K_S} \approx 0.18.
\eeq
The $\eta$ multiplicity agrees with measurements of Refs.~\cite{Aguilar-Benitez:1991hzq,AxialFieldSpectrometer:1986awq} (albeit at different beam energies); the $\eta'$ multiplicity is consistent with scaling the $\eta$ rate by $\sin^2\theta_{\eta\eta'}$ as a naive estimate. The kaon multiplicities qualitatively match the results of Ref.~\cite{Anticic:2010yg} (see their Fig. 130).

In order to avoid dealing with the attenuation of the proton beam we will 
focus on collisions in the first nuclear collision length of the target. 
Given a total number of protons on target, $\Npot$, the total number of 
mesons produced is then 
\beq
N_M = n_M \Npot \sigma_{pA} n_A T^{(p)}, 
\eeq
where $n_M$ is the number of meson of type $M$ produced per interaction,  $T^{(p)} = 16.77$ cm is a nuclear interaction length in iron and $\sigma_{pA}$ is given in Eq.~\ref{eq:tot_proton_nucleus_xsec}.
For the iron target we have 
\beq
\Npot \sigma_{pA} n_A T^{(p)} = 1.55\times 10^{18}
\;\left(\frac{\Npot}{10^{18}}\right)\left(\frac{T^{(p)}}{16.77\;\mathrm{cm}}\right).
\eeq
The total number of ALPs produced is then
\beq
N_a = N_M \br(M \to a + X).
\eeq
We calculate the branching fractions $\br(M \to a + X)$ using the matrix elements discussed in Sec.~\ref{sec:interactions_with_mesons_and_photons}: 
see Eqs.~\ref{eq:eta_to_a_pi0_pi0} through~\ref{eq:etap_to_a_pip_pim} for $\eta^{(\prime)}$ and Eqs.~\ref{eq:KS_to_pi0_a} to~\ref{eq:Kpm_to_pipm_a} for kaons.
The three-body matrix elements for $\eta^{(\prime)}$ decay only depend on the invariant mass $M_{23}^2 \equiv s$ (we identify particles 2 and 3 with the final state pions) so the partial widths are given by
\beq
\Gamma(\eta^{(\prime)}\to a \pi\pi )
= \frac{1}{256 \pi^3 m_{\eta^{(\prime)}} S} \int_{4m_\pi^2}^{(m_{\eta^{(\prime)}}-m_a)^2} ds \beta_1 \beta_{23} |\mathcal{A}(\eta^{(\prime)}\to a \pi\pi )|^2,
\eeq
where $S$ is a symmetry factor ($=1$ if the final state mesons are distinguishable, and $=2$ if they are not); $\beta_i$ are 
\beq
\beta_1 = \beta(m_a^2/m_{\eta^{(\prime)}}^2,s/m_{\eta^{(\prime)}}^2), 
\beta_{23} = \beta(m_\pi^2/s,m_\pi^2/s)
\eeq
with 
\beq
\beta(x,y) = \sqrt{1 - 2(x+y) + (x-y)^2}.
\eeq
The partial widths for the two-body decays of kaons are simply
\beq
\Gamma(K \to \pi  a) = \frac{1}{8\pi} |\mathcal{A}(K\to \pi a)|^2 \frac{|\mathbf{p}_a|}{m_K^2},
\eeq
where $|\mathbf{p}_a|$ is the magnitude of the ALP three-momentum
\beq
|\mathbf{p}_a|  = \frac{m_K}{2}\beta(m_a^2 / m_K^2,m_\pi^2/m_K^2).
%\frac{1}{2m_K} \left[\left(m_K^2 - (m_\pi+m_a)^2\right)  \left(m_K^2 - (m_a-m_\pi)^2\right)\right]^{1/2}.
\eeq

We simulate ALP production by sampling the integrands in the above expressions, reconstructing the full ALP four-vector and boosting it to the lab frame specified by the meson four-vector from \pythia. The ALPs are displaced and decayed to photons. A typical kinematic distribution of the daughter photons from $\eta'$-produced 
ALPs is shown in the left panel of Fig.~\ref{fig:decay_photon_distribution_gluon_coupled_alp} (other mesons give similar distributions). The angular photon 
separation is shown in Fig.~\ref{fig:decay_photon_separation_gluon_coupled_alp}.

\subsection{Proton Bremsstrahlung: \texorpdfstring{$p A \to a + X$}{p A to X a}}
\label{sec:proton_brem}
In this section we consider a model for ALP bremsstrahlung off protons based on the results of Ref.~\cite{Foroughi-Abari:2021zbm} (see also Ref.~\cite{Boiarska:2019jym}); a similar calculation of gluon-coupled pseudoscalar production geared towards LHC/Forward Physics Facility energies can be found in Ref.~\cite{Foroughi-Abari:2022aaa}. 
A key advantage of this process is that it allows us to simply use the physical, $\kappa$-independent ALP-proton interactions obtained in Sec.~\ref{sec:interactions_with_nucleons}.
The idea is to express the rate for $pp \to X + a$ in terms of the cross-section for a well measured process like
$pp \to X$ times a splitting function that encodes ALP radiation (we will generalize the discussion to proton-nucleus collisions at the end of this subsection). 
The $pp$ cross-section is dominated by low-momentum transfer reactions 
involving non-perturbative physics that can be modelled using Pomeron and meson exchange. Since we 
are considering relatively light ALPs, $pp \to X + a$ should still be dominated by similar processes. 
The $pp$ processes can be divided into elastic non-diffractive (both protons remain intact), 
single-diffractive (SD, one proton is dissociated, producing a 
multiparticle hadronic state $X$) or double-diffractive (DD, both protons dissociate). It was emphasized in Ref.~\cite{Foroughi-Abari:2021zbm} that if the underlying reaction is elastic or SD, the radiation of any light state from initial and final state protons 
interferes, leading to a severe cancellations in the emission rate.\footnote{Experimentally, the elastic, SD and DD processes 
are identified by the rapidity gap between final state hadrons.} 
As a result, we expect the 
production rate to be dominated by non-elastic, non-single-diffractive events (NSD) in which the initial state radiation of the ALP 
from the beam proton (emission from the target is suppressed -- see Ref.~\cite{Foroughi-Abari:2021zbm}) does not interfere with other possible 
contributions.

Under certain kinematic conditions (roughly small momentum transfers, i.e. large beam energy, small ALP mass and forward production) 
ALP bremsstrahlung can be factorized from the underlying $pp$ scattering, an approximation termed ``quasi-real'' in Ref.~\cite{Foroughi-Abari:2021zbm} because the 
slightly-off-shell intermediate proton is approximated as on-shell. We summarize the main results here, with a detailed derivation 
given in Appendix~\ref{sec:proton_brem_details}.
We start with the following interaction
\beq
\mathscr{L} \supset g_{pa}(\partial_\mu a) \bar{p} \gamma^\mu \gamma_5 p
\eeq
where $g_{pa}$ is computed in Sec.~\ref{sec:interactions_with_nucleons}. 
Under certain conditions the spin-summed and averaged matrix element can be written as
\beq
\overline{|\mathcal{A}(pp \to a + X)|^2} =  g_{pa}^2\left(\frac{z}{H}\right)^2 \left[\frac{\left(m_a^2 + p_T^2 -m_a^2 z - m_p^2 z^2\right)^2}{z^2(1-z)} + \frac{4 m_p^2 p_T^2}{1-z}\right] \overline{|\mathcal{A}(pp\to X)|^2},
\label{eq:brem_amp_partial}
\eeq
where $z$ is momentum fraction of the initial beam carried away by the ALP, $p_T$ is the ALP transverse momentum, $H = p_T^2 + z^2 m_p^2 + (1-z)m_a^2$ is related to the invariant mass of the intermediate (off-shell) proton and $\mathcal{A}(pp\to X)$ is the amplitude for the underlying hadronic interaction.
The differential cross-section can then be written as
\beq
\frac{d\sigma(pp \to a + X)}{dp_T^2 dz} \approx w_a(z,p_T^2)  \sigma(s')  
\label{eq:differential_brem_xsec}
\eeq
where the ALP splitting function $w_a$ is
\beq
w_a(z,p_T^2) = \frac{g_{pa}^2}{16\pi^2 }\frac{(1-z)z}{H^2} \left[\frac{\left(m_a^2 + p_T^2 -m_a^2 z - m_p^2 z^2\right)^2}{z^2(1-z)} + \frac{4 m_p^2 p_T^2}{1-z}\right]
\label{eq:alp_brem_splitting_function}
\eeq
and $\sigma(s')$ is the cross-section of the underlying hadronic process $pp \to X$, evaluated at a slightly different center-of-mass energy with $s' \approx 2m_p p_p (1-z) $ with $p_p$ the beam momentum.
This expression is not yet complete as we need to dress it with form-factors to account for the non-point-like nature of the beam particle; we discuss this 
in the following subsection. The factorization of ALP radiation in Eq.~\ref{eq:differential_brem_xsec} requires several assumptions that limit its range of validity; 
these assumptions are detailed in Ref.~\cite{Foroughi-Abari:2021zbm} and discussed in Appendix~\ref{sec:proton_brem_details}. Their physical content is to require the ALP, beam and recoil proton to be ultrarelativistic. This limits the range of $z$, $p_T$ and $m_a$ one can consider in this approximation. These conditions are achieved when the beam energy is much larger than other energy scales in the process; this is a good assumption for the DarkQuest configuration with a 120 GeV beam and a forward detector (limiting the range of relevant $p_T$) searching for ALPs with $m_a \lesssim \;\GeV$.

The above discussion followed closely Ref.~\cite{Foroughi-Abari:2021zbm} and focused on $pp$ collisions. 
However, it is easy to translate it to proton-nucleus collisions which are of more direct 
relevance for DarkQuest. The only modification is to replace the $pp$ NSD cross-section $\sigma(s')$ by an equivalent quantity for $pA$ scattering. 
In order to avoid the initial/final-state radiation interference discussed in~\cite{Foroughi-Abari:2021zbm} we focus on processes where the beam proton, or both beam proton and target nucleus are disrupted. This corresponds to any inelastic scattering process except for the ones where only the target nucleus is excited (termed target single diffractive, TSD, in~\cite{Carvalho:2003pza}).
We use fits to data for inelastic and TSD cross-sections from Ref.~\cite{Carvalho:2003pza}. The resulting cross-section is 
\begin{subequations}
\begin{align}
  \sigma(s)  & = \sigma_{\mathrm{inel}} - \sigma_{\mathrm{TSD}} = 43.55\;\mathrm{mb}\; A^{0.7111} - 3.84\;\mathrm{mb}\; A^{0.35} \\
& \approx 762\;\mathrm{mb}\;\left(\frac{A}{56}\right)^{0.7111}\left[1 - 0.021\left(\frac{56}{A}\right)^{0.36}\right]
\end{align}
\label{eq:proton_nucleus_nsd_xsec}
\end{subequations}
At the relatively low $\sqrt{s}\lesssim 15\;\GeV$ of interest, these 
cross-sections are very weakly dependent on $s$ (see, e.g., Ref.~\cite{Carroll:1978hc}) and 
we treat them as constant.

We simulate ALP production and compute the rate by sampling Eq.~\ref{eq:differential_brem_xsec} (with the cross-section in Eq.~\ref{eq:proton_nucleus_nsd_xsec} and the form-factors discussed in the next subsection) using \texttt{vegas}~\cite{Lepage:1977sw,peter_lepage_2020_3897199} and reconstructing kinematics using Eq.~\ref{eq:brem_kinematics}. The resulting bremsstrahlung cross-section is compared other processes in Fig.~\ref{fig:cross_sections_from_gluon_coupling}.
The ALPs are displaced and decayed to photons. A typical kinematic distribution of the daughter photons is shown in the right panel of Fig.~\ref{fig:decay_photon_distribution_gluon_coupled_alp}. The angular photon separation is shown in Fig.~\ref{fig:decay_photon_separation_gluon_coupled_alp}. We note that 
bremsstrahlung is able to produce heavier ALPs than rare meson decays, and these ALPs tend to be much more boosted. As a result, the decay photons are more collimated.

\subsubsection{Form Factors}
The previous calculation assumed that the protons and mesons are point-like. In order to account for their finite size and internal structure we include form factors following Ref.~\cite{Foroughi-Abari:2021zbm}. This means that each ALP-proton vertex should be multiplied 
by a scalar function of momenta; this can depend on any Lorentz invariant quantity relevant for the vertex, but momentum conservation implies that without loss of generality we can take 
\beq
F = F(p^2, (p-k)^2, k^2), 
\eeq
where $p$ and $p-k$ are the proton momenta before and after ALP radiation, and $k$ is the ALP momentum. In most discussions of proton form factors one assumes that the protons are on-shell, in which case the only non-trivial argument, $k$, is space-like. We are, however, interested in the situation where $p'$ is slightly off-shell and $k^2 = m_a > 0$. There is no data available for such a time-like form-factor, so we must construct a model. Following Ref.~\cite{Foroughi-Abari:2021zbm}, we take the incoming proton to be on-shell, and write the total form factor as 
\beq
F = F_1(k^2) F_{pp^*}((p-k)^2).
\eeq
We will refer to $F_1$ as the time-like form factor; $F_{pp^*}$ is inserted to control the factorization of the ALP bremsstrahlung cross-section into the form of Eq.~\ref{eq:differential_brem_xsec}, so it is chosen to be equal to $1$ when $p-k$ is on-shell and falls off as it becomes off-shell. 
As in Ref.~\cite{Foroughi-Abari:2021zbm} we use  
\beq
F_{pp^*}((p-k)^2) = \frac{\Lambda^4}{\Lambda^4 + ((p-k)^2 - m_p^2)^2},
\eeq
where $\Lambda$ is an unknown parameter $\mathcal{O}(\GeV)$. We will take $\Lambda = 1\;\GeV$. This form-factor has 
been successfully used to fit a variety of experimental data involving nucleon-nucleon-meson interactions~\cite{Feuster:1998cj,Penner:2002md} with $\Lambda \approx 1 \;\GeV$; while these 
datasets have $\sqrt{s}\lesssim 2\;\GeV$, the typical intermediate off-shell nucleon invariant mass squared is just $s$ for their $2\to 2$ reactions. 
For the bremsstrahlung process, $(p-k)^2 - m_p^2 = -H/z$ where $H$ is given below Eq.~\ref{eq:brem_amp_partial}; the form-factor therefore 
suppresses soft and high $p_T$ ALP emission, both regimes where the ``quasi-real'' factorization is spoiled~\cite{Foroughi-Abari:2021zbm}.
It would be interesting to fully validate this approach in a kinematic region similar to DarkQuest, with $\sqrt{s} \sim 15\;\GeV$; for example, the same 
approach can be used to study the forward emission of single $\pi^0$ or $\eta$. Unfortunately we are not aware of such a data set.

We now turn to the time-like form factor $F_1$. 
There are few studies on axial vector time-like form factors and no data in the relevant kinematic region. Ref.~\cite{Adamuscin:2007fk} and references therein describe how to analytically continue a space-like axial vector form-factor to the time-like region. In order to do this one needs to know the singularity structure of the form-factor; this is not captured by the usual ``dipole'' form factors that are fit or data, or to lattice results (see, e.g., \cite{Alexandrou:2020okk,Alexandrou:2021wzv}). Thus, Ref.~\cite{Adamuscin:2007fk} uses a more physical ``two-component'' model in the space-like region and analytically continues that. This model is 
\beq
F_a(k^2) |_\text{spacelike} = \frac{1}{(1-\gamma k^2)^2}\left[1 - \alpha + \alpha \frac{m_{a_1}^2}{m_{a_1}^2 - k^2}\right].
\eeq
The first factor is meant to describe the ``intrinsic structure'' of the nucleon consisting of three valence quarks. The second factor describes the contribution of resonances. We are specializing to the axial vector isovector current coupling for now; the lightest meson with the right quantum numbers $I^G (J^{PC}) = 1^- (1^{++})$ is the $a_1(1260)$ with $m_{a_1} = 1.23\;\GeV$ and $\Gamma_{a_1} = 400\;\MeV$. $\gamma$ and $\alpha$ are parameters, while $m_{a_1}$ is fixed. $\gamma$ is fit to electromagnetic scattering data with the result $\gamma = 0.515\;\GeV^{-2}$, leaving only $\alpha$ to fit to axial data in the space-like region. Ref.~\cite{Adamuscin:2007fk} obtained $\alpha = 0.95$ for a joint fit to multiple data sets with $-k^2 \lesssim 2 \;\GeV^2$ . The authors then proceed to extend this two-component model into the time-like domain; since there are singularities along $k^2 > 0$, phases arise, requiring the introduction of an additional parameter, $\delta$:
\beq
F_a(k^2) |_\text{timelike} = \frac{1}{(1-\gamma e^{i\delta} k^2)^2}\left[1 - \alpha + \alpha \frac{m_{a_1}^2(m_{a_1}^2 - k^2 + i m_{a_1} \Gamma_{a_1})}{(m_{a_1}^2 - k^2)^2 + m_{a_1}^2 \Gamma_{a_1}^2}\right].
\label{eq:timelike_isovector_ff_model}
\eeq
$\delta =  0.397$ was found in Ref.~\cite{Bijker:2004yu}. We emphasize that this form-factor is an extrapolation into the time-like region, where there is no data; some proposed measurements are discussed in Ref.~\cite{Adamuscin:2007fk}.

The previous discussion focused on the isovector axial form-factor. 
However, in the physical basis the ALP-proton coupling has both axial isovector and isoscalar pieces (this is easily seen in Eq.~\ref{eq:pseudoscalar_proton_lang_2f}). 
We have not found construction of a time-like form-factor for the axial isoscalar coupling similar to Eq.~\ref{eq:timelike_isovector_ff_model}, so we develop one here. We will use two lucky coincidences. First, in the space-like region the lattice results which provide dipole fits for the isovector~\cite{Alexandrou:2020okk} and isoscalar~\cite{Alexandrou:2021wzv} couplings find roughly the same ``axial mass'' (fitting parameter to the dipole form-factor which determines its $Q^2$ dependence) within uncertainties ($m_A = 1.169(72)(27)\;\GeV$ vs $m_A = 1.261\pm 0.188\;\GeV$). The second coincidence is that the relevant meson for axial isoscalar coupling with $I^G (J^{PC}) = 0^- (0^{+-})$ is the $h_1$ with a mass $m_{h_1} = 1.166\;\GeV$ and $\Gamma_{h_1} = 0.375\;\GeV$. Luckily these numbers are essentially the same as for the $a_1$ and the isovector case! The remaining parameters are $\alpha$ and $\delta$. $\delta$ is associated with the intrinsic form-factor of the proton (i.e., the valence quark distribution), so we can again use $\delta = 0.397$; $\alpha$ would need to be re-fit to space-like data. However, the lattice results above produce similar $Q^2$ dependence in the isoscalar and isovector form-factors, so we expect that the isoscalar $\alpha$ would be similar to the isovector one; therefore we use $\alpha\approx 0.95$.

To summarize, we argued that a reasonable first approximation for the time-like form-factors is to use the same momentum dependent form factor for both axial isovector and axial isoscalar couplings of the ALP to protons (the normalizations of these form factors are still different, and captured by the values of $g_A$ and $g_0$, or, equivalently, by $D$, $F$ and $D_s$). 
In any case, the time-like form factor is evaluated at $k^2 = m_a^2$ and can be factored out from the overall rate. So if better models become available, projections or experimental constraints on the ALP coupling can be simply rescaled.

\subsection{Other Channels}
Production of heavier ALPs is no longer coherent over the proton or the target nucleus. Moreover, at larger masses some of the assumptions used 
in the bremsstrahlung calculation can become invalid, as does the use of chiral perturbation theory. We therefore estimate 
the ALP production cross-section in gluon fusion at larger ALP masses. Since the inclusive ALP production rate should be continuous as a function of mass 
this calculation also serves as a sanity check for the bremsstrahlung cross-section in mass range $m_a \sim 1 - 2\;\GeV$. The inclusive rate for gluon-gluon fusion can be written as 
\beq
\sigma(gg \to a  + X) = \frac{\cGG^2 \alpha_s^2 m_a^2}{64 \pi f^2 s}
\int^1_{m_a^2/s} \frac{dz}{z} f_g(z, m_a^2) f_g(\frac{m_a^2}{s z},m_a^2)
\eeq
where $s\approx 2 m_p E_\mathrm{beam}$ ($\sqrt{s} \approx 15\;\GeV$ for DarkQuest) and $f_g(x,Q^2)$ is the gluon PDF.
We use the \texttt{CT18NLO} PDFs~\cite{Hou:2019efy} via \texttt{lhapdf}~\cite{Buckley:2014ana} to 
evaluate the cross-section which is shown in Fig.~\ref{fig:cross_sections_from_gluon_coupling}. 
We have scaled the cross-section by $A \approx 56$ to account for the number of nucleon targets per iron nucleus.
This gluon fusion line terminates at $m_a \approx 1.3\;\GeV$ since the PDFs cannot be evaluated at lower $Q$ values. We see that our estimate of the bremsstrahlung cross-section clearly does not saturate the hadronic cross-section at $m_a \gtrsim 2\;\GeV$. 
This is acceptable for our estimate of the DarkQuest sensitivity which does not extend much above a $\GeV$.
It might be possible to improve agreement at lower masses by including ALP mixing with heavier pseudoscalar resonances.

\section{Sensitivity Projections and Existing Constraints}
\label{sec:sensitivity}
In this section we use the production mechanisms from Secs.~\ref{sec:alp_production_from_photons} and~\ref{sec:alp_production_from_gluons} 
and the experimental set-up described in Sec.~\ref{sec:dq_overview} to estimate the sensitivity of DarkQuest to photon- and gluon-coupled ALPs. In both scenarios we consider only the decays $a\to \gamma\gamma$; in the gluon-coupled case sensitivity can be improved further by considering hadronic decay modes as well. In Sec.~\ref{sec:bg} we argued that additional shielding 
is likely needed to suppress SM processes that can mimic the signal; following Ref.~\cite{Berlin:2018pwi} we therefore require ALP decays to occur between $z = 7$ and $8$ m after the front of FMAG. We will select events where both photons from $a\to\gamma \gamma$ hit the $2 \text{ m} \times 2\text{ m}$ ECAL placed at $z = 19$ m with $E_\gamma \geq 1\;\GeV$. The ECAL modules have transverse granularity of $5.5$ cm, so we will also consider a requirement 
on minimum photon separation $\geq 5.5$ cm~\cite{Aphecetche:2003zr}. This can be used as an additional handle for rejecting SM processes with more than 2 photons as briefly mentioned in Sec.~\ref{sec:bg}. 
If DarkQuest is further enhanced with a preshower detector in front of the ECAL, an 
invariant mass measurement may be possible. Finally we will study the sensitivity of two phases of DarkQuest with $\Npot = 10^{18}$ and $10^{20}$. 

We only consider ALP production in the front of the FMAG. For photon-coupled ALPs this means the first 
interaction length for $\gamma$ fusion and the production of secondary photons in the first interaction length followed by 
the Primakoff process within the first radiation length -- see Sec.~\ref{sec:alp_production_from_photons}. 
For gluon-coupled ALPs this means we assume $\eta^{(\prime)}$ production and $pA$ bremsstrahlung occur in the first interaction length.
This simplification enables us to consider a mono-energetic $120$ GeV proton beam and results 
in a conservative estimate of the signal yield.

\subsection{Photon coupling}
\begin{figure}
    \centering
    \includegraphics[width=0.47\textwidth]{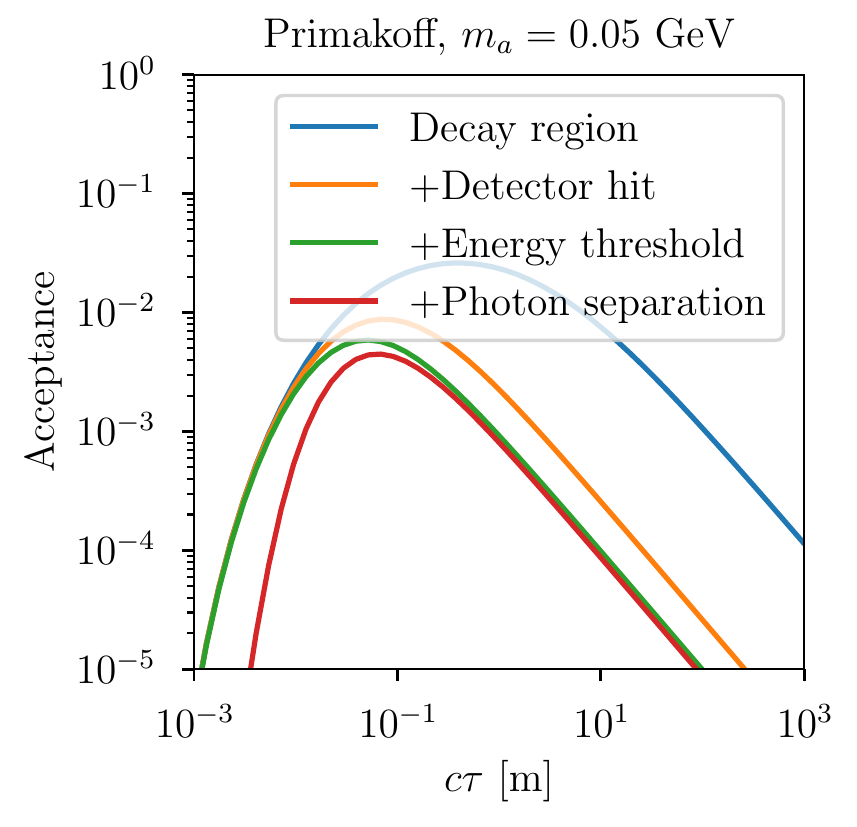}
    \includegraphics[width=0.47\textwidth]{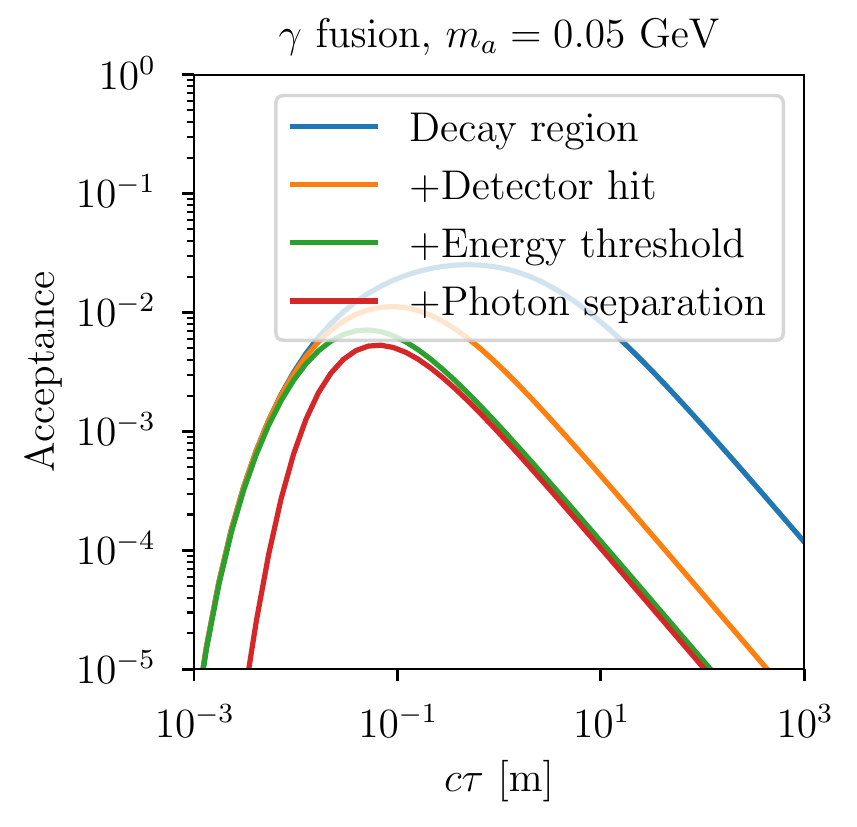}\\
    \includegraphics[width=0.47\textwidth]{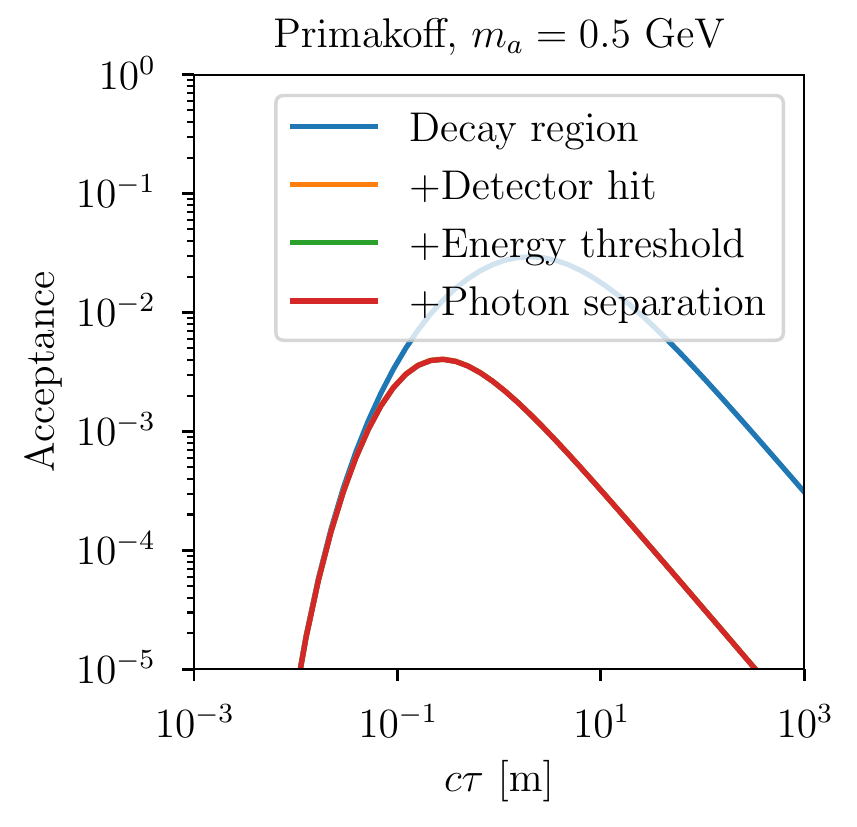}
    \includegraphics[width=0.47\textwidth]{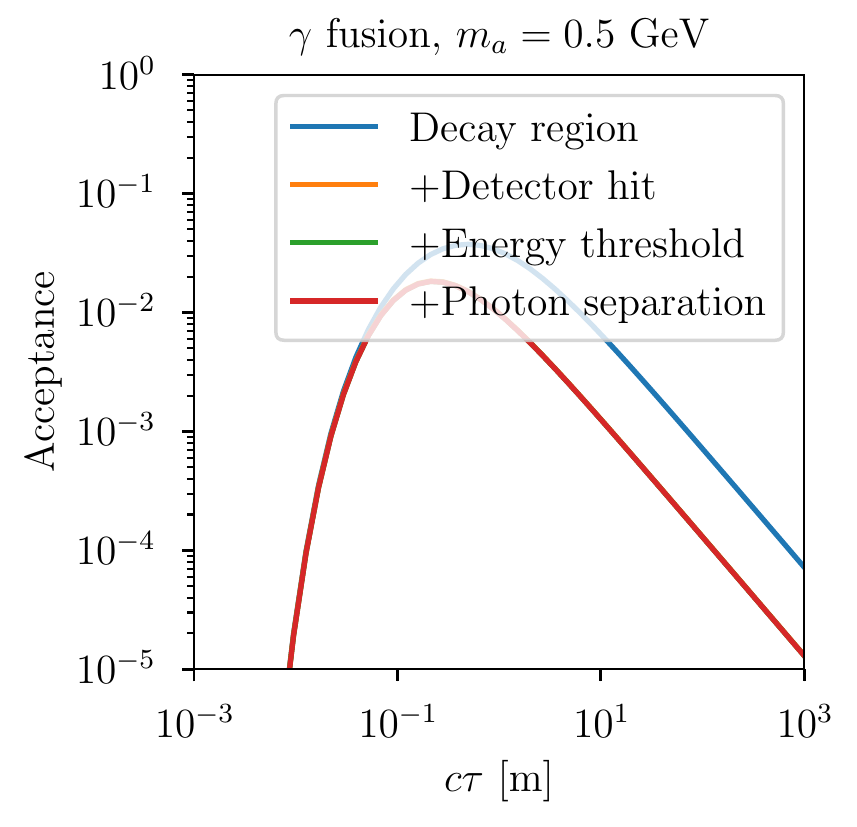}
    \caption{
    Accepted event fraction in the photon-coupled ALP model for different $m_a$, production mechanisms and 
    analysis selections. The upper row corresponds to $m_a = 0.05$ GeV, while the 
    lower row has $m_a = 0.5\;\GeV$. The left (right) column shows the acceptance for 
    ALPs produced via Primakoff ($\gamma$ fusion) processes. Each panel shows the 
    cut flow for ALPs decaying in the fiducial region, decay photons hitting 
    the ECAL, and satisfying the energy threshold and photon separation requirements.}
    \label{fig:acceptance_cut_flow_photon_production}
\end{figure}

The effect of the selections outlined above on event acceptance is 
shown in Fig.~\ref{fig:acceptance_cut_flow_photon_production} for two benchmark masses and the two main photon-coupled ALP production mechanisms. We see that geometric requirements on the decay vertex and 
decay photons are the dominant sources of signal ``loss''. At small masses, the 
requirement of well-separated photons limits the sensitivity to short lifetimes since 
events more boosted ALPs also give more collimated photons.
We combine these acceptances with the total cross sections shown in Fig.~\ref{fig:cross_sections} to derive the sensitivity of DarkQuest to photon-coupled ALPs. 

Figures~\ref{fig:dq_reach_primakoff_vs_fusion},~\ref{fig:dq_reach} and~\ref{fig:dq_reach_and_other_projections} are the main results of this work for the photon-coupled ALP and show that DarkQuest can significantly improve on existing constraints, shown as shaded gray in these plots (these are briefly described in the following subsection). 

As was pointed out in Ref.~\cite{Berlin:2018pwi}, ALP production via the Primakoff mechanism from secondary photons is the dominant process. However, despite having a smaller cross-section, $\gamma^*\gamma^*$ fusion produces somewhat more boosted ALPs making this channel nearly as powerful in the short lifetime regime. This is evident in Fig.~\ref{fig:dq_reach_primakoff_vs_fusion} where we show the sensitivity of each channel separately. In this and following figures we define the sensitivity contours as 10 signal events following Ref.~\cite{Berlin:2018pwi}.

Next we consider the impact of the photon separation selection in the ECAL. 
The sensitivity with and without this cut is shown in the left panel of Fig.~\ref{fig:dq_reach}. The requirement of observing well-separated photons is not stringent in most of the parameter space. It only starts to penalize the reach for $m_a \lesssim 0.05\;\GeV$. In the right panel of that figure we compare the sensitivity 
of the two proposed phases. Both phase 1 and phase 2 are sensitive to new parameter space. Phase 2 does not substantially improve the reach in the short lifetime/large coupling 
regime because of the exponentially falling acceptance in that direction. However, 
phase 2 gains almost a factor of 2 in the mass reach over phase 1. 

Finally, we compare the sensitivity of DarkQuest phase 1 to other near-future 
accelerator efforts in Fig.~\ref{fig:dq_reach_and_other_projections}. There we show projections for phase 1 of FASER~\cite{Feng:2018pew} with 300 fb$^{-1}$; NA62 with $10^{18}$ POT~\cite{Dobrich:2019dxc}; NA64 with $5\times 10^{12}$ EOT~\cite{Dusaev:2020gxi}; phase 0 of LUXE-NPOD~\cite{Bai:2021dgm}; Belle II with 20 fb$^{-1}$~\cite{Dolan:2017osp}; and a reanalysis of existing PrimEx data~\cite{Aloni:2019ruo}. These experiments are either 
under construction, are already built or even have completed running, so results 
can be expected on a timescale comparable to the DarkQuest timeline of $\sim 5$ years.
We see that DarkQuest's unique baseline, intensity and beam energy make it extremely competitive and complimentary to other experiments. In fact, DarkQuest is able to probe masses and couplings that are not accessible at any other experiment already in phase 1. 
On a longer timescale, the photon-coupled ALP parameter space will be further probed by future phases of FASER~\cite{FASER:2018eoc} and 
LUXE-NPOD~\cite{Bai:2021dgm}, and by Belle II~\cite{Dolan:2017osp} and DUNE~\cite{Brdar:2020dpr}. Heavier photon-coupled ALPs can be produced 
and detected at LEP~\cite{Abbiendi:2002je}, ATLAS~\cite{Aad:2014ioa,Aad:2015bua}, CMS~\cite{Chatrchyan:2012tv},\footnote{These searches were cast in the photon-coupled ALP parameter space in Ref.~\cite{Knapen:2016moh}.} and in heavy ion~\cite{Knapen:2016moh} and electron-ion collisions~\cite{Liu:2021lan}.

\subsubsection{Existing Constraints}
\begin{figure}
    \centering
    \includegraphics{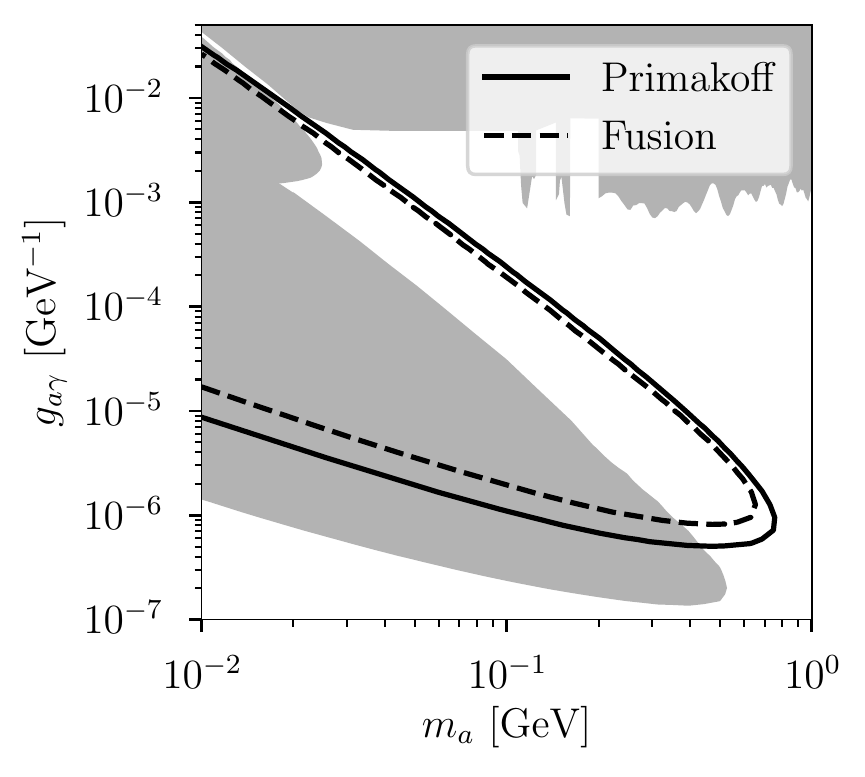}
    \caption{Sensitivity of phase 2 of DarkQuest broken down by ALP production channel. Despite having a smaller cross-section, $\gamma^*\gamma^*$ fusion produces 
    ALPs with a slightly larger boost, leading to competitive sensitivity in the short 
    lifetime regime. Both processes are included in the final projections shown 
    in Figs.~\ref{fig:dq_reach} and~\ref{fig:dq_reach_and_other_projections}. 
    Shaded regions are existing constraints described in the text.}
    \label{fig:dq_reach_primakoff_vs_fusion}
\end{figure}

\begin{figure}
    \centering
    \includegraphics[width=0.47\textwidth]{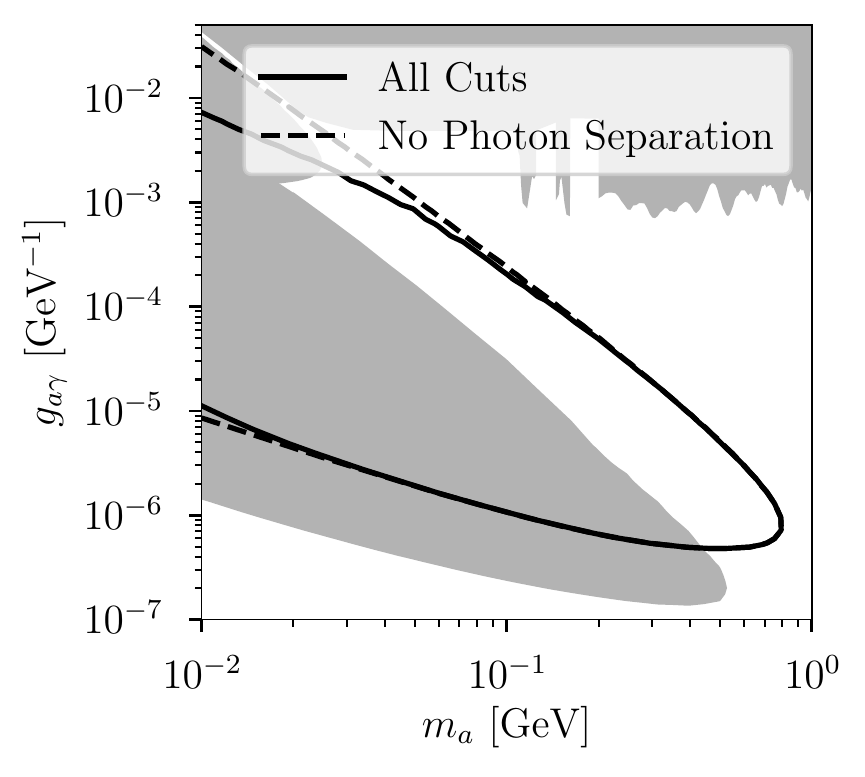}
    \includegraphics[width=0.47\textwidth]{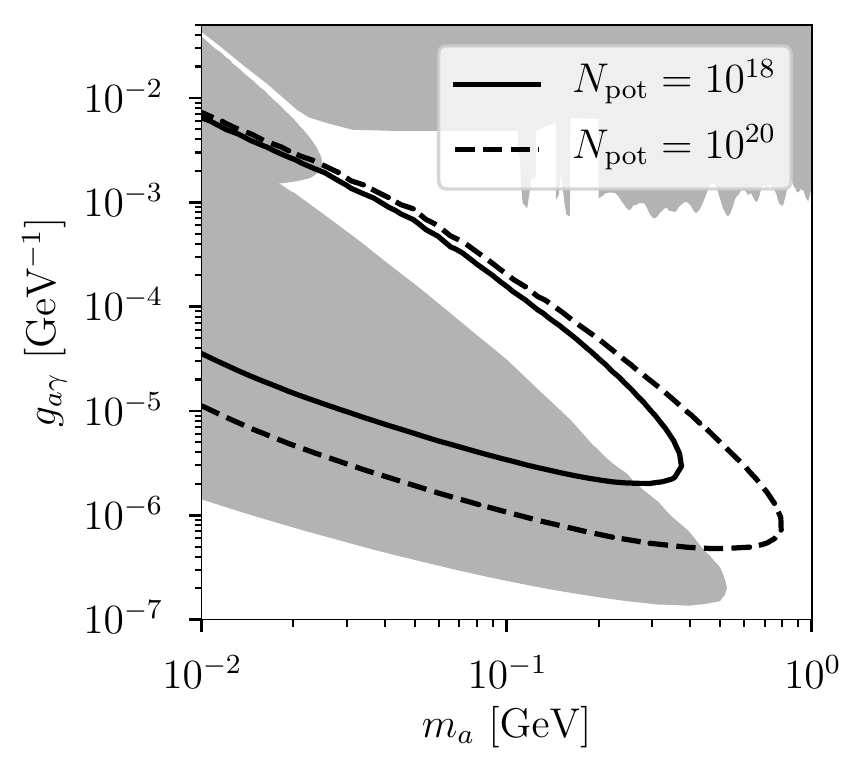}
    \caption{Comparison of DarkQuest sensitivity for different event selections (left panel) and the two proposed phases (right panel). The left plot shows the 
    impact of requiring well-separated photons in the ECAL, which slightly degrades the 
    reach at low masses. The right plot shows the improvement in DarkQuest reach for a second phase with two orders of magnitude more luminosity. Shaded regions are existing constraints described in the text.}
    \label{fig:dq_reach}
\end{figure}

\begin{figure}
    \centering
    \includegraphics{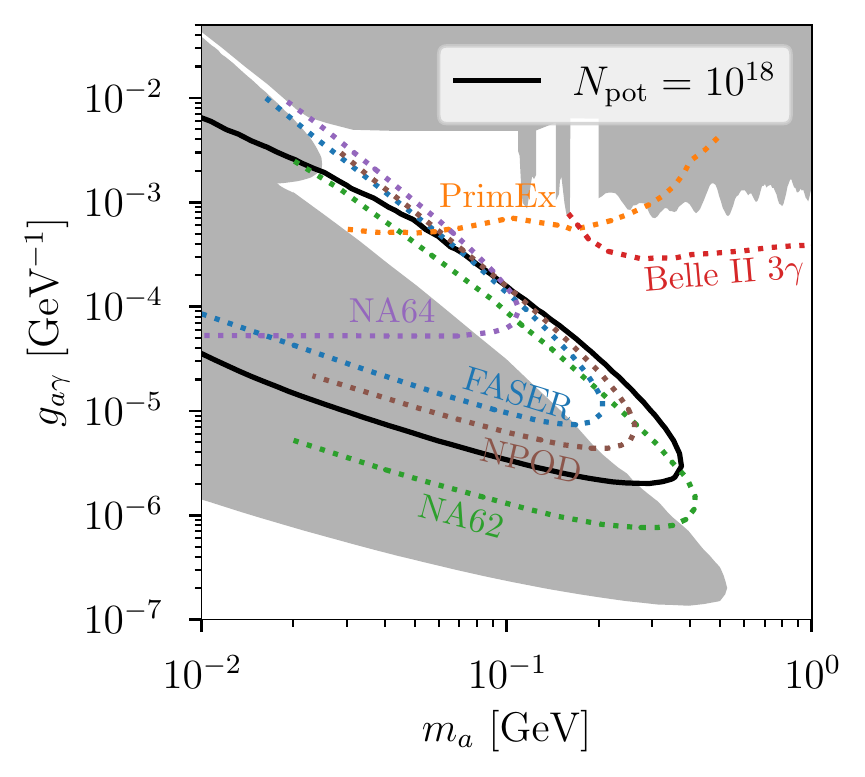}
    \caption{Comparison of phase 1 DarkQuest sensitivity (solid black line) to other near-future accelerator experiments (colored dotted lines). These projections are 
    for phase 1 of FASER~\cite{Feng:2018pew} with 300 fb$^{-1}$, NA62 with $10^{18}$ POT~\cite{Dobrich:2019dxc}, 
    NA64 with $5\times 10^{12}$ EOT~\cite{Dusaev:2020gxi}, phase 0 of LUXE-NPOD~\cite{Bai:2021dgm}, Belle II with 20 fb$^{-1}$~\cite{Dolan:2017osp} and a reanalysis 
    of existing PrimEx data~\cite{Aloni:2019ruo}. Shaded regions are existing constraints described in the text.}
    \label{fig:dq_reach_and_other_projections}
\end{figure}

The constraints on the photon-coupled ALPs are not subject to the same kind of theoretical consistency conditions as the gluon-coupled case, so we 
can make use of existing results. The bounds come from reanalyses of electron beam dump experiments E137 and E141~\cite{Dolan:2017osp}, proton beam dump experiments CHARM and $\nu$CAL~\cite{Dobrich:2019dxc}, $e^+e^-$ colliders LEP~\cite{Jaeckel:2015jla}, BaBar~\cite{Dolan:2017osp} and Belle II~\cite{BelleII:2020fag}, and from the photon-beam experiment PrimEx~\cite{Aloni:2019ruo}. A breakdown of the various constraints is shown in Ref.~\cite{Dolan:2017osp}. We combined all of these bounds in the gray regions of Figs.~\ref{fig:dq_reach_primakoff_vs_fusion},~\ref{fig:dq_reach} and~\ref{fig:dq_reach_and_other_projections}.

\subsection{Gluon coupling}
\begin{figure}
    \centering
    \includegraphics[width=0.47\textwidth]{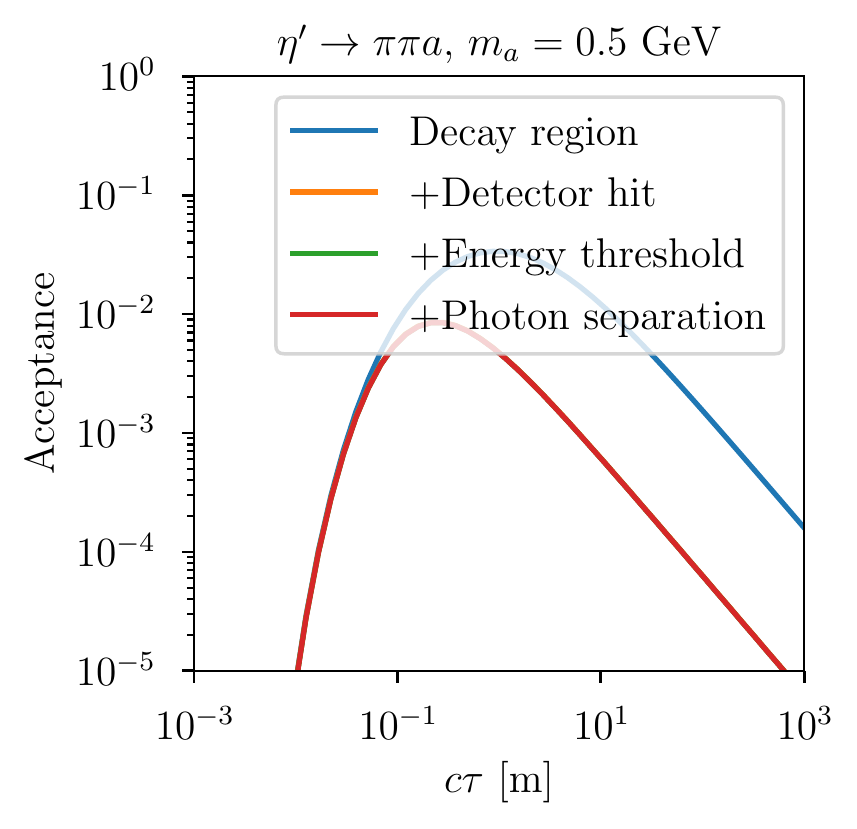}
    \includegraphics[width=0.47\textwidth]{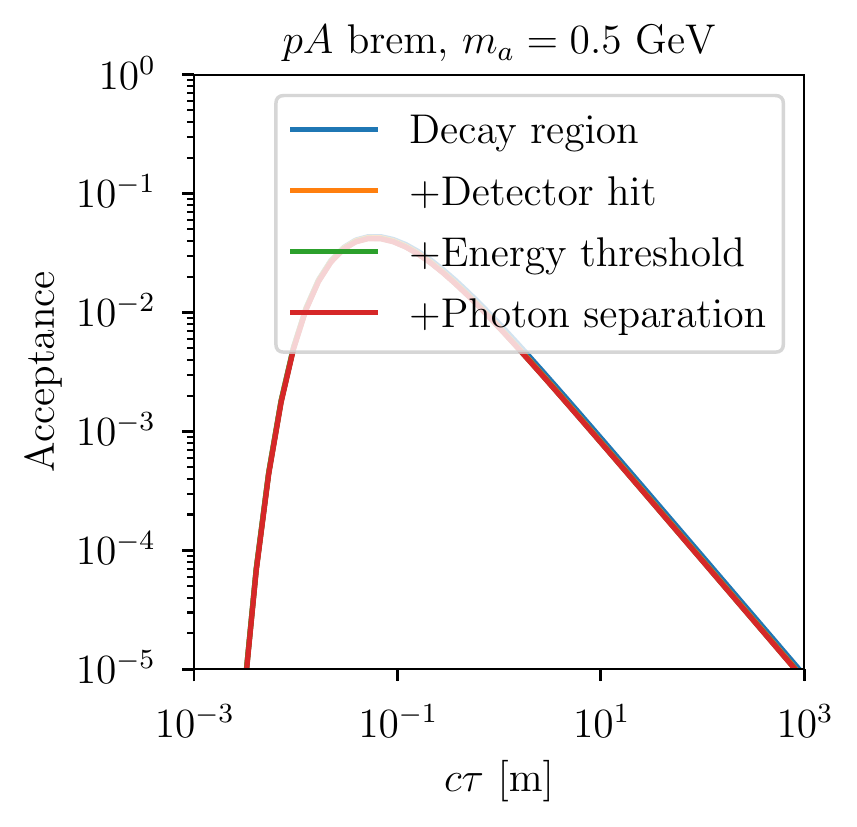}
    \caption{Accepted event fraction in the gluon-coupled ALP model with $m_a = 0.5\;\GeV$ for ALPs produced in rare 
      $\eta^\prime$ decays (left panel) and in proton bremsstrahlung (right panel).    
      In each panel the different-coloured lines correspond to different sets of event selections.
    Each panel shows the 
    cut flow for ALPs decaying in the fiducial region, decay photons hitting 
    the ECAL, and satisfying the energy threshold and photon separation requirements. In the right panel 
  all lines approximately overlap.}
    \label{fig:acceptance_cut_flow_gluon_production}
\end{figure}

The effect of experimental selections outlined above on event acceptance is 
shown in Fig.~\ref{fig:acceptance_cut_flow_gluon_production} for $m_a = 0.5\;\GeV$ and two gluon-coupled ALP production mechanisms (we will see 
that DarkQuest can cover new parameter space around this mass; lower masses are more constrained). 

We see that geometric requirements on the decay vertex and 
decay photons are the dominant sources of acceptance loss. At this mass, the 
requirement of well-separated photons does not limit the sensitivity. 
It is interesting to note that ALPs arising from $pA$ bremsstrahlung are very forward, so their acceptance 
is only dictated by the probability to decay within the fiducial volume: additional selections have no impact on the acceptance.
We combine these acceptances with the total cross sections shown in Fig.~\ref{fig:cross_sections_from_gluon_coupling} to derive the sensitivity of DarkQuest to ALPs. 

\begin{figure}
    \centering
    \includegraphics[width=0.47\textwidth]{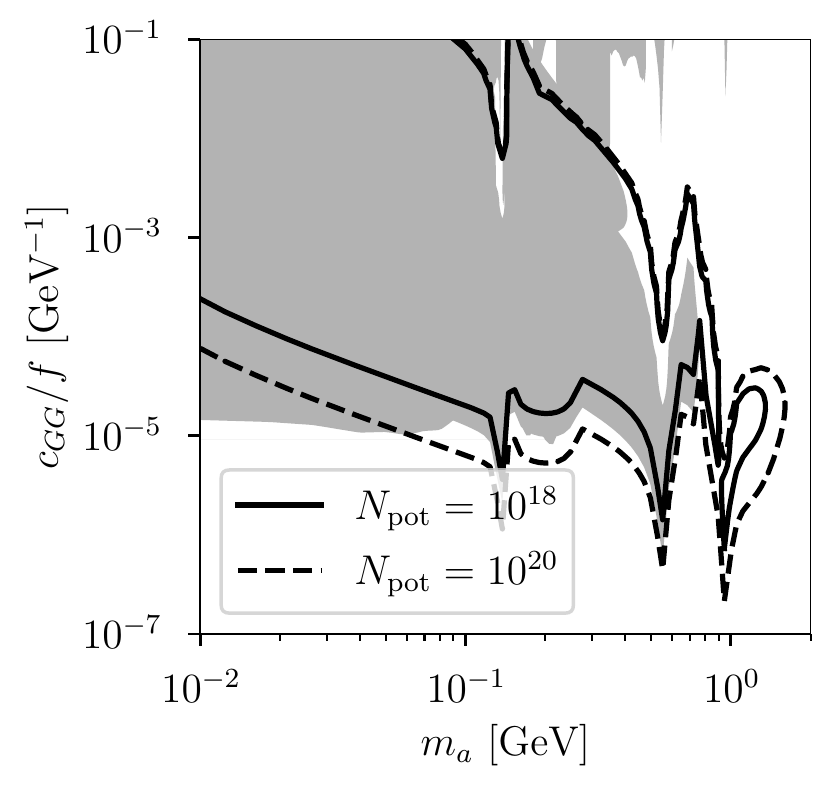}
    \includegraphics[width=0.47\textwidth]{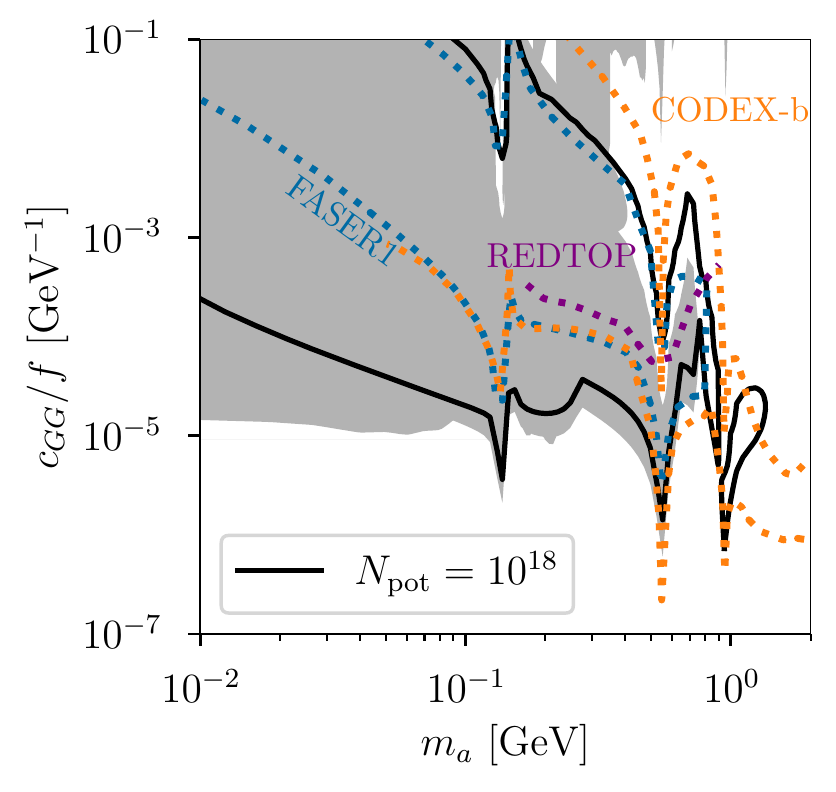}
    \caption{DarkQuest sensitivity to gluon-coupled ALPs decaying to photons. In the left panel we compare 
      the reach of phase 1 and 2 of DarkQuest (black solid and dashed lines, respectively) to existing constraints 
      in gray (these are described in Sec.~\ref{sec:existing_gluon_constraints}). In the right panel 
      we compare phase 1 of DarkQuest to other near-term prospects from FASER (150 fb$^{-1}$)~\cite{FASER:2018eoc}, CODEX-b (300 fb$^{-1}$)~\cite{Aielli:2019ivi} and REDTOP ($10^{17}$ POT)~\cite{Beacham:2019nyx}.}
    \label{fig:dq_gluon_reach}
\end{figure}

Figure~\ref{fig:dq_gluon_reach} is the main result of this work for the gluon-coupled ALP and shows that DarkQuest can improve on existing constraints shown as shaded gray (these are discussed in the following subsection). In particular DarkQuest can access new parameter space with $m_a \gtrsim 0.4\;\GeV$, where it is competing against $\nu$CAL. This improvement is mainly enabled by DarkQuest's shorter baseline. The ALP branching fraction to photons becomes suppressed in this mass range (see Fig.~\ref{fig:gluon_coupled_alp_lifetime_and_branching_fractions}); this means that sensitivity can be somewhat improved by considering other ALP decay channels, such as $a\to 3\pi$ and $a\to 2\pi+\gamma$.

In the left panel of Fig.~\ref{fig:dq_gluon_reach} we show the projected sensitivity for the two phases of DarkQuest. The futuristic phase 2 offers higher mass and lower coupling reach in the long-decay length regime (the lower part of the sensitivity contours).

Finally, we compare the sensitivity of DarkQuest phase 1 to other near-future proton accelerator efforts in the right panel of Fig.~\ref{fig:dq_gluon_reach}. 
We show the projections for phase 1 of FASER (150 fb$^{-1}$)~\cite{FASER:2018eoc}, CODEX-b (300 fb$^{-1}$)~\cite{Aielli:2019ivi} and REDTOP ($10^{17}$ POT)~\cite{Beacham:2019nyx}.\footnote{Note that we have not re-evaluated these projections in the $\kappa$-independent formalism.}
As for the photon-coupled ALP, we see that DarkQuest's unique baseline, intensity and beam energy make it competitive and complimentary to other experiments. Longer-term prospects for discovering gluon-coupled ALPs include phase 2 of FASER~\cite{FASER:2018eoc}, MATHUSLA~\cite{Chou:2016lxi,Beacham:2019nyx} and DUNE~\cite{Kelly:2020dda}. Heavier gluon-coupled ALPs can be produced and detected at Belle II~\cite{Bertholet:2021hjl}, LHCb~\cite{CidVidal:2018blh}, ATLAS~\cite{Aad:2014ioa,Mariotti:2017vtv} and CMS~\cite{CMS-PAS-HIG-14-037,CMS-PAS-HIG-17-013,CMS:2017dcz,Mariotti:2017vtv}, especially using ``triggerless'' techniques~\cite{Knapen:2021elo}.

\subsubsection{Existing Constraints}
\label{sec:existing_gluon_constraints}
\begin{figure}
    \centering
    \includegraphics{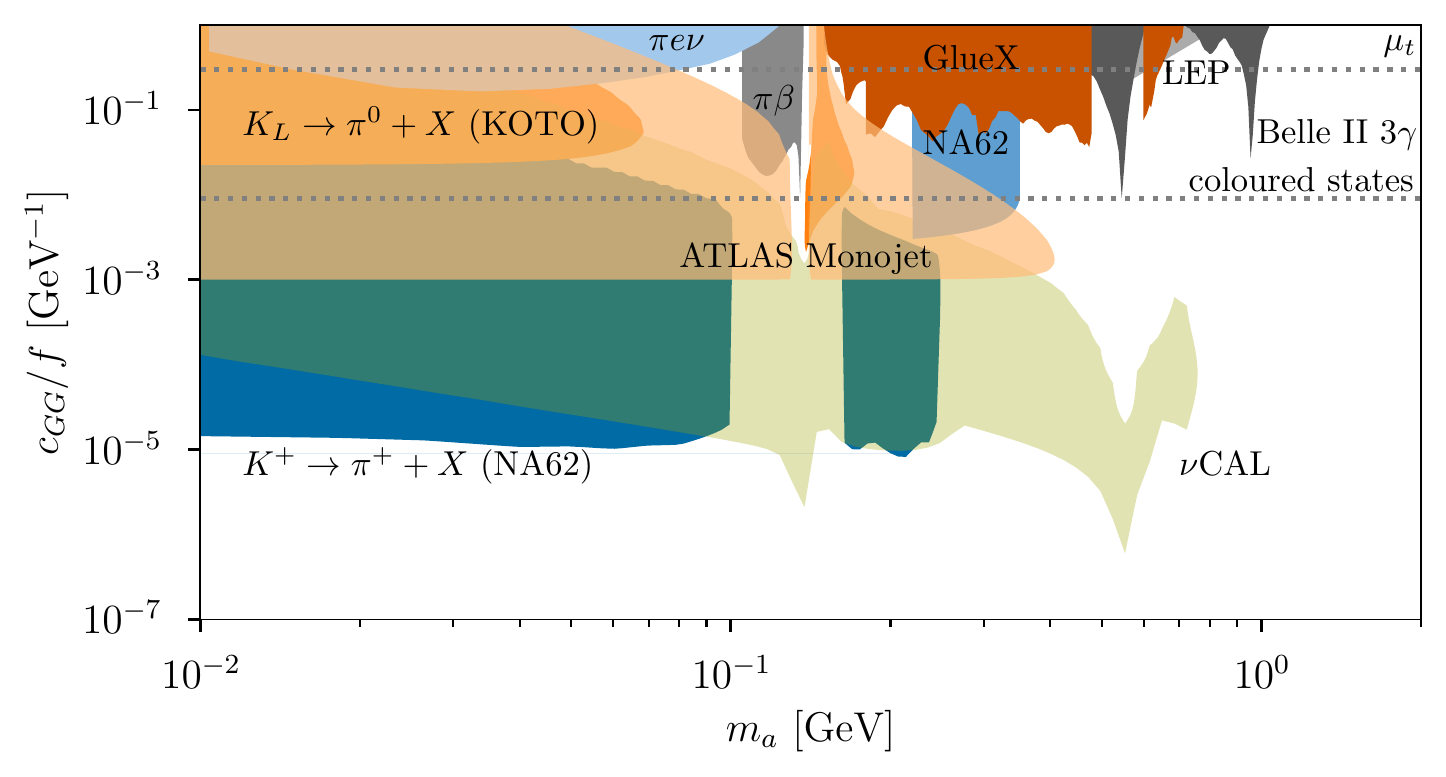}
    \caption{Constraints on the gluon-coupled ALPs evaluated in this work. The recasting procedure for each bound is described in Sec.~\ref{sec:existing_gluon_constraints}. Note that we are considering the model where only the gluon coupling is present at low scales. If we defined our theory at a high scale, RG evolution would 
    generate a multitude of other interactions leading to additional constraints~\cite{Chakraborty:2021wda,Bauer:2021mvw}. The dashed lines indicate the presence of new 
  QCD-charged matter at the TeV scale (lower dashed line) and a chromomagnetic dipole moment of the top quark (upper dashed line).}
    \label{fig:gluon_coupled_alp_constraints_only}
\end{figure}

Re-interpretation of existing searches is sensitive to the assumed model for ALP production and decay. 
This is particularly important for the gluon-coupled ALP because of the theoretical consistency issues relating to the cancellation of 
unphysical parameters. 
It is therefore beneficial to consider the entire set of existing searches within the same, consistent framework. 
While a comprehensive recasting of all gluon-coupled ALP constraints is beyond the scope of this work, we have performed simplified 
re-analyses of several experimental results, which provide the leading constraints in the parameter space relevant for DarkQuest. 
The complementarity of various experimental probes is shown in Fig.~\ref{fig:gluon_coupled_alp_constraints_only}; we discuss these results below.
For an exhaustive compilation of various bounds, see, e.g., Refs.~\cite{Chakraborty:2021wda,Bauer:2021mvw}.

We consider the following experimental results:
\begin{itemize}
  \item KOTO search for $K_L \to \pi^0 +\xspace \mathrm{invisible}$~\cite{KOTO:2018dsc}: This result was considered in the context of ALPs in Refs.~\cite{Gori:2020xvq,Bauer:2021mvw}. We follow the analysis procedure of Ref.~\cite{Gori:2020xvq} and generate a sample of $K_L$'s from the observed momentum distribution~\cite{Masuda:2015eta}. We then decay the $K_L$'s into ALPs, and compute their probability to decay outside of the detector volume, while requiring the $\pi^0$ to be in the detector acceptance. We use these weights to construct an effective invisible branching fraction for $K_L \to \pi^0 a$, which is compared to the mass-dependent upper limit of Fig. 4 of Ref.~\cite{KOTO:2018dsc}.
  \item NA62 search for $K^+ \to \pi^+ + \xspace \mathrm{invisible}$~\cite{NA62:2021zjw} was also considered in Refs.~\cite{Gori:2020xvq,Bauer:2021mvw}. In particular, our result for $\Gamma(K^+ \to \pi^+ a)$ matches that of Ref.~\cite{Bauer:2021mvw}; however, we find that their excluded region does not quite match the limits reported by NA62~\cite{NA62:2021zjw} (the experimental result features a gap in sensitivity for $0.1\;\GeV \lesssim m_a \lesssim 0.16\;\GeV$ which is absent in the theory paper). Ref.~\cite{NA62:2021zjw} provides branching fraction limits as a function of lifetime, which we interpolate and translate into the ALP parameter space using the total ALP width.

  \item NA62 search for $K^+ \to \pi^+ a(\gamma\gamma)$~\cite{NA62:2014ybm} constrains ALPs in the mass range $0.22\;\GeV \lesssim m_a \lesssim 0.35\;\GeV$~\cite{Gori:2020xvq}. In the parameter space accessible to this search the ALP decays promptly, so we simply exclude the region where $\br(K^+ \to \pi^+ a)\br(a\to\gamma\gamma) > 1.3\times 10^{-6}$ (corresponding to the $2\sigma$ upper bound on the observed $\br(K^+ \to \pi^+ \gamma\gamma)$). This constraint is labelled ``NA62'' in Fig.~\ref{fig:gluon_coupled_alp_constraints_only} at $m_a\sim 0.3\;\GeV$.

  \item ATLAS Monojet search~\cite{ATLAS:2021kxv} provides an interpretation of their results in terms of a model where the ALP is produced via the gluon coupling but decays invisibly. As for the KOTO search, we estimate an effective invisible branching fraction from the kinematics of the produced ALPs. We simulate $pp\to a + j$ in \texttt{MadGraph} v3.1.1~\cite{Alwall:2014hca} at $\sqrt{s} = 13\;\TeV$, finding a median ALP energy of $\sim 700\;\GeV$ for the event selections of~\cite{ATLAS:2021kxv}.
    For each simulated event we find the probability of the ALP to decay outside of the detector which we take to have a radius of $12.5$ meters (this coarse model of the detector acceptance should be improved in future studies). We then rescale the ATLAS bounds for invisibly-decaying ALPs, $\cGG/f \lesssim 10^{-3}\;\GeV^{-1}$,\footnote{We converted the ATLAS result, $\cGG/f < 8\times 10^{-6}\;\GeV^{-1}$ from the notation of Refs.~\cite{Mimasu:2014nea,Brivio:2017ije} into our convention for the gluon coupling normalization.} by the effective invisible branching fraction.

  \item The old proton beam-dump searches $\nu$CAL~\cite{Blumlein:1990ay,Blumlein:1991xh} and CHARM~\cite{CHARM:1985anb} have been used to constrain a multitude of long-lived particles, including dark photons~\cite{Blumlein:2013cua,Gninenko:2012eq} and inelastic dark matter~\cite{Tsai:2019buq}. We simulated the production of gluon-coupled ALPs for both experiments in processes described in Sec.~\ref{sec:alp_production_from_gluons} and decayed them into photons. We follow the same procedure for estimating the acceptance and signal rate as for DarkQuest, taking into account the different experimental geometries and event selections. We find that $\nu$CAL has superior reach throughout the relevant parameter space, so we only show the $\nu$CAL result.
  \item In Ref.~\cite{Altmannshofer:2019yji} the authors used experimental results from PIENU~\cite{PIENU:2017wbj} and PIBETA~\cite{Pocanic:2003pf} experiments to constrain rare pion decays $\pi^+ \to a e\nu$, where the coupling of the ALP is described by a mixing angle with $\pi^0$, $\sin\theta$. We 
    recast their results into the $\cGG/f$ parameter space by using the following $\kappa$-independent proxy for the mixing angle: 
    \beq
    \sin\theta = \frac{\mathcal{A}(\pi^+ \to a + e^+ \nu)}{\mathcal{A}(\pi^+ \to \pi^0 + e^+ \nu)} = \frac{2}{3} \frac{\cGG f_\pi m_\pi^2 \delta_I}{f (m_\pi^2-m_a^2)},
    \eeq
    which we evaluated using the electroweak chiral Lagrangian in the limit of $m_{\eta^{\prime}} \to \infty$. The resulting 
    constraints are labelled ``$\pi e \nu$'' and ``$\pi \beta$'' in Fig.~\ref{fig:gluon_coupled_alp_constraints_only}.
  \item We rescale the GlueX~\cite{GlueX:2021myx} constraint (based on the analysis of Ref.~\cite{Aloni:2019ruo}) into our gluon coupling normalization (we do not recompute the limit in the $\kappa$-independent formalism).
 
\end{itemize}

We also translated the bounds on the photon coupled-ALPs into the $\cGG/f$ parameter space using the induced photon coupling in Eq.~\ref{eq:two_photon_amp_3f} and accounting for the reduced branching fraction of $a\to \gamma\gamma$ at larger $m_a$. 
Only two photon-only results cover new parameter space compared to the hadronic experiments above. These are the Belle II~\cite{BelleII:2020fag} and LEP~\cite{Jaeckel:2015jla} searches for $e^+ e^- \to \gamma a(\gamma\gamma)$. 

The complementarity of various experimental probes is shown in Fig.~\ref{fig:gluon_coupled_alp_constraints_only}. In addition to direct searches for gluon 
and the (induced) photon coupling, we highlight parameter space that may feature UV-dependent bounds coming from 
additional QCD-charged particles below the TeV scale (the parameter space above the lower dashed line) and the chromomagnetic dipole moment of the top quark $\mu_t$ (the parameter space above the upper dashed line)~\cite{Bauer:2021mvw}. The former line is calculated in the minimal KSVZ model~\cite{Kim:1979if,Shifman:1979if} with a single pair of vector-like quarks with no non-QCD charges; this model generates $\cGG = 1$ and the quark mass is bounded above by $4\pi f /\sqrt{2}$. Note that it is not trivial to apply an existing collider search to derive a stringent constraint on $f$, since the exotic quarks are stable due to a conserved $U(1)$~\cite{DiLuzio:2016sbl,DiLuzio:2017pfr}. Therefore any collider analysis must make specific assumptions about their decays. We leave such an exploration to future work. Similarly, the constraint from the chromomagnetic moment of the top quark is also sensitive to the direct coupling of the ALP to $t$, which we have set to zero. We chose not to shade the parameter space highlighted by the dashed lines; we combined the remaining bounds in the gray regions of Fig.~\ref{fig:dq_gluon_reach}.

As mentioned above, a consistent UV treatment, such as that adopted in Ref.~\cite{Bauer:2021wjo} where the gluon coupling is defined at a high scale, generates a multitude of other couplings that result in additional bounds in the parameter space of Fig.~\ref{fig:gluon_coupled_alp_constraints_only} -- see, e.g., Refs.~\cite{Chakraborty:2021wda,Bauer:2021mvw}. 

\section{Conclusion}
\label{sec:conclusion}
The dimension-five ALP couplings are some of the leading candidates for BSM physics 
interacting with SM particles from the perspective of effective field theory. It is therefore important to continue testing these models with existing and future experiments. 
In this work we showed that the proposed proton beam-dump DarkQuest, will be able to probe new parameter space in models where the ALP 
couples dominantly to photons and gluons. We expect that similar results can be attained in more general scenarios, e.g., in which 
the ALP has interactions with quarks or leptons.

Our analysis contained several new aspects of ALP production and decays. In computing amplitudes involving ALPs in
chiral perturbation theory we carefully tracked their dependence on unphysical parameters, ensuring their cancellation in the final result. 
We also provided new calculations of ALP coupling to nucleons in the three-flavour theory, and used it to estimate ALP emission rate in proton-nucleus 
bremsstrahlung. It would be interesting to compare this mechanism to other methods used to evaluate ALP production in hadronic interactions, such as 
emission in a parton shower.

We made several simplifying assumptions in estimating the sensitivity of proton beam-dump experiments to ALPs, mainly relating to the production 
of mesons and photons in beam-target collisions. We took the interactions to happen in the first interaction or radiation length, neglecting the possible production 
deeper in the target with the attenuated beam. Clearly this yields a conservative estimate of the total yield, so it would be useful to study ALP production in more sophisticated simulations to see whether the true sensitivity is appreciably stronger. 
We also focused on ALP decays to photons; while in the photon-coupled case this is the only decay channel, gluon-coupled ALPs have a multitude of available channels in the parameter space to which DarkQuest is sensitive. These final states can also be looked for to recover the branching fraction penalty associated with $a\to \gamma\gamma$-only searches. 

Finally, our compilation of bounds on the gluon-coupled ALP in Fig.~\ref{fig:gluon_coupled_alp_constraints_only} revealed several gaps in experimental coverage. It is interesting to think of existing or future observations that can be used to test these in a model-agnostic way. All of these occur at fairly large couplings so naively this should be straightforward. One obvious possibility are indirect bounds from searches for QCD-coupled states that are needed to generate the dimension-five ALP gluon coupling. However, in minimal models these particles carry a new baryon number, making them potentially collider-stable. As a result, any collider search is necessarily sensitive to the operator mediating their decay, which requires specifying yet more unknown UV physics. Clearly, it would be more satisfying to close these gaps with observables entirely within the ALP effective theory.

\acknowledgments
We thank Patrick Draper, Stefania Gori, Andreas Kronfeld, Jessica Turner, Shirley Li, Simon Knapen, Gustavo Marques Tavares, Michael Williams and Nhan Tran for helpful discussions.
We are grateful to Dmitriy Kirpichnikov and Yunjie Yang for sharing the NA64 and GlueX projections and exclusions with us.
NB was supported in part by NSERC, Canada. This manuscript has been authored by Fermi Research Alliance,
LLC under Contract No. DE-AC02-07CH11359 with the U.S. Department of Energy, Office of Science, Office of High Energy Physics. We used \texttt{Package-X}~\cite{Patel:2015tea} to derive and check several analytic results in this work. Our numerical work was enabled by~\texttt{numpy}~\cite{harris2020array}, \texttt{scipy}~\cite{2020SciPy-NMeth} and~\texttt{matplotlib}~\cite{Hunter:2007}. 

\appendix
\section{Three-Flavour Mixing}
\label{sec:three_flavour_mixing}
Our goal is to derive the ALP-meson mixing in the three-flavour 
regime, so we must include $\pi^0$, $\eta$ and $\eta'$. 
Let the chiral basis states be $\pi^0$, $\eta_8$ and $\eta_1$. 
The starting Lagrangian is then 
\begin{subequations}
\begin{align}
    \mathscr{L} & \supset \frac{1}{2}(\partial a)^2 - \frac{m_{a,0}^2}{2}a^2  \\ 
&+  \frac{f_\pi^2}{4} \tr (\partial_\mu \Sigma) (\partial^\mu \Sigma)^\dagger
+ \frac{f_\pi^2}{2}B \tr \left(\Sigma \bm{m_q}^\dagger(a) + \bm{m_q}(a)\Sigma^\dagger\right) \\
& + \frac{i f_\pi^2}{2} \frac{\partial_\mu a}{2f}\tr \hat{\bm{c}}_{qq} \left(\Sigma \partial^\mu \Sigma^\dagger- \Sigma^\dagger \partial^\mu \Sigma\right) \\
& + \frac{f_\pi^2}{4 N_c} \frac{N_c}{3}\left(m_{\eta_1}^2 - 2\frac{2m_\pi^2(m_u + m_d + m_s)}{3(m_u+m_d)}\right)
\left(-i \tr 2 i \Pi/f_\pi\right)^2\label{eq:eta1_mass_term},
\end{align}
\label{eq:chiral_lagrangian_with_etap}
\end{subequations}
where we used $\ln \det \Sigma  = \tr 2 i \Pi/f_\pi$.
For now we assume that the full chiral symmetry is $U(3)_L\times U(3)_R$; this is softly broken by the mass term and by the anomaly as we will discuss below. The pion field that follows from this assumption is 
\beq
\Sigma = \exp(2 i \Pi /f_\pi),\; \Pi =\frac{1}{2} \begin{pmatrix}
  \pi_0 + \frac{1}{\sqrt{3}}\eta_8 & \sqrt{2}\pi^+ & \sqrt{2} K^+ \\
  \sqrt{2}\pi^- & \frac{1}{\sqrt{3}}\eta_8 - \pi_0 & \sqrt{2} K_0 \\
  \sqrt{2} K^- & \sqrt{2}\bar{K} & -\frac{2}{\sqrt{3}} \eta_8
\end{pmatrix} + \frac{1}{\sqrt{6}}\mathbb{1}\eta_1,
\label{eq:sigma_and_pi_def_3flav}
\eeq
The coefficient in front of $\eta_1$ is chosen to ensure the correct normalization of its kinetic term; it also means that the corresponding generator of the $U(1)_A$ transformation has the same normalization as the non-Abelian ones (in the Peskin \& Schroeder convention~\cite{Peskin:1995ev}):
\beq
\tr \left(\frac{1}{\sqrt{2N_f}}\mathbb{1}_{N_f\times N_f}\right)\left(\frac{1}{\sqrt{2N_f}}\mathbb{1}_{N_f\times N_f}\right) = \frac{1}{2}.
\eeq

The anomaly contribution in Eq.~\ref{eq:eta1_mass_term} is the leading term that breaks $U(3)_L\times U(3)_R$ to $SU(3)_L\times SU(3)_R\times U(1)_V$ and is consistent with large-$N$ arguments~\cite{Witten:1980sp}. Its coefficient is chosen such that if the mass term preserved the $U(1)$ (i.e., it had vanishing trace), it sets the mass of the $\eta_1$ state. It is also important to note that the anomaly term above is the one \emph{after} doing the field redefinition to eliminate the $G\tilde{G}$ coupling of the ALP. If we did not perform this transformation the ALP would appear in the combination
\beq
\left(\frac{2\cGG a}{f}-i \tr 2 i \Pi/f_\pi\right)^2.
\eeq
The relative coefficient here can be fixed by demanding that the ALP is a spurion of the $U(1)_A$ transformation. The rotation that eliminates $G\tilde{G}$ coupling is $\Sigma \to e^{2i\alpha}\Sigma$, where $\alpha = -\bm{\kappa}\cGG a /f$; one can check that the same transformation eliminates the ALP from the anomaly term. 

To leading order in $a/f$, the ALP-dependent quark mass matrix above is 
\beq
\bm{m_q}(a) \approx \begin{pmatrix}
m_u  &  0 & 0 \\
0 & m_d & 0 \\
0 & 0 & m_s 
\end{pmatrix}
\begin{pmatrix}
1 - 2 i \cGG \kappa_u a/f & 0 & 0 \\
0 & 1 - 2 i \cGG \kappa_d a/f & 0 \\
0 & 0 & 1 - 2 i \cGG \kappa_s a/f
\end{pmatrix}. 
\eeq
The kinetic mixing term $\hat{\bm{c}}_{qq}$ receives contributions both from ``fundamental'' ALP-quark couplings and terms $\propto \bm{\kappa}$; since we are working in the simplified case in which the ALP only couples to gluons, only the latter terms are present. We will not specialize to this case until very end though, we will only assume that $\hat{\bm{c}}_{qq}$ is diagonal:
\beq
\hat{\bm{c}}_{qq} = \begin{pmatrix}
c_{uu} & 0 & 0 \\
0 & c_{dd} & 0 \\
0 & 0 & c_{ss}
\end{pmatrix}.
\eeq

Our goal is to bring the ALP kinetic term into a canonical form and then diagonalize the resulting 
ALP-meson mass matrix.
We can write the ALP-pion Lagrangian as
\beq
\mathscr{L}\supset \frac{1}{2}(\partial\varphi)^T Z (\partial \varphi)
- \frac{1}{2}\varphi^T M^2 \varphi 
\eeq
where $\varphi = (\pi^0,\; \eta_8,\; \eta_0,\; a)^T$ and 
\beq
Z = \begin{pmatrix}
1 & 0 & 0 & \frac{(c_{uu}-c_{dd})f_\pi}{2 f}\\
0 & 1 & 0 & \frac{(c_{dd}-2c_{ss}+c_{uu})f_\pi}{2 \sqrt{3}f} \\
0 & 0 & 1 & \frac{(c_{dd}+c_{ss}+c_{uu})f_\pi}{ \sqrt{6}f} \\
\frac{(c_{uu}-c_{dd})f_\pi}{2 f} & \frac{(c_{dd}-2c_{ss}+c_{uu})f_\pi}{2\sqrt{3} f} & \frac{(c_{dd}+c_{ss}+c_{uu})f_\pi}{ \sqrt{6}f} & 1
\end{pmatrix},
\eeq
\beq
M^2 =\begin{pmatrix}
m_\pi^2 & - \frac{1}{\sqrt{3}}\delta_I m_\pi^2 & - \sqrt{\frac{2}{3}}\delta_I m_\pi^2 & -\frac{\cGG f_\pi}{f}((1+\delta_I)\kappa_d - (1-\delta_I)\kappa_u)m_\pi^2\\
 \cdot & m_{\eta_8}^2 & -\frac{\sqrt{2}(m_s - \hat{m})}{3 \hat{m}} m_\pi^2 & \frac{\cGG f_\pi (\hat{m}(\kappa_d(1+\delta_I) + \kappa_u(1-\delta_I)) - 2m_s \kappa_s)}{\sqrt{3}f\hat{m}}m_\pi^2 &  \\
\cdot & \cdot & m_{\eta_1}^2 & 
 \frac{\sqrt{2}\cGG f_\pi(\hat{m}(\kappa_d(1+\delta_I) + \kappa_u (1-\delta_I))+m_s\kappa_s)}{\sqrt{3}f \hat{m}}m_\pi^2 \\
\cdot & \cdot & \cdot & m_a^2
\end{pmatrix}.
\eeq
In the above we define variables that help to take the isospin limit:
\beq
\delta_I = \frac{m_d - m_u}{m_u+m_d},\;\;\; 
\hat{m} = \frac{1}{2}(m_u + m_d).
\eeq
The isospin limit corresponds to taking $\delta_I \to 0$. We have also defined tree-level meson masses
\beq
m_\pi^2 = B \hat{m},\; m_{K^\pm}^2 = B (m_u+m_s),\; m_{K^0}^2 = B (m_d + m_s),\; m_{\eta_8}^2 = \frac{1}{3}B(m_u + m_d + 4 m_s).
\eeq
which agree with, e.g., Ref.~\cite{Donoghue:1992dd}.

Perturbatively diagonalizing and normalizing $Z$ yields a modified mass matrix
\beq
\begin{pmatrix}
m_\pi^2 & M^2_{\pi^0 \eta_8}\delta_I & M^2_{\pi^0 \eta_1}\delta_I & M^2_{\pi^0 a} \epsilon \\
\cdot & m_{\eta_8}^2 & M^2_{\eta_8 \eta_1} & M^2_{\eta_8 a}\epsilon \\
\cdot & \cdot & m_{\eta_1}^2 & M^2_{\eta_1 a}\epsilon \\
\cdot & \cdot & \cdot & m_{a}^2
\end{pmatrix}
\eeq
where following Ref.~\cite{Aloni:2018vki}, we have defined shorthands for the various elements in this matrix that make explicit dependence on 
small quantities $\delta_I$ and $\epsilon = f_\pi/f$ (the $m_a^2$ entry also gets shifted, but only at $\mathcal{O}(f_\pi^2/f^2)$, so we dropped that term). The explicit expressions for the matrix entries are not very illuminating. The remaining meson mixings 
$\eta_8 - \eta_1$, $\pi_0 - \eta$, $\pi_0 - \eta'$ are removed by perturbative 
pairwise diagonalization. For the $\eta_8 - \eta_1$ system we find the mixing angle 
\beq
\tan 2\theta_{\eta\eta'} = \frac{2M_{\eta_8 \eta_1}^2}{m_{\eta_1}^2 - m_{\eta_8}^2}.
\label{eq:eta_etap_mixing_angle}
\eeq
In practice this angle is experimentally determined from the ratio of $\eta$ and $\eta'$ partial widths into photons (see the Quark Model review in Ref.~\cite{Tanabashi:2018oca}); the approximation of Ref.~\cite{Aloni:2018vki} uses instead $\sin \theta_{\eta \eta'} = -1/3$ for simplicity, which we use throughout this work.

The mass matrix after this transformation can be written in terms of new shorthand variables:
\beq
\begin{pmatrix}
m_\pi^2 & M^2_{\pi^0 \eta}\delta_I & M^2_{\pi^0 \eta'}\delta_I & M^2_{\pi^0 a} \epsilon \\
\cdot & M_{\eta \eta}^2 & 0 & M^2_{\eta a}\epsilon \\
\cdot & \cdot & M_{\eta'\eta'}^2 & M^2_{\eta' a}\epsilon \\
\cdot & \cdot & \cdot & m_{a}^2
\end{pmatrix}.
\eeq

Next we remove the $\pi^0 -\eta$ and $\pi^0 -\eta'$ mixings. The matrix entries that mix these states are proportional to $m_\pi^2$ and the isospin breaking parameter $\delta_I$, so the mixing angles will scale as $m_\pi^2 \delta_I / m_\eta^2 \sim 0.02\; (0.007)$ for 
$\eta$ and $\eta'$ respectively; we will therefore solve for these mixing angles perturbatively and drop higher order terms in these quantities. 
We find 
\beq
\sin \theta_{\pi^0 \eta} \approx \frac{M^2_{\pi^0 \eta}\delta_I}{M^2_{\eta\eta} - m_\pi^2}, 
\eeq
and 
\beq
\sin \theta_{\pi^0 \eta'} \approx \frac{M^2_{\pi^0 \eta'}\delta_I}{M^2_{\eta'\eta'} - m_\pi^2}.
\eeq

After these transformations, the kinetic term is canonical and only meson-ALP mixings remain in the mass matrix.
To leading order in small quantities (i.e., dropping terms like $\mathcal{O}(m_\pi^2\delta_I/m_{\eta'}^2)^2$ and $\mathcal{O}(\epsilon m_\pi^2\delta_I/m_{\eta'}^2)$) the mass matrix is
\beq
\begin{pmatrix}
m_\pi^2 & 0 & 0 & M^2_{\pi^0 a} \epsilon \\
\cdot & m_\eta^2 & 0 & M^2_{\eta a}\epsilon \\
\cdot & \cdot & m_{\eta'}^2 & M^2_{\eta' a}\epsilon \\
\cdot & \cdot & \cdot & m_a^2 
\end{pmatrix},
\eeq
where the diagonal entries now represent the physical masses 
of the mesons (the mixing with the ALP will change this only by $\mathcal{O}(f_\pi/f)$).

We can now finally diagonalize the ALP-meson mixing. As before, we do this by a series of three pair-wise rotations to eliminate the mixing 
of the ALP with each of the mesons. Solving for the mixing angles to remove off-diagonal terms in the 
mass matrix we find:
\begin{subequations}
\begin{align}
    \sin\theta_{\pi^0 a} & \approx - \frac{\left(M^2_{\pi^0 a} {\color{blue}- M^2_{\eta a}\sin\theta_{\pi \eta} - M^2_{\eta' a}\sin\theta_{\pi\eta'}}\right)\epsilon}{m_\pi^2 - m_{a}^2}\\
    \sin\theta_{\eta a} & \approx - \frac{\left(M^2_{\eta a} {\color{blue}+ M^2_{\pi a}\sin\theta_{\pi \eta}}\right)\epsilon}{m_\eta^2 - m_{a}^2}\\
    \sin\theta_{\eta' a} & \approx - \frac{\left(M^2_{\eta' a} {\color{blue}+M^2_{\pi a}\sin\theta_{\pi\eta'}}\right)\epsilon}{m_{\eta'}^2 - m_{a}^2}
    \label{eq:meson_alp_mixing_angles}
\end{align}
\end{subequations}
The terms highlighted in blue here are proportional to $\delta_I$ and therefore represent the leading isospin-breaking corrections to the mixing angles; note that these are not suppressed by the small ratio $m_\pi^2/m_\eta^2$. 

We can now write down the Lagrangian fields in terms of the physical mass eigenstates 
as 
\begin{subequations}
\begin{align}
    \pi^0 & = \pi^{0,\mathrm{phys}}  + \vev{\pi a} a^\mathrm{phys} + \sin \theta_{\pi\eta} \eta^\mathrm{phys}  + \sin\theta_{\pi\eta'}\eta^{\prime,\mathrm{phys}}\\
    \eta_8 & = \cos\theta_{\eta\eta'}\eta^\mathrm{phys} + \vev{\eta_8 a}a^{\mathrm{phys}} + \sin\theta_{\eta\eta'}\eta^{\prime,\mathrm{phys}} \nonumber\\ 
    &- (\cos\theta_{\eta\eta'}\sin\theta_{\pi^0 \eta} + \sin\theta_{\eta\eta'}\sin\theta_{\pi^0 \eta'})\pi^{0,\mathrm{phys}} \\
    \eta_1 & = \cos\theta_{\eta\eta'}\eta^{\prime,\mathrm{phys}} + 
    \vev{\eta_1 a} a^{\mathrm{phys}} - \sin\theta_{\eta\eta'}\eta^{\mathrm{phys}} \nonumber \\
    & - \left(\cos\theta_{\eta\eta'}\sin\theta_{\pi^0 \eta'} - \sin\theta_{\eta\eta'}\sin\theta_{\pi^0 \eta}\right)\pi^{0,\mathrm{phys}}\\
    a & = a^\mathrm{phys} + \mathcal{O}(f_\pi/f),
\end{align}
\label{eq:mesons_from_physical_fields}
\end{subequations}
where
\begin{subequations}
\begin{align}
    \vev{\pi a} & = \sin\theta_{\pi^0 a}-\frac{(c_{uu}-c_{dd})f_\pi}{2f} \\%{\color{red}+ \sin\theta_{\eta a}\sin\theta_{\pi \eta} + \sin\theta_{\eta' a} \sin\theta_{\pi\eta'}}\\
    \vev{\eta_8 a} & = \cos\theta_{\eta\eta'}\sin\theta_{\eta a} + \sin\theta_{\eta\eta'}\sin\theta_{\eta' a} -\frac{(c_{uu} -2c_{ss} +c_{dd})f_\pi}{2\sqrt{3}f}\\
   % & {\color{red}- \sin\theta_{\pi a}(\cos\theta_{\eta\eta'}\sin\theta_{\pi\eta}+\sin\theta_{\eta\eta'}\sin\theta_{\pi\eta'}) - \sin\theta_{\eta' a}\cos\theta_{\eta\eta'}\sin\theta_{\pi\eta} \sin\theta_{\pi\eta'}}\\
    \vev{\eta_1 a} & =\cos\theta_{\eta\eta'}\sin\theta_{\eta'a} -\sin\theta_{\eta\eta'}\sin\theta_{\eta a} - \frac{(c_{uu}+c_{dd} + c_{ss})f_\pi}{\sqrt{6}f}%\\
    %& {\color{red} + \sin\theta_{\pi a}(\sin\theta_{\eta\eta'} \sin{\theta_{\pi\eta}}-\cos\theta_{\eta\eta'}\sin\theta_{\pi\eta'})+\sin\theta_{\eta' a}\sin\theta_{\eta\eta'} \sin\theta_{\pi\eta} \sin\theta_{\pi\eta'}}
\end{align}
\label{eq:mixing_elements}
\end{subequations}
In writing the above, we dropped terms proportional to $\sim \delta_I m_\pi^2/m_{\eta^{(\prime)}}^{2}$.

\section{ALP Decays Into Photons in Three-Flavour \texorpdfstring{$\chi$PT}{chiPT}}
\label{sec:alp_to_two_photons_in_3f_chipt}
The three-flavour equivalent of the expression Eq.~\ref{eq:two_photon_amp_2f} is 
\begin{subequations}
\begin{align}
i\mathcal{A} & \propto 
%\frac{\alpha}{4\pi}
\left(\frac{i\hat{c}_{\gamma\gamma}}{f} + \left(\frac{-i}{f_\pi}\right)\left[\vev{\pi a} 
+ \frac{1}{\sqrt{3}}\vev{\eta a} + 2\frac{2}{3}\vev{\eta' a}  \right]\right) \\
& = \frac{2i \cGG m_0^2 \cos2\theta_{\eta\eta'}\left(8(m_a^2 - m_{\eta_8}^2)\cos2\theta_{\eta\eta'} + \sqrt{2}(m_{\eta_1}^2-m_{\eta_8}^2)\sin2\theta_{\eta\eta'}\right)}
{3f((m_{\eta_8}^2 + m_{\eta_1}^2 -2m_a^2)^2\cos^2 2\theta_{\eta\eta'} - (m_{\eta_1}^2 - m_{\eta_8}^2)^2)}\\
& + \frac{8i\cGG m_{\eta_1}^2 m_\pi^2 (m_{\eta_8}^2 - m_\pi^2)\delta_I}{3f(m_a^2 -m_\pi^2)((m_{\eta_1}^2+m_{\eta_8}^2 - 2m_\pi^2)^2-(m_{\eta_1}^2-m_{\eta_8}^2)^2)}
\end{align}
\label{eq:two_photon_amp_3f}
\end{subequations}
where the coefficients in front of the mixing angles in the first line are ratios of the electromagnetic anomaly for each meson relative to the $\pi^0$. In the second line made use of the explicit expressions for $\hat{c}_{\gamma\gamma}$ and $\hat{c}_{qq}$ in terms of $\cGG$ and $\kappa_q$ (Eq.~\ref{eq:hat_constant_defs}), the expressions for the mixing angles in terms of tree-level meson masses, and the meson masses in terms of the quark masses. The third line arises from the leading isospin-violating contributions that appear in the mixing angles 
in Eqs.~\ref{eq:meson_alp_mixing_angles} (other isospin violating contributions at the same order in $\delta_I$ are suppressed by $m_\pi^2/m_{\eta^{(\prime)}}^2$). 
Note that we have not expressed the $\eta_8$ and $\eta_1$ masses in terms of the physical $\eta$ and $\eta^\prime$ mass since this facilitates 
taking interesting limits below.

This expression is remarkable in that all of the unphysical $\kappa_q$ dependence has cancelled; the result (in the $\delta_I \to 0$ limit) is proportional to $m_0$, the coefficient in front of the anomaly term in the chiral Lagrangian, Eq.~\ref{eq:chiral_lagrangian_with_etap},
\beq
m_0^2 = m_{\eta_1}^2 - 2\frac{2m_\pi^2(m_u + m_d + m_s)}{3(m_u+m_d)}.
\eeq

In addition to the vanishing of the $\kappa_q$ dependence we can perform another consistency check by taking the two-flavour limit to see if we reproduce both terms of Eq.~\ref{eq:two_photon_amp_2f}. 
In order to do this, we can express $m_{\eta_1}$, $m_{\eta_8}$ and $\tan\theta_{\eta\eta'}$ in terms for the quark masses and $m_0$; one can then take the limits $m_s\to \infty$ and $m_0 \to \infty$. As long as the mixing angle is self-consistently chosen (i.e., via Eq.~\ref{eq:eta_etap_mixing_angle}), the order of the limits does not matter.
In the limit $m_s,m_0\to\infty$, Eq.~\ref{eq:two_photon_amp_3f} gives
\beq
  -\frac{i\cGG}{f}\left(
\frac{5}{3}
+\frac{ m_\pi^2\delta_I}{m_\pi^2-m_a^2}\right),
\eeq
matching exactly Eq.~\ref{eq:two_photon_amp_2f}.

\section{Rare Eta Decays}
\label{sec:rare_eta_decays}
$\eta$ and $\eta'$ decays can be computed from the Lagrangian in Eq.~\ref{eq:chiral_lagrangian_with_etap} by using the expressions of Eq.~\ref{eq:mesons_from_physical_fields}. For $\eta^{(\prime)}\to \pi^0\pi^0 a$ decays
the only terms that contribute come from the mass terms; for the decays involving charged pions in the final state the kinetic mixing terms also contribute. For generic 
$\eta-\eta'$ mixing angles, the expressions for the amplitudes corresponding to these processes are long, so we will present the limit $\sin \theta_{\eta\eta'} = -1/3$ which significantly simplifies the expressions; this limit was also used in Refs.~\cite{Aloni:2018vki,Gan:2020aco}. The measured mixing angle differs from this by a few percent (see the discussion in Ref.~\cite{Donoghue:1992dd}), but this is sufficient accuracy for BSM calculations. The other approximation is to drop sub-leading isospin violating terms $\sim \delta_I m_\pi^2/m_{\eta}^2$ -- these are terms that are proportional to $\sin\theta_{\pi\eta^{(\prime)}}$ that are not multiplied by $m_\eta^2$.

Under these assumptions we find the following expressions before plugging in the explicit expressions for the mixing matrix elements (Eq.~\ref{eq:mixing_elements}):
\begin{subequations}
\begin{align}
\mathcal{A}(\eta\to\pi^0 \pi^0 a) & =\frac{m_\pi^2}{f_\pi^2}
\left[\frac{\sqrt{2}f_\pi \cGG}{\sqrt{3}f}(\kappa_u - \kappa_u \delta_I + \kappa_d + \delta_I \kappa_D) -\sqrt{\frac{2}{3}}\delta_I \vev{\pi a} + \frac{\sqrt{2}}{3}\vev{\eta_8 a} + \frac{2}{3}\vev{\eta_1 a}\right] \\
\mathcal{A}(\eta^\prime\to\pi^0 \pi^0 a) & =\frac{m_\pi^2}{f_\pi^2}
\left[\frac{f_\pi \cGG}{\sqrt{3}f}(\kappa_u - \kappa_u \delta_I + \kappa_d + \delta_I \kappa_D) -\frac{1}{\sqrt{3}}\delta_I \vev{\pi a} + \frac{1}{3}\vev{\eta_8 a} + \frac{\sqrt{2}}{3}\vev{\eta_1 a}\right]
\end{align}
\end{subequations}
At this point it is clear that there must be some non-trivial cancellations among the $\kappa_q$ between the various terms. Plugging in the mixing elements in Eq.~\ref{eq:mixing_elements} we find at leading order in isospin violating parameter $\delta_I$ 
\begin{subequations}
\begin{align}
\mathcal{A}(\eta\to\pi^0 \pi^0 a) & = \frac{\sqrt{\frac{2}{3}} \cGG m_\pi^2 \left(3 m_a^2-2 m_{\eta}^2-m_{\eta'}^2\right) \left(2 m_{\eta}^2-5 m_{\eta'}^2+3
   m_\pi^2\right)}{27 f f_\pi \left(m_a^2-m_{\eta}^2\right) \left(m_a^2-m_{\eta'}^2\right)} \\%\nonumber\\  &
   %-\frac{\sqrt{\frac{2}{3}}
   %\cGG \delta_I^2 m_\pi^4 \left(2 m_{\eta}^2+m_{\eta'}^2-3 m_\pi^2\right) \left(2 m_{\eta}^2-5 m_{\eta'}^2+3
   %m_\pi^2\right)}{27 f f_\pi \left(m_a^2-m_\pi^2\right) \left(m_\pi^2-m_{\eta}^2\right)
   %\left(m_\pi^2-m_{\eta'}^2\right)}\\
\mathcal{A}(\eta^\prime \to\pi^0 \pi^0 a) & = \frac{ \cGG m_\pi^2 \left(3 m_a^2-2 m_{\eta}^2-m_{\eta'}^2\right) \left(2 m_{\eta}^2-5 m_{\eta'}^2+3
   m_\pi^2\right)}{27 \sqrt{3}f f_\pi \left(m_a^2-m_{\eta}^2\right) \left(m_a^2-m_{\eta'}^2\right)} \\%\nonumber\\ &
   %-\frac{\cGG \delta_I^2 m_\pi^4 \left(2 m_{\eta}^2+m_{\eta'}^2-3 m_\pi^2\right) \left(2 m_{\eta}^2-5 m_{\eta'}^2+3
   %m_\pi^2\right)}{27\sqrt{3} f f_\pi \left(m_a^2-m_\pi^2\right) \left(m_\pi^2-m_{\eta}^2\right)
   %\left(m_\pi^2-m_{\eta'}^2\right)}\\
\mathcal{A}(\eta\to\pi^+ \pi^- a) & = \frac{\sqrt{\frac{2}{3}} \cGG m_\pi^2 \left(3 m_a^2-2 m_{\eta}^2-m_{\eta'}^2\right) \left(2 m_{\eta}^2-5 m_{\eta'}^2+3
   m_\pi^2\right)}{27 f f_\pi \left(m_a^2-m_{\eta}^2\right) \left(m_a^2-m_{\eta'}^2\right)}\\%\nonumber\\
   %& -\frac{\sqrt{\frac{2}{3}}
   %\cGG \delta_I^2 m_\pi^4 \left(m_a^2+3 m_\pi^2-3 s\right) \left(2 m_{\eta}^2-5 m_{\eta'}^2+3 m_\pi^2\right)
   %\left(-2 m_{\eta}^2-m_{\eta'}^2+3 m_\pi^2\right)}{81 f f_\pi \left(m_\pi^2-m_a^2\right)
   %\left(m_{\eta}^2-m_\pi^2\right)^2 \left(m_\pi^2-m_{\eta'}^2\right)}\\
\mathcal{A}(\eta^\prime \to\pi^+ \pi^- a) & =\frac{\cGG m_\pi^2 \left(3 m_a^2-2 m_{\eta}^2-m_{\eta'}^2\right) \left(2 m_{\eta}^2-5 m_{\eta'}^2+3 m_\pi^2\right)}{27
   \sqrt{3} f f_\pi \left(m_a^2-m_{\eta}^2\right) \left(m_a^2-m_{\eta'}^2\right)}%\nonumber\\
   %& - \frac{\cGG \delta_I^2
   %m_\pi^4 \left(m_a^2+3 m_\pi^2-3 s\right) \left(2 m_{\eta}^2+m_{\eta'}^2-3 m_\pi^2\right) \left(-2 m_{\eta}^2+5
   %m_{\eta'}^2-3 m_\pi^2\right)}{81 \sqrt{3} f f_\pi \left(m_\pi^2-m_a^2\right) \left(m_\pi^2-m_{\eta}^2\right)
   %(m_{\eta'}^2-m_\pi^2)^2}
\end{align}
\end{subequations}
An important observation is that all $\kappa_q$ dependence has disappeared; in the $\pi^+\pi^-$ final states this requires the combination of contributions both from the kinetic and the mass terms of the chiral Lagrangian. Moreover, this cancellation is not specific to the $\sin\theta_{\eta\eta'} = -1/3$ choice; we have checked that it is true for an arbitrary mixing angle. The isospin breaking contributions arise at $\mathcal{O}(\delta_I^2)$ because $\vev{\pi a} \propto \delta_I$ is also multiplied by $\delta_I$ in the amplitudes above. 

\section{Rare Kaon Decays}
\label{sec:rare_kaon_decays}
For brevity we present here the amplitudes in the limit of decoupled singlet meson and to leading order in isospin violation, which matches the assumptions in, e.g., Ref.~\cite{Bauer:2021wjo}. 
In our numerics we use the full results that include $\eta-\eta'$ mixing.
The charged kaon decay amplitudes are
\begin{align}
  \mathcal{A}(K^- \to \pi^- a)  & = \frac{8 i \cGG f_\pi^2 G_8 \left(m_a^2-m_K^2\right) \left(m_K^2-m_\pi^2\right)}{f \left(3 m_a^2-4
  m_K^2+m_\pi^2\right)}-\frac{i \cGG \delta_I f_\pi^2 G_8 m_\pi^2}{f}.
\end{align}
The corresponding $K^+$ amplitude has the opposite sign. The $\delta_I \to 0$ limit of this expression matches the result of 
Ref.~\cite{Bauer:2021wjo} once the difference in $f_\pi$ conventions is taken into account.

For the rare neutral kaon decays we first compute the amplitude for the neutral strong-interaction eigenstates $K^0$ and $\overline{K^0}$:
\begin{subequations}
\begin{align}
  \mathcal{A}(K^0 \to \pi^0 a) & \approx \frac{4 i \sqrt{2} \cGG f_\pi^2 G_8 \left(m_a^2-m_K^2\right) \left(m_K^2-m_\pi^2\right)}{f \left(3 m_a^2-4 m_K^2+m_\pi^2\right)}\\
  & +\frac{i \cGG \delta_I f_\pi^2 G_8 m_\pi^2 \left(5 m_a^4+2 m_a^2 \left(m_\pi^2-6 m_K^2\right)+8 m_K^4-4 m_K^2 m_\pi^2+m_\pi^4\right)}{\sqrt{2} f \left(m_\pi^2-m_a^2\right) \left(3 m_a^2-4 m_K^2+m_\pi^2\right)}. 
\end{align}
\end{subequations}
The corresponding $\overline{K^0}$ amplitude has the opposite sign. We find that the leading isospin-violating contribution is important at small ALP masses, 
and changes the amplitude by up to $\sim 50\%$.
The amplitudes of the physical weak eigenstates $K_{S,L}$ follow from the definitions $K_S = K_1 + \epsilon_K K_2$, $K_L = K_2 - \epsilon_K K_1$, 
where $K_{1,2} = (K^0 - \overline{K^0})/\sqrt{2}$ are the CP eigenstates and $\epsilon_K \approx 2.23\times 10^{-3}$ is the CP violation parameter:
\begin{align}
\mathcal{A}(K_S \to \pi^0 a) & \approx \sqrt{2} \mathcal{A}(K^0 \to \pi^0 a) \\
\mathcal{A}(K_L \to \pi^0 a) & \approx - \epsilon_K \mathcal{A}(K_S \to \pi^0 a).
\end{align}

\section{Proton Bremsstrahlung}
\label{sec:proton_brem_details}
We follow the procedure outlined in Ref.~\cite{Foroughi-Abari:2021zbm} and 
consider the process $p(p) + p(p_t) \to a(k) + f(p_f)$ where the four-momenta labels are given in the 
parentheses; $f$ represents an unspecified hadronic final state. The intermediate beam proton has momentum $p'$. We parametrize the four-momenta as 
\begin{subequations}
\begin{align}
  p & = \left(p_p + \frac{m_p^2}{2p_p},0,0,p_p\right) \\
  p_t & = (m_p,0,0,0)\\
  k & = \left(z p_p + \frac{p_T^2 + m_a^2}{2z p_p},(p_T)_x,(p_T)_y,z p_p\right)\\
  p' & = \left((1-z)p_p + \frac{p_T^2 +m_p^2}{2(1-z)p_p},-(p_T)_x,-(p_T)_y,(1-z)p_p\right),
\end{align}
  \label{eq:brem_kinematics}
\end{subequations}
where $z$ is the longitudinal momentum fraction of the original beam proton inherited by the ALP and $p_T$ is the 
magnitude of the transverse momentum.
These parametrizations assume that $(p_T^2 + m_a^2) /(z^2 p_p^2),\; (p_T^2 + m_p^2) /((1-z)^2 p_p^2)\ll 1$; the zeroth components 
of each four vector are chosen to enforce the on-shell conditions, not momentum conservation. 
In reality, the intermediate proton is off-shell and we can quantify this by considering 
\beq
(p-k)^2 = m_p^2 - \frac{H}{z},\;\; H \approx p_T^2 + z^2 m_p^2 + (1-z)m_a^2,
\label{eq:proton_offshellness}
\eeq
where we dropped terms higher order in $p_T/(z p_p)$ and $m_a/(z p_p)$. Thus we see that the ``off-shell-ness'' 
of the intermediate proton is characterized by $z$, $p_T$, $m_p$ and $m_a$.

The amplitude of ALP emission off the proton with helicity $r$ is then
\begin{subequations}
\begin{align}
  i\mathcal{M}^{pp_t\to a f}_r & = i g_{pa} \mathcal{A}(p-k,p_f) \frac{i(\slashed{p}-\slashed{k} + m_p)}{(p-k)^2 - m_p^2} (i\slashed{k}) \gamma_5 u_r(p)  \\
  & = i^2 g_{pa} \mathcal{A}(p-k,p_f) \left(\frac{-iz}{H}\right)\sum_{r'} u_{r'}(p-k) \bar{u}_{r'}(p-k) \slashed{k} \gamma_5 u_r(p) \\
  & \equiv \sum_{r'} \mathcal{M}_{r'}^{pp\to X} \left(\frac{z}{H}\right) V_{r' r},\;\; V_{r'r} = i g_{pa} \bar{u}_{r'} \slashed{k} \gamma_5 u_r.
\end{align}
\end{subequations}
where $\mathcal{A}$ is part of the amplitude associated with the $pp \to X$ interaction; the last line defines $\mathcal{M}_{r'}^{pp\to X}$, which is the actual 
amplitude for this sub-process, albeit with a different initial proton four-momentum ($p'$ instead of $p$, i.e. we replaced $p-k$ in $\mathcal{A}$ by the 
on-shell momentum $p'$). In order to evaluate the squared matrix element, we will need explicit expressions for the vertex function $V_{r'r}$ 
which can be obtained by using explicit spinor expressions; we use the ones from Peskin \& Schroeder~\cite{Peskin:1995ev}, while Ref.~\cite{Foroughi-Abari:2021zbm} uses a different phase 
conventions (as a result, intermediate results will be slightly different, but the final answer must be the same):
\beq
u_r (p) = \begin{pmatrix}\sqrt{p\cdot\sigma}\xi_r \\ \sqrt{p\cdot \bar\sigma} \xi_r\end{pmatrix},
\eeq
where $\xi_r$ is a two-component object with $\xi_{+1} = (1,0)$ and $\xi_{-1} = (0,1)$.
We find 
\beq
V_{r'r}/g_{pa} = \frac{r}{z\sqrt{1-z}} \left(m_a^2 + p_T^2 -m_a^2 z - m_p^2 z^2\right)\delta_{r',r}  - \frac{2m_p p_T}{\sqrt{1-z}}e^{ri\phi}\delta_{r',-r}.
\eeq
The matrix element squared involves the combination 
\begin{subequations}
\begin{align}
V_{r'r}(V_{r''r})^*/g_{pa}^2 & = \left[\frac{\left(m_a^2 + p_T^2 -m_a^2 z - m_p^2 z^2\right)^2}{z^2(1-z)} \delta_{r',r} + \frac{4 m_p^2 p_T^2}{1-z}\delta_{r',-r}\right]\delta_{r'',r'}\\
& + \frac{2 m_p p_T r'}{z(1-z)}e^{-r'i\phi}(\delta_{r',-r}-\delta_{r',r})\delta_{r'',-r'},
\end{align}
\end{subequations}
where no sum is implied over repeated indices. 
The squared and initial-proton-spin-averaged 
matrix element is 
\begin{subequations}
\begin{align}
\overline{|\mathcal{M}^{pp_t \to a f}|^2}  & = \frac{1}{2}\sum_{r,r',r''} V_{r'r}(V_{r''r})^*   \left(\frac{z}{H}\right)^2 \mathcal{M}_{r'}^{pp\to X} (\mathcal{M}_{r''}^{pp\to X})^* \\ 
& =  g_{pa}^2\left(\frac{z}{H}\right)^2 \left[\frac{\left(m_a^2 + p_T^2 -m_a^2 z - m_p^2 z^2\right)^2}{z^2(1-z)} + \frac{4 m_p^2 p_T^2}{1-z}\right] \overline{|\mathcal{M}^{pp\to X}|^2}.
\end{align}
\end{subequations}
Note that the imaginary (and spinor-phase-dependent terms) have vanished, as they must have.
Finally, the differential cross-section is 
\begin{subequations}
\begin{align}
d\sigma^{pp_t \to a f} & = 
\frac{(2\pi)^4 \delta^4(p+p_t - k - \sum_f p_f)}{4\sqrt{(p\cdot p_t)^2 - m_p^4}} \frac{d^3 k}{(2\pi)^3 2E_k} \prod_f \frac{d^3 p_f}{(2\pi)^3 2E_f} 
\overline{|\mathcal{M}^{pp_t \to a f}|^2} \\
& \equiv w_a(z,p_T^2) dp_T^2 dz \sigma(s'),
\end{align}
\end{subequations}
where $s' = (p' + p_t)^2$ and
\beq
w_a(z,p_T^2) \approx \frac{g_a^2}{16\pi^2 z}\frac{E_{p'}}{E_p} \left(\frac{z}{H}\right)^2 \left[\frac{\left(m_a^2 + p_T^2 -m_a^2 z - m_p^2 z^2\right)^2}{z^2(1-z)} + \frac{4 m_p^2 p_T^2}{1-z}\right].
\eeq
In order to write the differential cross-section for the 
ALP bremsstrahlung as a product of the hadronic cross-section and the ``splitting function'' $w_a$ we had to 
approximate $E_p - E_{k} \approx E_{p'}$ (this corresponds to neglecting terms like $p_T/(z p_p)$) in the delta function, as well as expand the ratio of square roots for $E_{p,p'} \gg m_p$; we also expressed 
the Lorentz-invariant phase-space $d^3 k/E_k$ in terms of $dz d(p_T^2)$. 
The final result for the $pp$ differential cross-section is 
\begin{subequations}
\begin{align}
d\sigma^{pp_t \to a f} & \approx w_a(z,p_T^2) dp_T^2 dz \sigma(s')\\  
w_a(z,p_T^2) &= \frac{g_{pa}^2}{16\pi^2 }\frac{(1-z)z}{H^2} \left[\frac{\left(m_a^2 + p_T^2 -m_a^2 z - m_p^2 z^2\right)^2}{z^2(1-z)} + \frac{4 m_p^2 p_T^2}{1-z}\right]
\end{align}
\end{subequations}
The same result can be obtained in a somewhat faster way following the appendix of Ref.~\cite{Boiarska:2019jym} while remembering that in our case the fast majority of radiation comes from the beam particle and not the target (that reference accounted for both by an additional factor of 2 in $w_a$).

The underlying hadronic cross-section, $\sigma(s')$, must be chosen appropriately for fixed-target proton-nucleus collisions; the above expression must also be dressed with form-factors that encode the non-point-like nature of the beam particle. These issues are discussed in Sec.~\ref{sec:proton_brem}.

\bibliographystyle{JHEP}
\bibliography{biblio}

\providecommand{\href}[2]{#2}\begingroup\raggedright\begin{thebibliography}{100}

\bibitem{Peccei:1977hh}
R.~D. Peccei and H.~R. Quinn, \emph{{CP Conservation in the Presence of
  Instantons}},
  \href{http://dx.doi.org/10.1103/PhysRevLett.38.1440}{\emph{Phys. Rev. Lett.}
  {\bf 38} (1977) 1440--1443}.

\bibitem{Peccei:1977ur}
R.~D. Peccei and H.~R. Quinn, \emph{{Constraints Imposed by CP Conservation in
  the Presence of Instantons}},
  \href{http://dx.doi.org/10.1103/PhysRevD.16.1791}{\emph{Phys. Rev. D} {\bf
  16} (1977) 1791--1797}.

\bibitem{Wilczek:1977pj}
F.~Wilczek, \emph{{Problem of Strong $P$ and $T$ Invariance in the Presence of
  Instantons}}, \href{http://dx.doi.org/10.1103/PhysRevLett.40.279}{\emph{Phys.
  Rev. Lett.} {\bf 40} (1978) 279--282}.

\bibitem{Weinberg:1977ma}
S.~Weinberg, \emph{{A New Light Boson?}},
  \href{http://dx.doi.org/10.1103/PhysRevLett.40.223}{\emph{Phys. Rev. Lett.}
  {\bf 40} (1978) 223--226}.

\bibitem{Arias:2012az}
P.~Arias, D.~Cadamuro, M.~Goodsell, J.~Jaeckel, J.~Redondo and A.~Ringwald,
  \emph{{WISPy Cold Dark Matter}},
  \href{http://dx.doi.org/10.1088/1475-7516/2012/06/013}{\emph{JCAP} {\bf 06}
  (2012) 013}, [\href{http://arxiv.org/abs/1201.5902}{{\tt 1201.5902}}].

\bibitem{Svrcek:2006yi}
P.~Svrcek and E.~Witten, \emph{{Axions In String Theory}},
  \href{http://dx.doi.org/10.1088/1126-6708/2006/06/051}{\emph{JHEP} {\bf 06}
  (2006) 051}, [\href{http://arxiv.org/abs/hep-th/0605206}{{\tt
  hep-th/0605206}}].

\bibitem{Arvanitaki:2009fg}
A.~Arvanitaki, S.~Dimopoulos, S.~Dubovsky, N.~Kaloper and J.~March-Russell,
  \emph{{String Axiverse}},
  \href{http://dx.doi.org/10.1103/PhysRevD.81.123530}{\emph{Phys. Rev. D} {\bf
  81} (2010) 123530}, [\href{http://arxiv.org/abs/0905.4720}{{\tt 0905.4720}}].

\bibitem{Cicoli:2012sz}
M.~Cicoli, M.~Goodsell and A.~Ringwald, \emph{{The type IIB string axiverse and
  its low-energy phenomenology}},
  \href{http://dx.doi.org/10.1007/JHEP10(2012)146}{\emph{JHEP} {\bf 10} (2012)
  146}, [\href{http://arxiv.org/abs/1206.0819}{{\tt 1206.0819}}].

\bibitem{Abbott:1982af}
L.~F. Abbott and P.~Sikivie, \emph{{A Cosmological Bound on the Invisible
  Axion}}, \href{http://dx.doi.org/10.1016/0370-2693(83)90638-X}{\emph{Phys.
  Lett. B} {\bf 120} (1983) 133--136}.

\bibitem{Dine:1982ah}
M.~Dine and W.~Fischler, \emph{{The Not So Harmless Axion}},
  \href{http://dx.doi.org/10.1016/0370-2693(83)90639-1}{\emph{Phys. Lett. B}
  {\bf 120} (1983) 137--141}.

\bibitem{Preskill:1982cy}
J.~Preskill, M.~B. Wise and F.~Wilczek, \emph{{Cosmology of the Invisible
  Axion}}, \href{http://dx.doi.org/10.1016/0370-2693(83)90637-8}{\emph{Phys.
  Lett. B} {\bf 120} (1983) 127--132}.

\bibitem{Ertas:2020xcc}
F.~Ertas and F.~Kahlhoefer, \emph{{On the interplay between astrophysical and
  laboratory probes of MeV-scale axion-like particles}},
  \href{http://dx.doi.org/10.1007/JHEP07(2020)050}{\emph{JHEP} {\bf 07} (2020)
  050}, [\href{http://arxiv.org/abs/2004.01193}{{\tt 2004.01193}}].

\bibitem{Agrawal:2021dbo}
P.~Agrawal et~al., \emph{{Feebly-interacting particles: FIPs 2020 workshop
  report}}, \href{http://dx.doi.org/10.1140/epjc/s10052-021-09703-7}{\emph{Eur.
  Phys. J. C} {\bf 81} (2021) 1015},
  [\href{http://arxiv.org/abs/2102.12143}{{\tt 2102.12143}}].

\bibitem{Gardner:2015wea}
S.~Gardner, R.~Holt and A.~Tadepalli, \emph{{New Prospects in Fixed Target
  Searches for Dark Forces with the SeaQuest Experiment at Fermilab}},
  \href{http://dx.doi.org/10.1103/PhysRevD.93.115015}{\emph{Phys. Rev. D} {\bf
  93} (2016) 115015}, [\href{http://arxiv.org/abs/1509.00050}{{\tt
  1509.00050}}].

\bibitem{Berlin:2018tvf}
A.~Berlin, N.~Blinov, S.~Gori, P.~Schuster and N.~Toro, \emph{{Cosmology and
  Accelerator Tests of Strongly Interacting Dark Matter}},
  \href{http://dx.doi.org/10.1103/PhysRevD.97.055033}{\emph{Phys. Rev. D} {\bf
  97} (2018) 055033}, [\href{http://arxiv.org/abs/1801.05805}{{\tt
  1801.05805}}].

\bibitem{Berlin:2018pwi}
A.~Berlin, S.~Gori, P.~Schuster and N.~Toro, \emph{{Dark Sectors at the
  Fermilab SeaQuest Experiment}},
  \href{http://dx.doi.org/10.1103/PhysRevD.98.035011}{\emph{Phys. Rev. D} {\bf
  98} (2018) 035011}, [\href{http://arxiv.org/abs/1804.00661}{{\tt
  1804.00661}}].

\bibitem{Choi:2019pos}
K.-Y. Choi, T.~Inami, K.~Kadota, I.~Park and O.~Seto, \emph{{Searching for
  Axino-Like Particle at Fixed Target Experiments}},
  \href{http://dx.doi.org/10.1016/j.dark.2020.100460}{\emph{Phys. Dark Univ.}
  {\bf 27} (2020) 100460}, [\href{http://arxiv.org/abs/1902.10475}{{\tt
  1902.10475}}].

\bibitem{Dobrich:2019dxc}
B.~D\"obrich, J.~Jaeckel and T.~Spadaro, \emph{{Light in the beam dump - ALP
  production from decay photons in proton beam-dumps}},
  \href{http://dx.doi.org/10.1007/JHEP05(2019)213}{\emph{JHEP} {\bf 05} (2019)
  213}, [\href{http://arxiv.org/abs/1904.02091}{{\tt 1904.02091}}].

\bibitem{Tsai:2019buq}
Y.-D. Tsai, P.~deNiverville and M.~X. Liu, \emph{{Dark Photon and Muon $g-2$
  Inspired Inelastic Dark Matter Models at the High-Energy Intensity
  Frontier}},
  \href{http://dx.doi.org/10.1103/PhysRevLett.126.181801}{\emph{Phys. Rev.
  Lett.} {\bf 126} (2021) 181801}, [\href{http://arxiv.org/abs/1908.07525}{{\tt
  1908.07525}}].

\bibitem{Darme:2020ral}
L.~Darm\'e, S.~A.~R. Ellis and T.~You, \emph{{Light Dark Sectors through the
  Fermion Portal}},
  \href{http://dx.doi.org/10.1007/JHEP07(2020)053}{\emph{JHEP} {\bf 07} (2020)
  053}, [\href{http://arxiv.org/abs/2001.01490}{{\tt 2001.01490}}].

\bibitem{Batell:2020vqn}
B.~Batell, J.~A. Evans, S.~Gori and M.~Rai, \emph{{Dark Scalars and Heavy
  Neutral Leptons at DarkQuest}},  \href{http://arxiv.org/abs/2008.08108}{{\tt
  2008.08108}}.

\bibitem{Bauer:2020jbp}
M.~Bauer, M.~Neubert, S.~Renner, M.~Schnubel and A.~Thamm, \emph{{The
  Low-Energy Effective Theory of Axions and ALPs}},
  \href{http://dx.doi.org/10.1007/JHEP04(2021)063}{\emph{JHEP} {\bf 04} (2021)
  063}, [\href{http://arxiv.org/abs/2012.12272}{{\tt 2012.12272}}].

\bibitem{Bauer:2021wjo}
M.~Bauer, M.~Neubert, S.~Renner, M.~Schnubel and A.~Thamm, \emph{{Consistent
  Treatment of Axions in the Weak Chiral Lagrangian}},
  \href{http://dx.doi.org/10.1103/PhysRevLett.127.081803}{\emph{Phys. Rev.
  Lett.} {\bf 127} (2021) 081803}, [\href{http://arxiv.org/abs/2102.13112}{{\tt
  2102.13112}}].

\bibitem{SeaQuest:2017kjt}
{\scshape SeaQuest} collaboration, C.~A. Aidala et~al., \emph{{The SeaQuest
  Spectrometer at Fermilab}},
  \href{http://dx.doi.org/10.1016/j.nima.2019.03.039}{\emph{Nucl. Instrum.
  Meth. A} {\bf 930} (2019) 49--63},
  [\href{http://arxiv.org/abs/1706.09990}{{\tt 1706.09990}}].

\bibitem{Aphecetche:2003zr}
{\scshape PHENIX} collaboration, L.~Aphecetche et~al., \emph{{PHENIX
  calorimeter}},
  \href{http://dx.doi.org/10.1016/S0168-9002(02)01954-X}{\emph{Nucl. Instrum.
  Meth. A} {\bf 499} (2003) 521--536}.

\bibitem{Shiltsev:2017mle}
V.~Shiltsev, \emph{{Fermilab Proton Accelerator Complex Status and Improvement
  Plans}}, \href{http://dx.doi.org/10.1142/S0217732317300129}{\emph{Mod. Phys.
  Lett. A} {\bf 32} (2017) 1730012},
  [\href{http://arxiv.org/abs/1705.03075}{{\tt 1705.03075}}].

\bibitem{Anticic:2010yg}
{\scshape NA49} collaboration, T.~Anticic et~al., \emph{{Inclusive production
  of charged kaons in p+p collisions at 158 GeV/c beam momentum and a new
  evaluation of the energy dependence of kaon production up to collider
  energies}},
  \href{http://dx.doi.org/10.1140/epjc/s10052-010-1328-0}{\emph{Eur. Phys. J.
  C} {\bf 68} (2010) 1--73}, [\href{http://arxiv.org/abs/1004.1889}{{\tt
  1004.1889}}].

\bibitem{Aduszkiewicz:2015dmr}
{\scshape NA61/SHINE} collaboration, A.~Aduszkiewicz et~al., \emph{{Production
  of $\Lambda $ -hyperons in inelastic p+p interactions at 158
  ${\mathrm{GeV}}\!/\!c$}},
  \href{http://dx.doi.org/10.1140/epjc/s10052-016-4003-2}{\emph{Eur. Phys. J.
  C} {\bf 76} (2016) 198}, [\href{http://arxiv.org/abs/1510.03720}{{\tt
  1510.03720}}].

\bibitem{Bauer:2021mvw}
M.~Bauer, M.~Neubert, S.~Renner, M.~Schnubel and A.~Thamm, \emph{{Flavor probes
  of axion-like particles}},  \href{http://arxiv.org/abs/2110.10698}{{\tt
  2110.10698}}.

\bibitem{Farina:2016tgd}
M.~Farina, D.~Pappadopulo, F.~Rompineve and A.~Tesi, \emph{{The photo-philic
  QCD axion}}, \href{http://dx.doi.org/10.1007/JHEP01(2017)095}{\emph{JHEP}
  {\bf 01} (2017) 095}, [\href{http://arxiv.org/abs/1611.09855}{{\tt
  1611.09855}}].

\bibitem{Hook:2018dlk}
A.~Hook, \emph{{TASI Lectures on the Strong CP Problem and Axions}},
  {\emph{PoS} {\bf TASI2018} (2019) 004},
  [\href{http://arxiv.org/abs/1812.02669}{{\tt 1812.02669}}].

\bibitem{Beacham:2019nyx}
J.~Beacham et~al., \emph{{Physics Beyond Colliders at CERN: Beyond the Standard
  Model Working Group Report}},
  \href{http://dx.doi.org/10.1088/1361-6471/ab4cd2}{\emph{J. Phys. G} {\bf 47}
  (2020) 010501}, [\href{http://arxiv.org/abs/1901.09966}{{\tt 1901.09966}}].

\bibitem{Scherer:2002tk}
S.~Scherer, \emph{{Introduction to chiral perturbation theory}}, {\emph{Adv.
  Nucl. Phys.} {\bf 27} (2003) 277},
  [\href{http://arxiv.org/abs/hep-ph/0210398}{{\tt hep-ph/0210398}}].

\bibitem{Scherer:2005ri}
S.~Scherer and M.~R. Schindler, \emph{{A Chiral perturbation theory primer}},
  \href{http://arxiv.org/abs/hep-ph/0505265}{{\tt hep-ph/0505265}}.

\bibitem{Aloni:2018vki}
D.~Aloni, Y.~Soreq and M.~Williams, \emph{{Coupling QCD-Scale Axionlike
  Particles to Gluons}},
  \href{http://dx.doi.org/10.1103/PhysRevLett.123.031803}{\emph{Phys. Rev.
  Lett.} {\bf 123} (2019) 031803}, [\href{http://arxiv.org/abs/1811.03474}{{\tt
  1811.03474}}].

\bibitem{Cheng:2021kjg}
H.-C. Cheng, L.~Li and E.~Salvioni, \emph{{A Theory of Dark Pions}},
  \href{http://arxiv.org/abs/2110.10691}{{\tt 2110.10691}}.

\bibitem{Harada:2003jx}
M.~Harada and K.~Yamawaki, \emph{{Hidden local symmetry at loop: A New
  perspective of composite gauge boson and chiral phase transition}},
  \href{http://dx.doi.org/10.1016/S0370-1573(03)00139-X}{\emph{Phys. Rept.}
  {\bf 381} (2003) 1--233}, [\href{http://arxiv.org/abs/hep-ph/0302103}{{\tt
  hep-ph/0302103}}].

\bibitem{Tanabashi:2018oca}
{\scshape Particle Data Group} collaboration, M.~Tanabashi et~al.,
  \emph{{Review of Particle Physics}},
  \href{http://dx.doi.org/10.1103/PhysRevD.98.030001}{\emph{Phys. Rev. D} {\bf
  98} (2018) 030001}.

\bibitem{Donoghue:1992dd}
J.~F. Donoghue, E.~Golowich and B.~R. Holstein, \emph{{Dynamics of the standard
  model}}, vol.~2.
\newblock CUP, 2014,
  \href{http://dx.doi.org/10.1017/CBO9780511524370}{10.1017/CBO9780511524370}.

\bibitem{Holstein:2001bt}
B.~R. Holstein, \emph{{Allowed eta decay modes and chiral symmetry}},
  \href{http://dx.doi.org/10.1238/Physica.Topical.099a00055}{\emph{Phys.
  Scripta T} {\bf 99} (2002) 55--67},
  [\href{http://arxiv.org/abs/hep-ph/0112150}{{\tt hep-ph/0112150}}].

\bibitem{Gori:2020xvq}
S.~Gori, G.~Perez and K.~Tobioka, \emph{{KOTO vs. NA62 Dark Scalar Searches}},
  \href{http://dx.doi.org/10.1007/JHEP08(2020)110}{\emph{JHEP} {\bf 08} (2020)
  110}, [\href{http://arxiv.org/abs/2005.05170}{{\tt 2005.05170}}].

\bibitem{GrillidiCortona:2015jxo}
G.~Grilli~di Cortona, E.~Hardy, J.~Pardo~Vega and G.~Villadoro, \emph{{The QCD
  axion, precisely}},
  \href{http://dx.doi.org/10.1007/JHEP01(2016)034}{\emph{JHEP} {\bf 01} (2016)
  034}, [\href{http://arxiv.org/abs/1511.02867}{{\tt 1511.02867}}].

\bibitem{Borasoy:1999nd}
B.~Borasoy, \emph{{The eta-prime in baryon chiral perturbation theory}},
  \href{http://dx.doi.org/10.1103/PhysRevD.61.014011}{\emph{Phys. Rev. D} {\bf
  61} (2000) 014011}, [\href{http://arxiv.org/abs/hep-ph/0001102}{{\tt
  hep-ph/0001102}}].

\bibitem{Bruns:2019fwi}
P.~C. Bruns and A.~Cieply, \emph{{Coupled channels approach to $\eta N$ and
  $\eta' N$ interactions}},
  \href{http://dx.doi.org/10.1016/j.nuclphysa.2019.121630}{\emph{Nucl. Phys. A}
  {\bf 992} (2019) 121630}, [\href{http://arxiv.org/abs/1903.10350}{{\tt
  1903.10350}}].

\bibitem{Close:1993mv}
F.~E. Close and R.~G. Roberts, \emph{{Consistent analysis of the spin content
  of the nucleon}},
  \href{http://dx.doi.org/10.1016/0370-2693(93)90673-6}{\emph{Phys. Lett. B}
  {\bf 316} (1993) 165--171}, [\href{http://arxiv.org/abs/hep-ph/9306289}{{\tt
  hep-ph/9306289}}].

\bibitem{Borasoy:1998pe}
B.~Borasoy, \emph{{Baryon axial currents}},
  \href{http://dx.doi.org/10.1103/PhysRevD.59.054021}{\emph{Phys. Rev. D} {\bf
  59} (1999) 054021}, [\href{http://arxiv.org/abs/hep-ph/9811411}{{\tt
  hep-ph/9811411}}].

\bibitem{Alexandrou:2020okk}
C.~Alexandrou et~al., \emph{{Nucleon axial and pseudoscalar form factors from
  lattice QCD at the physical point}},
  \href{http://dx.doi.org/10.1103/PhysRevD.103.034509}{\emph{Phys. Rev. D} {\bf
  103} (2021) 034509}, [\href{http://arxiv.org/abs/2011.13342}{{\tt
  2011.13342}}].

\bibitem{Alexandrou:2021wzv}
C.~Alexandrou, S.~Bacchio, M.~Constantinou, K.~Hadjiyiannakou, K.~Jansen and
  G.~Koutsou, \emph{{Quark flavor decomposition of the nucleon axial form
  factors}}, \href{http://dx.doi.org/10.1103/PhysRevD.104.074503}{\emph{Phys.
  Rev. D} {\bf 104} (2021) 074503},
  [\href{http://arxiv.org/abs/2106.13468}{{\tt 2106.13468}}].

\bibitem{Dobrich:2015jyk}
B.~D{\"o}brich, J.~Jaeckel, F.~Kahlhoefer, A.~Ringwald and K.~Schmidt-Hoberg,
  \emph{{ALPtraum: ALP production in proton beam dump experiments}},
  \href{http://dx.doi.org/10.1007/JHEP02(2016)018}{\emph{JHEP} {\bf 02} (2016)
  018}, [\href{http://arxiv.org/abs/1512.03069}{{\tt 1512.03069}}].

\bibitem{Aloni:2019ruo}
D.~Aloni, C.~Fanelli, Y.~Soreq and M.~Williams, \emph{{Photoproduction of
  Axionlike Particles}},
  \href{http://dx.doi.org/10.1103/PhysRevLett.123.071801}{\emph{Phys. Rev.
  Lett.} {\bf 123} (2019) 071801}, [\href{http://arxiv.org/abs/1903.03586}{{\tt
  1903.03586}}].

\bibitem{Bonesini:2001iz}
M.~Bonesini, A.~Marchionni, F.~Pietropaolo and T.~Tabarelli~de Fatis, \emph{{On
  Particle production for high-energy neutrino beams}},
  \href{http://dx.doi.org/10.1007/s100520100656}{\emph{Eur. Phys. J. C} {\bf
  20} (2001) 13--27}, [\href{http://arxiv.org/abs/hep-ph/0101163}{{\tt
  hep-ph/0101163}}].

\bibitem{Carvalho:2003pza}
J.~Carvalho, \emph{{Compilation of cross sections for proton nucleus
  interactions at the HERA energy}},
  \href{http://dx.doi.org/10.1016/S0375-9474(03)01597-5}{\emph{Nucl. Phys. A}
  {\bf 725} (2003) 269--275}.

\bibitem{RamanaMurthy:1975vfu}
P.~Ramana~Murthy, C.~A. Ayre, H.~Gustafson, L.~W. Jones and M.~J. Longo,
  \emph{{Neutron Total Cross-Sections on Nuclei at Fermilab Energies}},
  \href{http://dx.doi.org/10.1016/0550-3213(75)90182-0}{\emph{Nucl. Phys. B}
  {\bf 92} (1975) 269--308}.

\bibitem{Dusaev:2020gxi}
R.~R. Dusaev, D.~V. Kirpichnikov and M.~M. Kirsanov, \emph{{Photoproduction of
  axionlike particles in the NA64 experiment}},
  \href{http://dx.doi.org/10.1103/PhysRevD.102.055018}{\emph{Phys. Rev. D} {\bf
  102} (2020) 055018}, [\href{http://arxiv.org/abs/2004.04469}{{\tt
  2004.04469}}].

\bibitem{Lewin:1995rx}
J.~Lewin and P.~Smith, \emph{{Review of mathematics, numerical factors, and
  corrections for dark matter experiments based on elastic nuclear recoil}},
  \href{http://dx.doi.org/10.1016/S0927-6505(96)00047-3}{\emph{Astropart.
  Phys.} {\bf 6} (1996) 87--112}.

\bibitem{Harland-Lang:2019zur}
L.~Harland-Lang, J.~Jaeckel and M.~Spannowsky, \emph{{A fresh look at ALP
  searches in fixed target experiments}},
  \href{http://dx.doi.org/10.1016/j.physletb.2019.04.045}{\emph{Phys. Lett. B}
  {\bf 793} (2019) 281--289}, [\href{http://arxiv.org/abs/1902.04878}{{\tt
  1902.04878}}].

\bibitem{Budnev:1974de}
V.~Budnev, I.~Ginzburg, G.~Meledin and V.~Serbo, \emph{{The Two photon particle
  production mechanism. Physical problems. Applications. Equivalent photon
  approximation}},
  \href{http://dx.doi.org/10.1016/0370-1573(75)90009-5}{\emph{Phys. Rept.} {\bf
  15} (1975) 181--281}.

\bibitem{Bernauer:2013tpr}
{\scshape A1} collaboration, J.~Bernauer et~al., \emph{{Electric and magnetic
  form factors of the proton}},
  \href{http://dx.doi.org/10.1103/PhysRevC.90.015206}{\emph{Phys. Rev. C} {\bf
  90} (2014) 015206}, [\href{http://arxiv.org/abs/1307.6227}{{\tt 1307.6227}}].

\bibitem{Lepage:1977sw}
G.~Lepage, \emph{{A New Algorithm for Adaptive Multidimensional Integration}},
  \href{http://dx.doi.org/10.1016/0021-9991(78)90004-9}{\emph{J. Comput. Phys.}
  {\bf 27} (1978) 192}.

\bibitem{peter_lepage_2020_3897199}
P.~Lepage, \emph{gplepage/vegas: vegas version 3.4.5},  tech. rep., June, 2020,
  \url{https://doi.org/10.5281/zenodo.3897199}, 10.5281/zenodo.3897199.

\bibitem{Manohar:2016nzj}
A.~Manohar, P.~Nason, G.~P. Salam and G.~Zanderighi, \emph{{How bright is the
  proton? A precise determination of the photon parton distribution function}},
  \href{http://dx.doi.org/10.1103/PhysRevLett.117.242002}{\emph{Phys. Rev.
  Lett.} {\bf 117} (2016) 242002}, [\href{http://arxiv.org/abs/1607.04266}{{\tt
  1607.04266}}].

\bibitem{Manohar:2017eqh}
A.~V. Manohar, P.~Nason, G.~P. Salam and G.~Zanderighi, \emph{{The Photon
  Content of the Proton}},
  \href{http://dx.doi.org/10.1007/JHEP12(2017)046}{\emph{JHEP} {\bf 12} (2017)
  046}, [\href{http://arxiv.org/abs/1708.01256}{{\tt 1708.01256}}].

\bibitem{Alwall:2014hca}
J.~Alwall, R.~Frederix, S.~Frixione, V.~Hirschi, F.~Maltoni, O.~Mattelaer
  et~al., \emph{{The automated computation of tree-level and next-to-leading
  order differential cross sections, and their matching to parton shower
  simulations}}, \href{http://dx.doi.org/10.1007/JHEP07(2014)079}{\emph{JHEP}
  {\bf 07} (2014) 079}, [\href{http://arxiv.org/abs/1405.0301}{{\tt
  1405.0301}}].

\bibitem{Buckley:2014ana}
A.~Buckley, J.~Ferrando, S.~Lloyd, K.~Nordstr\"om, B.~Page, M.~R\"ufenacht
  et~al., \emph{{LHAPDF6: parton density access in the LHC precision era}},
  \href{http://dx.doi.org/10.1140/epjc/s10052-015-3318-8}{\emph{Eur. Phys. J.
  C} {\bf 75} (2015) 132}, [\href{http://arxiv.org/abs/1412.7420}{{\tt
  1412.7420}}].

\bibitem{Aielli:2019ivi}
G.~Aielli et~al., \emph{{Expression of interest for the CODEX-b detector}},
  \href{http://dx.doi.org/10.1140/epjc/s10052-020-08711-3}{\emph{Eur. Phys. J.
  C} {\bf 80} (2020) 1177}, [\href{http://arxiv.org/abs/1911.00481}{{\tt
  1911.00481}}].

\bibitem{Kelly:2020dda}
K.~J. Kelly, S.~Kumar and Z.~Liu, \emph{{Heavy axion opportunities at the DUNE
  near detector}},
  \href{http://dx.doi.org/10.1103/PhysRevD.103.095002}{\emph{Phys. Rev. D} {\bf
  103} (2021) 095002}, [\href{http://arxiv.org/abs/2011.05995}{{\tt
  2011.05995}}].

\bibitem{Foroughi-Abari:2021zbm}
S.~Foroughi-Abari and A.~Ritz, \emph{{Dark Sector Production via Proton
  Bremsstrahlung}},  \href{http://arxiv.org/abs/2108.05900}{{\tt 2108.05900}}.

\bibitem{Aguilar-Benitez:1991hzq}
M.~Aguilar-Benitez et~al., \emph{{Inclusive particle production in 400-GeV/c p
  p interactions}}, \href{http://dx.doi.org/10.1007/BF01551452}{\emph{Z. Phys.
  C} {\bf 50} (1991) 405--426}.

\bibitem{AxialFieldSpectrometer:1986awq}
{\scshape Axial Field Spectrometer} collaboration, T.~Akesson et~al.,
  \emph{{Inclusive $\eta$ Production at Low Transverse Momentum in 63-{GeV} $p
  p$ Collisions at the {CERN} Intersecting Storage Rings}},
  \href{http://dx.doi.org/10.1016/0370-2693(86)91409-7}{\emph{Phys. Lett. B}
  {\bf 178} (1986) 447}.

\bibitem{Boiarska:2019jym}
I.~Boiarska, K.~Bondarenko, A.~Boyarsky, V.~Gorkavenko, M.~Ovchynnikov and
  A.~Sokolenko, \emph{{Phenomenology of GeV-scale scalar portal}},
  \href{http://dx.doi.org/10.1007/JHEP11(2019)162}{\emph{JHEP} {\bf 11} (2019)
  162}, [\href{http://arxiv.org/abs/1904.10447}{{\tt 1904.10447}}].

\bibitem{Foroughi-Abari:2022aaa}
S.~Foroughi-Abari and A.~Ritz, \emph{{To Appear}}, .

\bibitem{Carroll:1978hc}
A.~S. Carroll et~al., \emph{{Absorption Cross-Sections of $\pi^{\pm}$,
  $K^{\pm}$, p and $\bar{p}$ on Nuclei Between 60 GeV/c and 280 GeV/c}},
  \href{http://dx.doi.org/10.1016/0370-2693(79)90226-0}{\emph{Phys. Lett. B}
  {\bf 80} (1979) 319--322}.

\bibitem{Feuster:1998cj}
T.~Feuster and U.~Mosel, \emph{{Photon and meson induced reactions on the
  nucleon}}, \href{http://dx.doi.org/10.1103/PhysRevC.59.460}{\emph{Phys. Rev.
  C} {\bf 59} (1999) 460--491},
  [\href{http://arxiv.org/abs/nucl-th/9803057}{{\tt nucl-th/9803057}}].

\bibitem{Penner:2002md}
G.~Penner and U.~Mosel, \emph{{Vector meson production and nucleon resonance
  analysis in a coupled channel approach for energies m(N) less than S**(1/2)
  less than 2-GeV. 2. Photon induced results}},
  \href{http://dx.doi.org/10.1103/PhysRevC.66.055212}{\emph{Phys. Rev. C} {\bf
  66} (2002) 055212}, [\href{http://arxiv.org/abs/nucl-th/0207069}{{\tt
  nucl-th/0207069}}].

\bibitem{Adamuscin:2007fk}
C.~Adamuscin, E.~Tomasi-Gustafsson, E.~Santopinto and R.~Bijker,
  \emph{{Two-component model for the axial form factor of the nucleon}},
  \href{http://dx.doi.org/10.1103/PhysRevC.78.035201}{\emph{Phys. Rev. C} {\bf
  78} (2008) 035201}, [\href{http://arxiv.org/abs/0706.3509}{{\tt 0706.3509}}].

\bibitem{Bijker:2004yu}
R.~Bijker and F.~Iachello, \emph{{Re-analysis of the nucleon space- and
  time-like electromagnetic form-factors in a two-component model}},
  \href{http://dx.doi.org/10.1103/PhysRevC.69.068201}{\emph{Phys. Rev. C} {\bf
  69} (2004) 068201}, [\href{http://arxiv.org/abs/nucl-th/0405028}{{\tt
  nucl-th/0405028}}].

\bibitem{Hou:2019efy}
T.-J. Hou et~al., \emph{{New CTEQ global analysis of quantum chromodynamics
  with high-precision data from the LHC}},
  \href{http://dx.doi.org/10.1103/PhysRevD.103.014013}{\emph{Phys. Rev. D} {\bf
  103} (2021) 014013}, [\href{http://arxiv.org/abs/1912.10053}{{\tt
  1912.10053}}].

\bibitem{Feng:2018pew}
J.~L. Feng, I.~Galon, F.~Kling and S.~Trojanowski, \emph{{Axionlike particles
  at FASER: The LHC as a photon beam dump}},
  \href{http://dx.doi.org/10.1103/PhysRevD.98.055021}{\emph{Phys. Rev. D} {\bf
  98} (2018) 055021}, [\href{http://arxiv.org/abs/1806.02348}{{\tt
  1806.02348}}].

\bibitem{Bai:2021dgm}
Z.~Bai et~al., \emph{{LUXE-NPOD: new physics searches with an optical dump at
  LUXE}},  \href{http://arxiv.org/abs/2107.13554}{{\tt 2107.13554}}.

\bibitem{Dolan:2017osp}
M.~J. Dolan, T.~Ferber, C.~Hearty, F.~Kahlhoefer and K.~Schmidt-Hoberg,
  \emph{{Revised constraints and Belle II sensitivity for visible and invisible
  axion-like particles}},
  \href{http://dx.doi.org/10.1007/JHEP12(2017)094}{\emph{JHEP} {\bf 12} (2017)
  094}, [\href{http://arxiv.org/abs/1709.00009}{{\tt 1709.00009}}].

\bibitem{FASER:2018eoc}
{\scshape FASER} collaboration, A.~Ariga et~al., \emph{{FASER\textquoteright{}s
  physics reach for long-lived particles}},
  \href{http://dx.doi.org/10.1103/PhysRevD.99.095011}{\emph{Phys. Rev. D} {\bf
  99} (2019) 095011}, [\href{http://arxiv.org/abs/1811.12522}{{\tt
  1811.12522}}].

\bibitem{Brdar:2020dpr}
V.~Brdar, B.~Dutta, W.~Jang, D.~Kim, I.~M. Shoemaker, Z.~Tabrizi et~al.,
  \emph{{Axionlike Particles at Future Neutrino Experiments: Closing the
  Cosmological Triangle}},
  \href{http://dx.doi.org/10.1103/PhysRevLett.126.201801}{\emph{Phys. Rev.
  Lett.} {\bf 126} (2021) 201801}, [\href{http://arxiv.org/abs/2011.07054}{{\tt
  2011.07054}}].

\bibitem{Abbiendi:2002je}
{\scshape OPAL} collaboration, G.~Abbiendi et~al., \emph{{Multiphoton
  production in e+ e- collisions at s**(1/2) = 181-GeV to 209-GeV}},
  \href{http://dx.doi.org/10.1140/epjc/s2002-01074-5}{\emph{Eur. Phys. J. C}
  {\bf 26} (2003) 331--344}, [\href{http://arxiv.org/abs/hep-ex/0210016}{{\tt
  hep-ex/0210016}}].

\bibitem{Aad:2014ioa}
{\scshape ATLAS} collaboration, G.~Aad et~al., \emph{{Search for Scalar
  Diphoton Resonances in the Mass Range $65-600$ GeV with the ATLAS Detector in
  $pp$ Collision Data at $\sqrt{s}$ = 8 $TeV$}},
  \href{http://dx.doi.org/10.1103/PhysRevLett.113.171801}{\emph{Phys. Rev.
  Lett.} {\bf 113} (2014) 171801}, [\href{http://arxiv.org/abs/1407.6583}{{\tt
  1407.6583}}].

\bibitem{Aad:2015bua}
{\scshape ATLAS} collaboration, G.~Aad et~al., \emph{{Search for new phenomena
  in events with at least three photons collected in $pp$ collisions at
  $\sqrt{s}$ = 8 TeV with the ATLAS detector}},
  \href{http://dx.doi.org/10.1140/epjc/s10052-016-4034-8}{\emph{Eur. Phys. J.
  C} {\bf 76} (2016) 210}, [\href{http://arxiv.org/abs/1509.05051}{{\tt
  1509.05051}}].

\bibitem{Chatrchyan:2012tv}
{\scshape CMS} collaboration, S.~Chatrchyan et~al., \emph{{Search for Exclusive
  or Semi-Exclusive Photon Pair Production and Observation of Exclusive and
  Semi-Exclusive Electron Pair Production in $pp$ Collisions at $\sqrt{s}=7$
  TeV}}, \href{http://dx.doi.org/10.1007/JHEP11(2012)080}{\emph{JHEP} {\bf 11}
  (2012) 080}, [\href{http://arxiv.org/abs/1209.1666}{{\tt 1209.1666}}].

\bibitem{Knapen:2016moh}
S.~Knapen, T.~Lin, H.~K. Lou and T.~Melia, \emph{{Searching for Axionlike
  Particles with Ultraperipheral Heavy-Ion Collisions}},
  \href{http://dx.doi.org/10.1103/PhysRevLett.118.171801}{\emph{Phys. Rev.
  Lett.} {\bf 118} (2017) 171801}, [\href{http://arxiv.org/abs/1607.06083}{{\tt
  1607.06083}}].

\bibitem{Liu:2021lan}
Y.~Liu and B.~Yan, \emph{{Searching for the axion-like particle at the EIC}},
  \href{http://arxiv.org/abs/2112.02477}{{\tt 2112.02477}}.

\bibitem{Jaeckel:2015jla}
J.~Jaeckel and M.~Spannowsky, \emph{{Probing MeV to 90 GeV axion-like particles
  with LEP and LHC}},
  \href{http://dx.doi.org/10.1016/j.physletb.2015.12.037}{\emph{Phys. Lett. B}
  {\bf 753} (2016) 482--487}, [\href{http://arxiv.org/abs/1509.00476}{{\tt
  1509.00476}}].

\bibitem{BelleII:2020fag}
{\scshape Belle-II} collaboration, F.~Abudinen, \emph{{Search for Axion-Like
  Particles produced in $e^+e^-$ collisions at Belle II}},
  \href{http://arxiv.org/abs/2007.13071}{{\tt 2007.13071}}.

\bibitem{Chou:2016lxi}
J.~P. Chou, D.~Curtin and H.~J. Lubatti, \emph{{New Detectors to Explore the
  Lifetime Frontier}},
  \href{http://dx.doi.org/10.1016/j.physletb.2017.01.043}{\emph{Phys. Lett. B}
  {\bf 767} (2017) 29--36}, [\href{http://arxiv.org/abs/1606.06298}{{\tt
  1606.06298}}].

\bibitem{Bertholet:2021hjl}
E.~Bertholet, S.~Chakraborty, V.~Loladze, T.~Okui, A.~Soffer and K.~Tobioka,
  \emph{{Heavy QCD Axion at Belle II: Displaced and Prompt Signals}},
  \href{http://arxiv.org/abs/2108.10331}{{\tt 2108.10331}}.

\bibitem{CidVidal:2018blh}
X.~Cid~Vidal, A.~Mariotti, D.~Redigolo, F.~Sala and K.~Tobioka, \emph{{New
  Axion Searches at Flavor Factories}},
  \href{http://dx.doi.org/10.1007/JHEP01(2019)113}{\emph{JHEP} {\bf 01} (2019)
  113}, [\href{http://arxiv.org/abs/1810.09452}{{\tt 1810.09452}}].

\bibitem{Mariotti:2017vtv}
A.~Mariotti, D.~Redigolo, F.~Sala and K.~Tobioka, \emph{{New LHC bound on
  low-mass diphoton resonances}},
  \href{http://dx.doi.org/10.1016/j.physletb.2018.06.039}{\emph{Phys. Lett. B}
  {\bf 783} (2018) 13--18}, [\href{http://arxiv.org/abs/1710.01743}{{\tt
  1710.01743}}].

\bibitem{CMS-PAS-HIG-14-037}
{\scshape CMS} collaboration, \emph{{Search for new resonances in the diphoton
  final state in the mass range between 80 and 110 GeV in pp collisions at
  $\sqrt{s}=8$ TeV}},  Tech. Rep. CMS-PAS-HIG-14-037, CERN, Geneva, 2015,
  \url{https://cds.cern.ch/record/2063739}.

\bibitem{CMS-PAS-HIG-17-013}
{\scshape CMS Collaboration} collaboration, \emph{{Search for new resonances in
  the diphoton final state in the mass range between 70 and 110 GeV in pp
  collisions at $\sqrt{s}=$ 8 and 13 TeV}},  Tech. Rep. CMS-PAS-HIG-17-013,
  CERN, Geneva, 2017, \url{https://cds.cern.ch/record/2285326}.

\bibitem{CMS:2017dcz}
{\scshape CMS} collaboration, A.~M. Sirunyan et~al., \emph{{Search for low mass
  vector resonances decaying into quark-antiquark pairs in proton-proton
  collisions at $ \sqrt{s}=13 $ TeV}},
  \href{http://dx.doi.org/10.1007/JHEP01(2018)097}{\emph{JHEP} {\bf 01} (2018)
  097}, [\href{http://arxiv.org/abs/1710.00159}{{\tt 1710.00159}}].

\bibitem{Knapen:2021elo}
S.~Knapen, S.~Kumar and D.~Redigolo, \emph{{Searching for axion-like particles
  with data scouting at ATLAS and CMS}},
  \href{http://arxiv.org/abs/2112.07720}{{\tt 2112.07720}}.

\bibitem{Chakraborty:2021wda}
S.~Chakraborty, M.~Kraus, V.~Loladze, T.~Okui and K.~Tobioka, \emph{{Heavy QCD
  axion in b\textrightarrow{}s transition: Enhanced limits and projections}},
  \href{http://dx.doi.org/10.1103/PhysRevD.104.055036}{\emph{Phys. Rev. D} {\bf
  104} (2021) 055036}, [\href{http://arxiv.org/abs/2102.04474}{{\tt
  2102.04474}}].

\bibitem{KOTO:2018dsc}
{\scshape KOTO} collaboration, J.~K. Ahn et~al., \emph{{Search for the $K_L
  \!\to\! \pi^0 \nu \overline{\nu}$ and $K_L \!\to\! \pi^0 X^0$ decays at the
  J-PARC KOTO experiment}},
  \href{http://dx.doi.org/10.1103/PhysRevLett.122.021802}{\emph{Phys. Rev.
  Lett.} {\bf 122} (2019) 021802}, [\href{http://arxiv.org/abs/1810.09655}{{\tt
  1810.09655}}].

\bibitem{Masuda:2015eta}
T.~Masuda et~al., \emph{{Long-lived neutral-kaon flux measurement for the KOTO
  experiment}}, \href{http://dx.doi.org/10.1093/ptep/ptv171}{\emph{PTEP} {\bf
  2016} (2016) 013C03}, [\href{http://arxiv.org/abs/1509.03386}{{\tt
  1509.03386}}].

\bibitem{NA62:2021zjw}
{\scshape NA62} collaboration, E.~Cortina~Gil et~al., \emph{{Measurement of the
  very rare K$^{+}$\textrightarrow{}$ {\pi}^{+}\nu \overline{\nu} $ decay}},
  \href{http://dx.doi.org/10.1007/JHEP06(2021)093}{\emph{JHEP} {\bf 06} (2021)
  093}, [\href{http://arxiv.org/abs/2103.15389}{{\tt 2103.15389}}].

\bibitem{NA62:2014ybm}
{\scshape NA62} collaboration, C.~Lazzeroni et~al., \emph{{Study of the
  $K^\pm\to\pi^\pm\gamma\gamma$ decay by the NA62 experiment}},
  \href{http://dx.doi.org/10.1016/j.physletb.2014.03.016}{\emph{Phys. Lett. B}
  {\bf 732} (2014) 65--74}, [\href{http://arxiv.org/abs/1402.4334}{{\tt
  1402.4334}}].

\bibitem{ATLAS:2021kxv}
{\scshape ATLAS} collaboration, G.~Aad et~al., \emph{{Search for new phenomena
  in events with an energetic jet and missing transverse momentum in $pp$
  collisions at $\sqrt {s}$ =13 TeV with the ATLAS detector}},
  \href{http://dx.doi.org/10.1103/PhysRevD.103.112006}{\emph{Phys. Rev. D} {\bf
  103} (2021) 112006}, [\href{http://arxiv.org/abs/2102.10874}{{\tt
  2102.10874}}].

\bibitem{Mimasu:2014nea}
K.~Mimasu and V.~Sanz, \emph{{ALPs at Colliders}},
  \href{http://dx.doi.org/10.1007/JHEP06(2015)173}{\emph{JHEP} {\bf 06} (2015)
  173}, [\href{http://arxiv.org/abs/1409.4792}{{\tt 1409.4792}}].

\bibitem{Brivio:2017ije}
I.~Brivio, M.~B. Gavela, L.~Merlo, K.~Mimasu, J.~M. No, R.~del Rey et~al.,
  \emph{{ALPs Effective Field Theory and Collider Signatures}},
  \href{http://dx.doi.org/10.1140/epjc/s10052-017-5111-3}{\emph{Eur. Phys. J.
  C} {\bf 77} (2017) 572}, [\href{http://arxiv.org/abs/1701.05379}{{\tt
  1701.05379}}].

\bibitem{Blumlein:1990ay}
J.~Blumlein et~al., \emph{{Limits on neutral light scalar and pseudoscalar
  particles in a proton beam dump experiment}},
  \href{http://dx.doi.org/10.1007/BF01548556}{\emph{Z. Phys. C} {\bf 51} (1991)
  341--350}.

\bibitem{Blumlein:1991xh}
J.~Blumlein et~al., \emph{{Limits on the mass of light (pseudo)scalar particles
  from Bethe-Heitler e+ e- and mu+ mu- pair production in a proton - iron beam
  dump experiment}},
  \href{http://dx.doi.org/10.1142/S0217751X9200171X}{\emph{Int. J. Mod. Phys.
  A} {\bf 7} (1992) 3835--3850}.

\bibitem{CHARM:1985anb}
{\scshape CHARM} collaboration, F.~Bergsma et~al., \emph{{Search for Axion Like
  Particle Production in 400-{GeV} Proton - Copper Interactions}},
  \href{http://dx.doi.org/10.1016/0370-2693(85)90400-9}{\emph{Phys. Lett. B}
  {\bf 157} (1985) 458--462}.

\bibitem{Blumlein:2013cua}
J.~Bl\"umlein and J.~Brunner, \emph{{New Exclusion Limits on Dark Gauge Forces
  from Proton Bremsstrahlung in Beam-Dump Data}},
  \href{http://dx.doi.org/10.1016/j.physletb.2014.02.029}{\emph{Phys. Lett. B}
  {\bf 731} (2014) 320--326}, [\href{http://arxiv.org/abs/1311.3870}{{\tt
  1311.3870}}].

\bibitem{Gninenko:2012eq}
S.~N. Gninenko, \emph{{Constraints on sub-GeV hidden sector gauge bosons from a
  search for heavy neutrino decays}},
  \href{http://dx.doi.org/10.1016/j.physletb.2012.06.002}{\emph{Phys. Lett. B}
  {\bf 713} (2012) 244--248}, [\href{http://arxiv.org/abs/1204.3583}{{\tt
  1204.3583}}].

\bibitem{Altmannshofer:2019yji}
W.~Altmannshofer, S.~Gori and D.~J. Robinson, \emph{{Constraining axionlike
  particles from rare pion decays}},
  \href{http://dx.doi.org/10.1103/PhysRevD.101.075002}{\emph{Phys. Rev. D} {\bf
  101} (2020) 075002}, [\href{http://arxiv.org/abs/1909.00005}{{\tt
  1909.00005}}].

\bibitem{PIENU:2017wbj}
{\scshape PIENU} collaboration, A.~Aguilar-Arevalo et~al., \emph{{Improved
  search for heavy neutrinos in the decay $\pi\rightarrow e\nu$}},
  \href{http://dx.doi.org/10.1103/PhysRevD.97.072012}{\emph{Phys. Rev. D} {\bf
  97} (2018) 072012}, [\href{http://arxiv.org/abs/1712.03275}{{\tt
  1712.03275}}].

\bibitem{Pocanic:2003pf}
D.~Pocanic et~al., \emph{{Precise measurement of the pi+ ---\ensuremath{>} pi0
  e+ nu branching ratio}},
  \href{http://dx.doi.org/10.1103/PhysRevLett.93.181803}{\emph{Phys. Rev.
  Lett.} {\bf 93} (2004) 181803},
  [\href{http://arxiv.org/abs/hep-ex/0312030}{{\tt hep-ex/0312030}}].

\bibitem{GlueX:2021myx}
{\scshape GlueX} collaboration, S.~Adhikari et~al., \emph{{Search for
  photoproduction of axion-like particles at GlueX}},
  \href{http://arxiv.org/abs/2109.13439}{{\tt 2109.13439}}.

\bibitem{Kim:1979if}
J.~E. Kim, \emph{{Weak Interaction Singlet and Strong CP Invariance}},
  \href{http://dx.doi.org/10.1103/PhysRevLett.43.103}{\emph{Phys. Rev. Lett.}
  {\bf 43} (1979) 103}.

\bibitem{Shifman:1979if}
M.~A. Shifman, A.~I. Vainshtein and V.~I. Zakharov, \emph{{Can Confinement
  Ensure Natural CP Invariance of Strong Interactions?}},
  \href{http://dx.doi.org/10.1016/0550-3213(80)90209-6}{\emph{Nucl. Phys. B}
  {\bf 166} (1980) 493--506}.

\bibitem{DiLuzio:2016sbl}
L.~Di~Luzio, F.~Mescia and E.~Nardi, \emph{{Redefining the Axion Window}},
  \href{http://dx.doi.org/10.1103/PhysRevLett.118.031801}{\emph{Phys. Rev.
  Lett.} {\bf 118} (2017) 031801}, [\href{http://arxiv.org/abs/1610.07593}{{\tt
  1610.07593}}].

\bibitem{DiLuzio:2017pfr}
L.~Di~Luzio, F.~Mescia and E.~Nardi, \emph{{Window for preferred axion
  models}}, \href{http://dx.doi.org/10.1103/PhysRevD.96.075003}{\emph{Phys.
  Rev. D} {\bf 96} (2017) 075003}, [\href{http://arxiv.org/abs/1705.05370}{{\tt
  1705.05370}}].

\bibitem{Patel:2015tea}
H.~H. Patel, \emph{{Package-X: A Mathematica package for the analytic
  calculation of one-loop integrals}},
  \href{http://dx.doi.org/10.1016/j.cpc.2015.08.017}{\emph{Comput. Phys.
  Commun.} {\bf 197} (2015) 276--290},
  [\href{http://arxiv.org/abs/1503.01469}{{\tt 1503.01469}}].

\bibitem{harris2020array}
C.~R. Harris, K.~J. Millman, S.~J. van~der Walt, R.~Gommers, P.~Virtanen,
  D.~Cournapeau et~al., \emph{Array programming with {NumPy}},
  \href{http://dx.doi.org/10.1038/s41586-020-2649-2}{\emph{Nature} {\bf 585}
  (Sept., 2020) 357--362}.

\bibitem{2020SciPy-NMeth}
P.~Virtanen, R.~Gommers, T.~E. Oliphant, M.~Haberland, T.~Reddy, D.~Cournapeau
  et~al., \emph{{{SciPy} 1.0: Fundamental Algorithms for Scientific Computing
  in Python}}, \href{http://dx.doi.org/10.1038/s41592-019-0686-2}{\emph{Nature
  Methods} {\bf 17} (2020) 261--272}.

\bibitem{Hunter:2007}
J.~D. Hunter, \emph{Matplotlib: A 2d graphics environment},
  \href{http://dx.doi.org/10.1109/MCSE.2007.55}{\emph{Computing in Science \&
  Engineering} {\bf 9} (2007) 90--95}.

\bibitem{Peskin:1995ev}
M.~E. Peskin and D.~V. Schroeder, \emph{{An Introduction to quantum field
  theory}}.
\newblock Addison-Wesley, Reading, USA, 1995.

\bibitem{Witten:1980sp}
E.~Witten, \emph{{Large N Chiral Dynamics}},
  \href{http://dx.doi.org/10.1016/0003-4916(80)90325-5}{\emph{Annals Phys.}
  {\bf 128} (1980) 363}.

\bibitem{Gan:2020aco}
L.~Gan, B.~Kubis, E.~Passemar and S.~Tulin, \emph{{Precision tests of
  fundamental physics with $\eta$ and $\eta^\prime$ mesons}},
  \href{http://arxiv.org/abs/2007.00664}{{\tt 2007.00664}}.

\end{thebibliography}\endgroup
\end{document}